\documentclass[twocolumn]{aastex631}
\usepackage{amsmath}
\usepackage{graphicx}
\usepackage{subfigure}
\usepackage{booktabs}
\DeclareUnicodeCharacter{0301}{\'{e}}

\usepackage{xcolor}

\received{March 12, 2024}
\revised{May 4, 2024}
\accepted{May 8, 2024}

\shorttitle{Free-floating planets}

\shortauthors{Yu and Lai}

\graphicspath{{./}{figures/}}

\begin{document}
\title{Free-Floating Planets, Survivor Planets, Captured Planets and Binary Planets from Stellar Flybys}
\correspondingauthor{Fangyuan Yu}
\email{yufangyuan@sjtu.edu.cn}
\correspondingauthor{Dong Lai}
\email{dong@astro.cornell.edu}

\author[0009-0004-6973-3955]{Fangyuan Yu}
\affiliation{Zhiyuan College, Shanghai Jiao Tong University, 800 Dongchuan Road, Minhang, Shanghai 200240, China}
\affiliation{Department of Astronomy, School of Physics and Astronomy, Shanghai Jiao Tong University, 800 Dongchuan Road, Minhang, Shanghai 200240, China}

\author[0000-0002-1934-6250]{Dong Lai}
\affiliation{Center for Astrophysics and Planetary Science, Department of Astronomy, Cornell University, Ithaca, NY 14853, USA}
\affiliation{Tsung-Dao Lee Institute, Shanghai Jiao Tong University, Shanghai, 520 Shengrong Road, 201210, China}

\begin{abstract}
In star clusters, close stellar encounters can strongly impact the architecture of a planetary system or even destroy it. We present a systematic study on the effects of stellar flybys on two-planet systems.
When such a system experiences flybys, one or both planets can be ejected, forming free-floating planets (FFPs), captured planets (CPs) around the flyby star, and free-floating binary planets (BPs); the remaining single-surviving-planets (SSPs) can have their orbital radii and eccentricities greatly changed.
Through numerical experiments, we calculate the formation fractions (or branching ratios) of FFPs, SSPs, CPs and BPs as a function of the pericenter distance of the flyby, and use them to derive analytical expressions for the formation rates of FFPs, SSPs, CPs and BPs in general cluster environments.
We find that the production rates of FFPs and SSPs are similar
(for initial planet semi-major axis ratio $a_1/a_2=0.6-0.8$),
while the rate for CPs is a few times smaller.
The formation fraction of BPs depends strongly on $a_1/a_2$ and on the planet masses.
For Jupiter-mass planets, the formation fraction of BPs is always less than $1\%$ (for $a_1/a_2=0.8$) and typically much smaller ($\lesssim 0.2\%$ for $a_1/a_2\lesssim 0.7$). The fraction remains less than $1\%$ when considering $4M_{\rm J}$ planets.
Overall, when averaging over all flybys, the production rate of BPs is less than $0.1\%$ of that for FFPs. 
We also derive the velocity distribution of FFPs produced by stellar flybys, and the orbital parameter distributions of SSPs, CPs and BPs. 
These results can be used in future studies of exotic planets (including FFPs) and planetary systems.
\end{abstract}


\section{Introduction}
\label{sec: intro}

Free-floating planets (FFPs) of a few Jupiter masses were first discovered more than 20 years ago in nearby star-forming regions \citep{Lucas2000MNRAS,Osorio2000Sci,Luhman2004ApJ}. 
Since then, many FFPs have been detected by imaging and spectroscopic observations in various young star clusters \citep{Scholz2012ApJ,Ramirez2012ApJ,Miret-Roig2022NatAstro,Bouy2022aap} and by microlensing surveys toward the Galactic bulge \citep{Mróz2020ApJL,Ryu2021AJ}.
The imaging observations probe FFPs of a Jupiter mass or more, while the microlensing surveys have found FFPs down to a few Earth masses.

In a recent near-infrared survey of the Trapezium Cluster in the inner Orion Nebula using JWST, \citet{Pearson2023arXiv01} discovered a large population of 540 FFP candidates.
These FFPs have estimated individual mass in the range of 0.7-13 Jupiter mass ($M_{\rm J}$), based on the cluster age ($\sim 1$ Myrs) and thermal evolution models of giant planets. 
Surprisingly, among the 540 objects, there are 40 binaries (called Jupiter-mass binary objects, or JuMBOs) and two triple systems. The binaries have separation in the plane of the sky between 25 and 390 au.

There are various possible production channels for the FFPs. The most natural one is the ejection channel. 
A multi-planet system with sufficiently small orbital spacings is dynamically unstable \citep{Gladman1993Icar,Zhou2007ApJ,Petit2018aap}. 
A possible outcome of this instability is the ejection of planets from the system due to strong gravitational scatterings;
ejection is most likely in the outer planetary region, where the orbital velocity is smaller than the escape speed from the planet \citep[e.g.,][]{Chatterjee2008ApJ,Jurić2008ApJ,Nagasawa2011ApJ,Petrovich2014ApJ,Anderson2020MNRAS,Li2021MNRAS}.

Alternatively, FFPs could in principle form as a scaled-down version of star formation, via gravitational collapse and fragmentation of molecular cloud cores \citep{Padoan2002ApJ,Hennebelle2008ApJ}. 
While this is likely for brown dwarfs of 10's of Jupiter masses, it is not clear whether the same collapse mechanisms can produce a large number of Jupiter-mass FFPs \citep[e.g.,][]{Low1976MNRAS,Whitworth2006aap}.

The claimed detection of a large fraction (9 percent) of JuMBOs among FFPs \citep{Pearson2023arXiv01} seems to suggest that core collapse and fragmentation (i.e. scaled-down star formation) channel plays an important role in producing FFPs down to Jupiter masses, since we do not expect the ejection channel to produce binary planets. 
On the other hand, \citet{Miret-Roig2022NatAstro} suggested that the observed abundance of FFPs in young star clusters significantly exceeds the core-collapse model predictions, indicating that ejections of giant planets must be frequent within the first 10 Myr of a planetary system's life.

A variant of the ejection scenario involves stellar flybys: In a dense star cluster, close encounters between a planetary system and other stars can be frequent, and these encounters change the system architecture and possibly lead to planet ejections \citep{Laughlin1998ApJ,Malmberg2011MNRAS,Parker2012MNRAS,Cai2017MNRAS,van2019aap,Li2020MNRAS,Rodet2021ApJ,Rodet2022MNRAS}.

Recently, \citet{Wang2024arXiv02} suggested that bound binary planets could be produced by stellar flybys. 
Based on a large suite of $N$-body simulations, they claimed
that the ratio of the cross-section of binary planet production $\sigma_{\rm BP}$ to that of single planet ejection $\sigma_{\rm FFP}$ can approach $\sim 20\%$ under certain conditions. In this paper we find this ratio to be much smaller [cf. \citep{Portegies2024arXiv01}].

This paper is devoted to a systematic study of the effects of stellar flybys on two-planet systems. 
While many previous works (see above) have considered planetary systems in `realistic' cluster environments, our work seeks to obtain, through numerical experiments, the scale-free ``microphysics" results for the effects of stellar flybys, so that these results can be used in generic situations. 
For a single-planet system, \citet{Rodet2021ApJ} has given some of the ``scale-free" results (e.g. the probability of planet ejection as a function of the dimensionless impact parameter of the flyby; see their Fig.~4). 
For a two-planet system experiencing stellar flybys, there are different possible outcomes, including the production of single free-floating planets (FFPs), (rare) free-floating binary planets (BPs), and captured planets (CPs) by the flyby star. 
We are interested in the probabilities (or branching ratios) of these different outcomes, as well as the questions such as
(a) what is the velocity distribution of single FFPs produced by stellar flybys?
(b) what are the orbital parameter distributions of the single-surviving-planet systems (after the other has been ejected), the captured planetary systems and the (rare) free-floating binary planet systems?

Our paper is organized as follows. 
In Section~\ref{sec: method} we describe the setup and method of our numerical experiments. 
In Section~\ref{sec: numerical results} we present the results of our fiducial runs -- some of the results can be rescaled to different parameters. 
In Section~\ref{sec: additional runs} we present some additional results that provide a (limited) survey of parameter space. 
In Section~\ref{sec: production rates} we apply the ``microphysics" results of Sections \ref{sec: numerical results}-\ref{sec: additional runs} to analytical expression for the production rates of FFPs, CPs and BPs, and we conclude in Section~\ref{sec: summary and discussion}.

\section{Method and Setup}
\label{sec: method}

We use the $N$-body code \texttt{REBOUND} with the \texttt{IAS15} integrator \citep{rebound,reboundias15} to simulate the close encounter between a planetary system and a flyby star.

\subsection{Simulation setup}
\label{sec: method_setup}

We consider a planetary system around star $M_1$ with two planets $m_1$ and $m_2$, initially on coplanar, circular orbits with semi-major axes $a_1$ and $a_2$. 
This system is perturbed by a flyby star $M_2$, on a slightly hyperbolic orbit relative to $M_1$.
The initial planetary orbits satisfy the stability condition \citep{Gladman1993Icar}:
\begin{equation}
a_2-a_1 > 2\sqrt{3}\,R_{\rm Hill},
\label{eq:stability}
\end{equation}
where $R_{\rm Hill}$ is the Hill radius:
\begin{equation}
R_{\rm Hill}=\frac{a_1+a_2}{2}\,\Bigl(\frac{m_1+m_2}{3M_1}\Bigr)^{1/3}.
\end{equation}

Throughout the paper, unless otherwise noted, we adopt the fiducial values for the masses $M_1 = M_2\equiv M= M_\odot$ and $m_1 = m_2 \equiv m= M_{\rm J}$, although many of our results can be scaled appropriately as long as $m\ll M$. 
We adopt $a_1/a_2=0.7$ in most of our simulations; this $a_1/a_2$ value satisfies the stability criterion (Eq.~(\ref{eq:stability})).
We also set the radii of the stars and the planets to $1\, R_\odot$ and $1\, R_{\rm J}$, respectively. 
These radius values do not affect our main results since the probabilities of physical collisions between planets and stars are very low.

The flyby trajectory is specified by the pericenter distance $q$ and eccentricity $e>1$.
For stars $M_1$ and $M_2$ approaching each other with relative velocity $V_\infty$ \citep[typically $\sim 1\, \rm km/s$ for young star clusters; see][]{Wright2018MNRAS}, the velocity at periastron $q$ is given by 
\begin{equation}
\begin{aligned}
& V_{\rm peri}^2=V_\infty^2+\frac{2GM_{12}}{q}\\
&=V_\infty^2+(8.4\,{\rm km\,s^{-1}})^2\Bigl(\frac{50\,\rm au}{q}\Bigr)\Bigl(\frac{M_{12}}{2M_\odot}\Bigr),
\end{aligned}
\end{equation}
where $M_{12}=M_1+M_2$. 
Under typical conditions, the gravitational focusing term dominates, and $V_{\rm peri}\gg V_\infty$. 
This implies that the stellar orbit is nearly parabolic with $e\simeq 1$. 
In our fiducial simulations, we adopt $e=1.1$ for all encounters (Section~\ref{sec: numerical results}).

The simulation covers a time span from 50\,$t_0$ before the periastron passage of the flyby (approximately 30\,$a_2$ away) to 500\,$t_0$ after (approximately 500\,$a_2$ away), where $t_0$ is defined as $t_0=\sqrt{a_2^3/(GM)}$. 
The simulation utilizes an adaptive time-step.

Aside from the mass ratio $m/M$, the semi-axis ratio $a_1/a_2$ and the dimensionless periastron distance ${\tilde q}\equiv q/a_2$, the outcome of a flyby depends on various angles, including the inclination angle ($i$) between the flyby orbital plane and the initial planetary orbital plane, the argument of periastron ($\omega$) of the flyby orbit, and the initial orbital phases ($\lambda_1$ and $\lambda_2$) of the two planets. 
Note that since the initial planetary orbits are circular, the outcome of an encounter does not depend on $\Omega$, the longitude of the node of the flyby orbit. 
In our simulations, we sample the dimensionless pericenter distance ${\tilde q} = q/a_2$ across a uniform grid ranging from $0.1$ to $2$, and include a detailed exploration of cases with $\tilde q<0.1$. 
For the four angles, we sample
\begin{enumerate}
    \item $i$, with a $\sin (i)$ probability distribution, between 0 and $\pi$, or equivalently, uniform distribution in $\cos (i)$,
    \item $\omega$, with a flat prior between 0 and 2$\pi$,
    \item $\lambda_1$, with a flat prior between 0 and 2$\pi$,
    \item $\lambda_2$, with a flat prior between 0 and 2$\pi$.
\end{enumerate}
[This approach is similar to \citet{Rodet2021ApJ}, who studied the effects of flybys on single-planet systems.]
For each value of $\tilde q$, our fiducial simulations sample $20 \times 20 \times 20 \times 20$ angles (for $i$, $\omega$, $\lambda_1$, and $\lambda_2$) to determine the distributions of outcomes for the post-flyby systems.

\subsection{Outcomes of planets after flyby and branching criteria}
\label{sec: method_criteria}

Following a stellar flyby, both the inner and outer planets may be ejected from their original star system, or they may remain bound within the original system. 
In the case of ejection, the expelled planet could either become a free-floating planet or be captured by the flyby star $M_2$.
If both planets are simultaneously ejected, it is also possible to form a pair of free-floating binary planets, also known as JuMBOs, which is of particular interest in our study. 
We are interested in the branching ratios of these different outcomes and the orbital properties of these various planetary systems after a stellar flyby.

At the end of each simulation run, all planets can be categorized into four types: 
free-floating planets (FFPs), planets captured by the flyby star (``captured planets", CPs), survivors that stay in the original system (``survivor planets", SPs), and free-floating binary planets (BPs). 
It is crucial for us to establish appropriate criteria for distinguishing between these categories. 
After extensive tests, we set our branching criteria at the end of each simulation run as follows:
\begin{itemize}
    \item {\it Free-floating planet} (FFP): the separations between the planet and both stars exceed 50\,$a_2$, and the total energy relative to both stars is positive, i.e.
    \begin{equation}
        \varepsilon_i\equiv \frac{1}{2} ({\bf v}-{\bf V_i})^2-\frac{GM_i}{r_i}>0,
    \end{equation}
    where ${\bf v}$ is the planet's velocity, ${\bf V}_i$ is the velocity of star $M_i$ ($i=1,2$), and $r_i$ is the separation between the planet and star $M_i$.
    \item {\it Captured planet} (CP): the separation between the planet and the initial star $M_1$ exceeds 50\,$a_2$, while the separation between the planet and the flyby $M_2$ is less than 50\,$a_2$.
    Its energy is positive relative to the original star ($\varepsilon_1>0$) and negative relative to the flyby star ($\varepsilon_2<0$).
    \item {\it Survivor planet} (SP): the separation between the planet and the initial star $M_1$ is less than 50\,$a_2$, and its energy is negative relative to the original star ($\varepsilon_1<0$). 
    We are particularly interested in the {\it single-surviving-planet} (SSP) system (i.e. one of the original planets is ejected from the original system, leaving behind a single survivor planet).
    \item {\it Free-floating binary planets} (BP): the separations between each planet and star exceed 50\,$a_2$; the separation between the planets is less than half of the ``instantaneous" Hill radius $r_{\rm Hill}=(2m/3M)^{1/3}\,d$, where $d$ is the minimum of the separations between each planet and star; the relative energy between the two planets is negative, i.e.
    \begin{equation}
        \frac{1}{2}{\bf v}_{12}^2-\frac{G(m_1+m_2)}{r_{12}}<0,
        \label{Eq:inertial}
    \end{equation}
    where ${\bf v}_{12}$ is the relative velocity between the two planets and $r_{12}$ is their separation.
\end{itemize}

We categorize the outcomes at 400\,$t_0$ after the periastron passage of the flyby and double-check the results at 500\,$t_0$. 
Only consistent results are recorded, and any discrepancies are identified as anomalies. The majority of these ``anomalies" involve two planets or a planet and the flyby star moving out at nearly the same time in similar directions 
-- Longer-term integration would be needed to determine the fate of these ``anomalies". Our criteria have been optimized to the extent that the probability of anomalies is negligible and can be disregarded
(e.g. even when the two ``anomaly" planets become bound at later times, 
their number is still much smaller than the recorded BPs).
There are also some simulations where star-planet collision and planet-planet collision occur (before 400\,$t_0$). The probability of these cases in our simulations is very low (less than 1\%), so we do not analyze them further.

\section{Numerical results: Fiducial Runs}
\label{sec: numerical results}

In this section, we present the results of our fiducial simulations (as described in Section~\ref{sec: method}) and the related analysis. 
(These fiducial runs have parameters $a_1=0.7a_2$, $e=1.1$ and $m_1=m_2=1 M_{\rm J}$. In Section~\ref{sec: additional runs} we will present some results for different parameters.)
In Section~\ref{sec: numerical results_branching ratio}, we give the fractions of different outcomes as a function of the dimensionless pericenter distance $\tilde q$, averaged over all phase angles and inclinations. 
In Section~\ref{sec: Numerical results_FFP}, we analyze the velocity distribution of the free-floating planets (FFPs). 
In Section~\ref{sec: Numerical results_SSP}, we discuss the properties of the single-surviving-planet (SSP) systems. 
In Section~\ref{sec: Numerical results_CP}, we analyze the orbital parameter distributions of the captured planets (CPs) around the flyby star. 
In Section~\ref{sec: Numerical results_BP}, we analyze the orbital parameter distributions of the free-floating binary planets (BPs) produced by flybys.

\subsection{Branching ratios}
\label{sec: numerical results_branching ratio}

\begin{figure*}[htbp]
    \centering
    \begin{minipage}[b]{0.49\linewidth}
        \includegraphics[width=\linewidth]{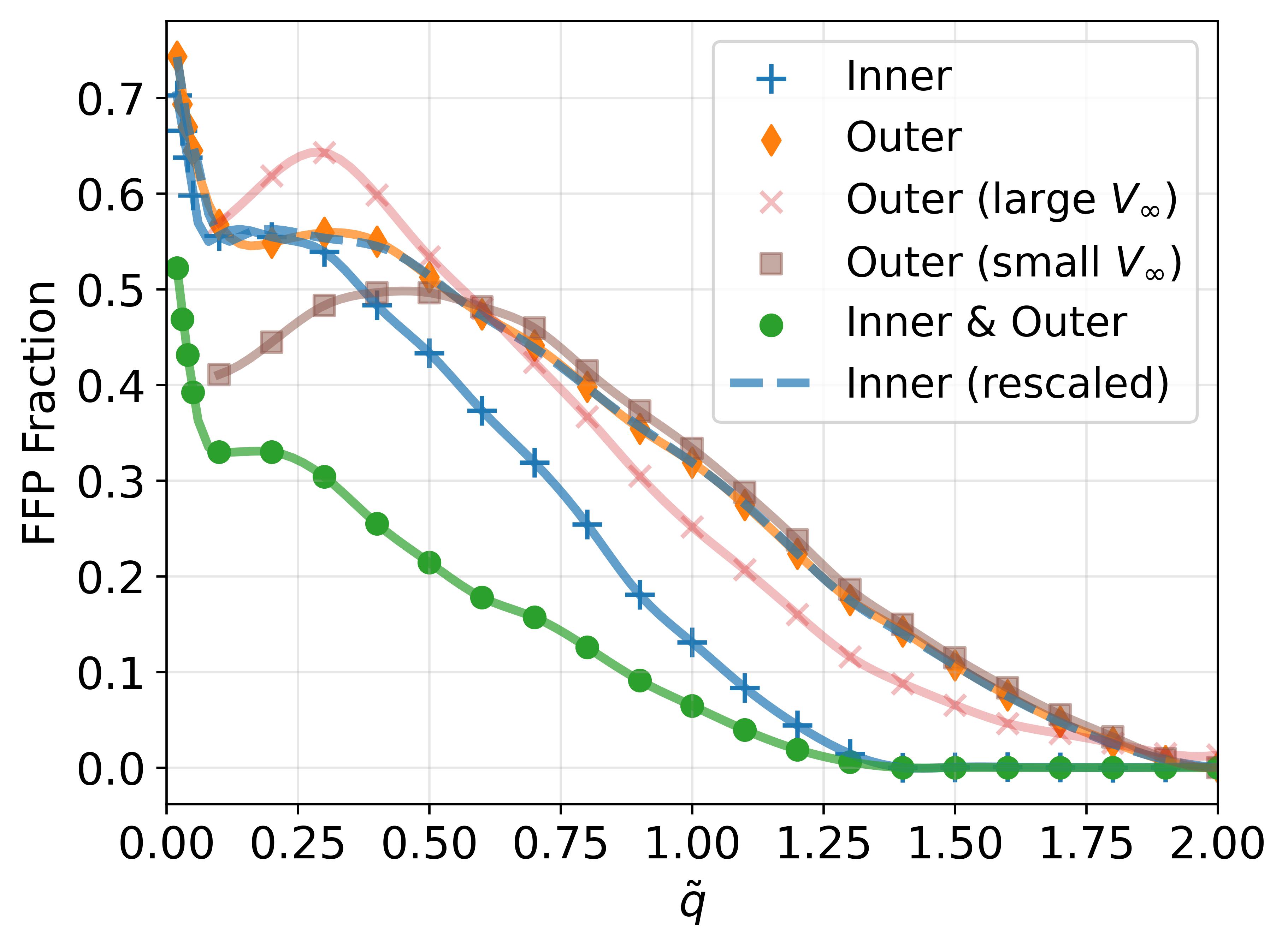}
        \label{The fraction of forming free-floating planets}
    \end{minipage}
    \begin{minipage}[b]{0.49\linewidth}
        \includegraphics[width=\linewidth]{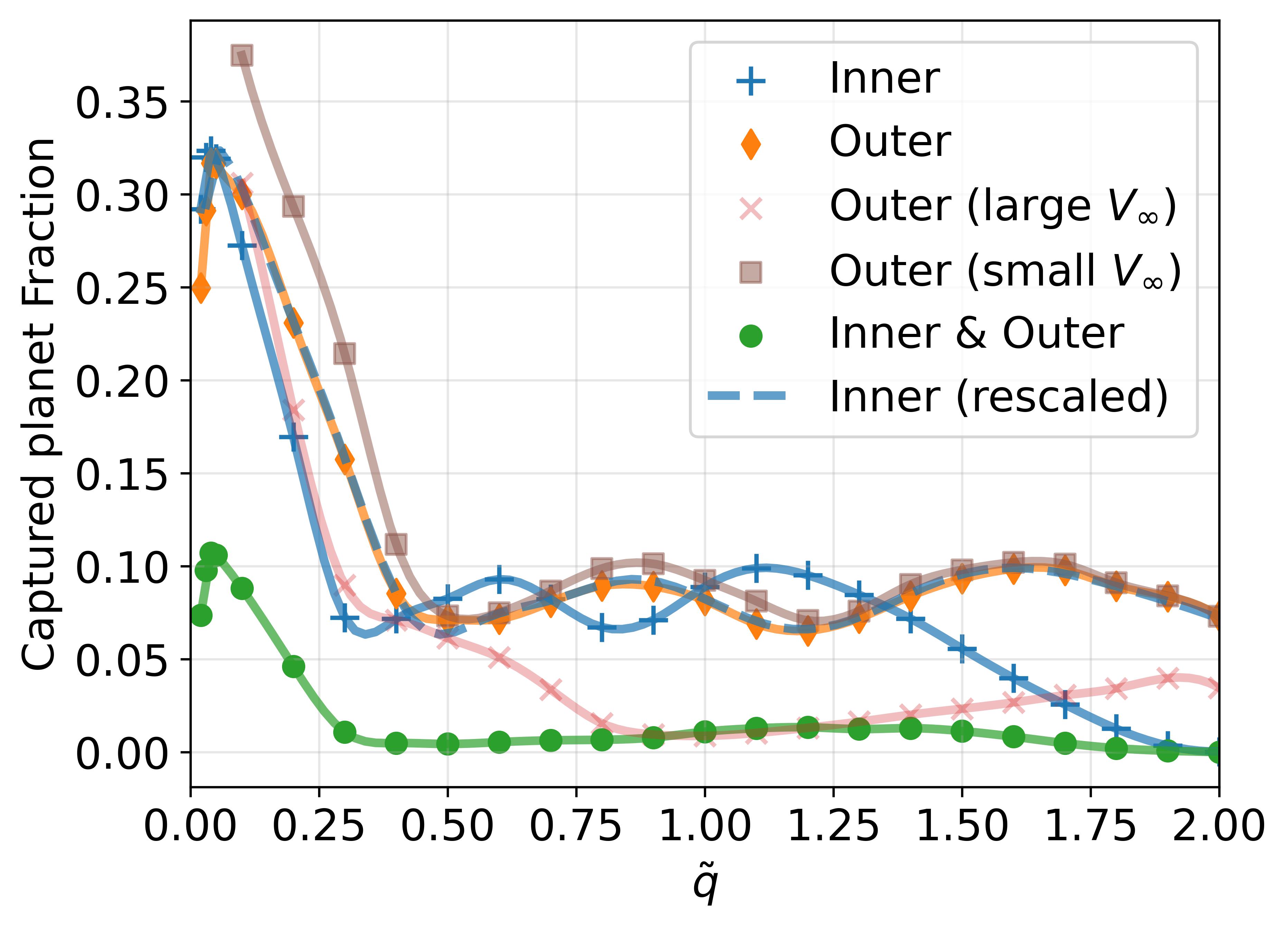}
        \label{The fraction of planet stolen}
    \end{minipage}

    \begin{minipage}[b]{0.49\linewidth}
        \includegraphics[width=\linewidth]{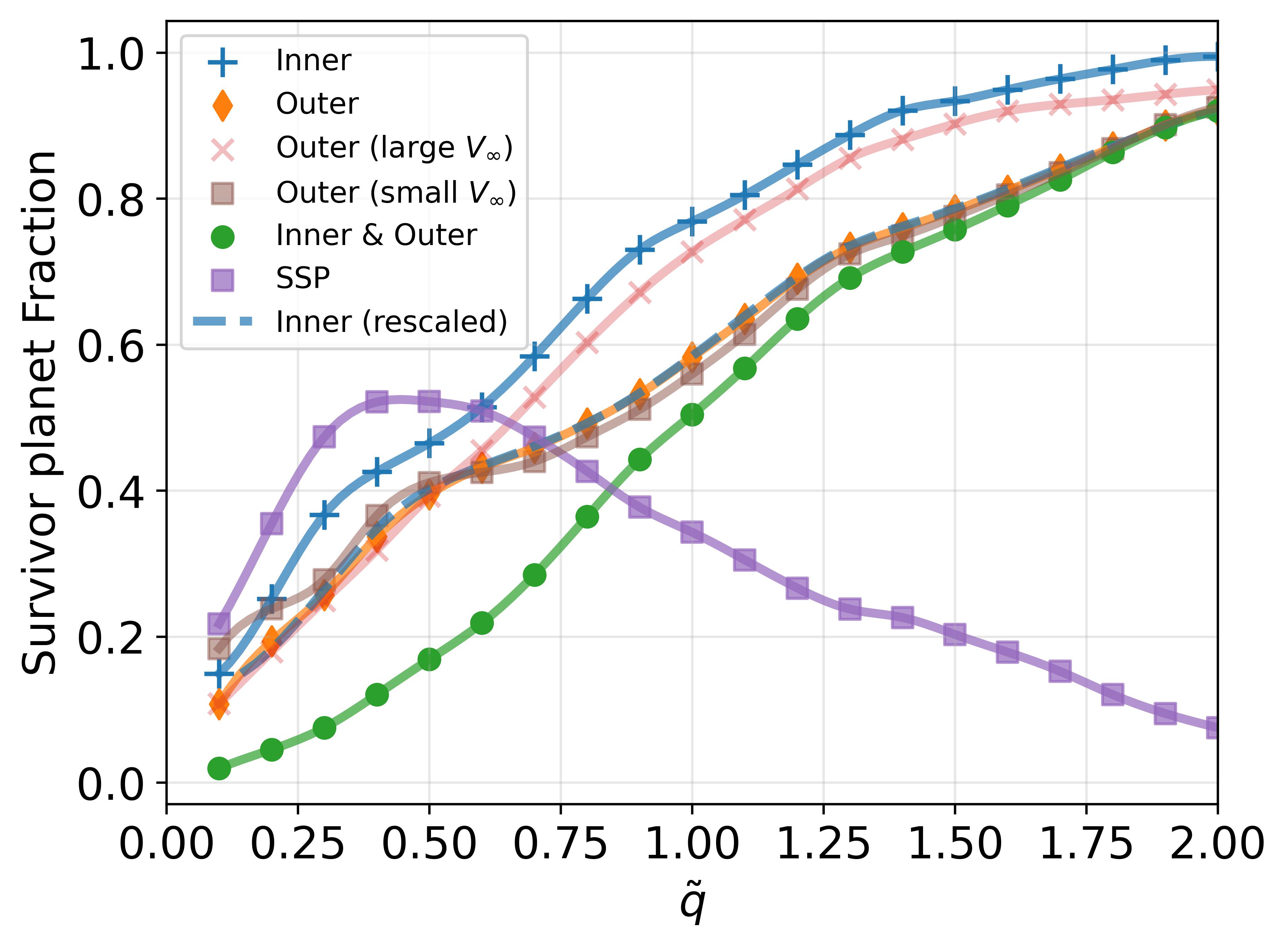}
        \label{The fraction of remaining bound}
    \end{minipage}
    \begin{minipage}[b]{0.49\linewidth}
        \includegraphics[width=\linewidth]{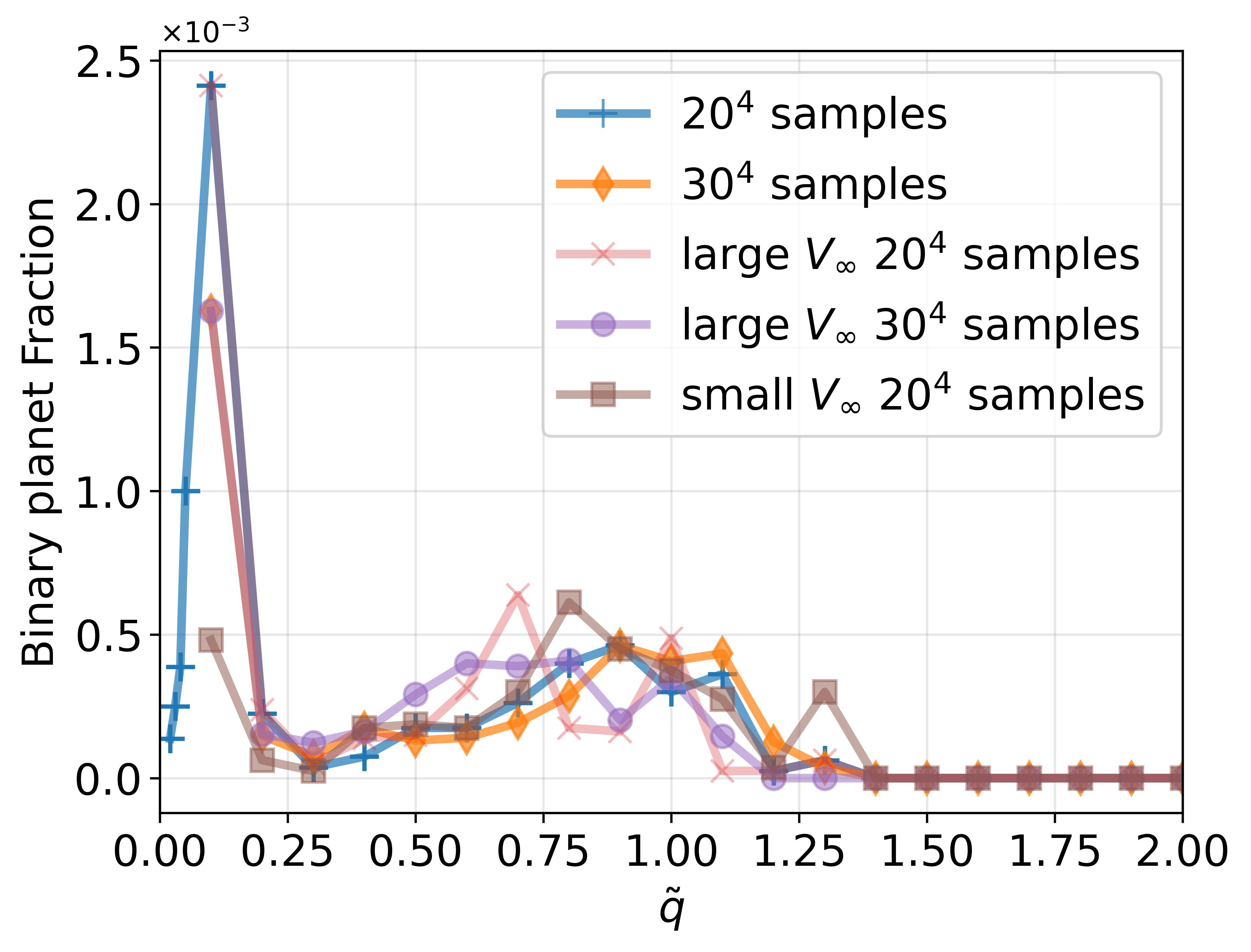}
        \label{The fraction of forming Jupiter Mass Binary Objects (JuMBOs)}
    \end{minipage}
    \caption{
    The fractions of different outcomes as a function of the dimensionless pericenter distance ${\tilde q} = q/a_2$ of the flyby, averaged over all phase angles and inclinations. The top left, top right, bottom left, and bottom right panels represent the fraction of free-floating planets (FFPs), captured planets (CPs), survivor planets (SPs), and free-floating binary planets (BPs), respectively.
    The stars have $M_1=M_2=M_\odot$, the two planets have $m_1=m_2=M_{\rm J}$ and initial $a_1/a_2=0.7$.
    {\bf Top left:} For FFPs, the solid blue and orange lines correspond to the fractions of the inner and outer planets that are ejected and become FFPs, respectively. 
    The solid green lines show the fraction of systems where both planets are ejected and become free-floating. 
    {\bf Top right:} For CPs, the solid blue and orange lines correspond to the fractions of the inner and outer planets that are captured by the flyby, respectively. 
    The solid green line shows the fraction of systems where both planets are captured by the flyby. 
    {\bf Bottom left:} For SPs, the solid blue and orange lines correspond to the fractions of the inner and outer planets that remain bound to the original star, respectively. 
    The solid green lines show the fraction of systems where both planets remain bound to the original star. 
    The orange line agrees with the gray line in Fig.~4 of \citep{Rodet2021ApJ}. 
    The purple line represents the fraction of surviving single-planets (SSPs) regardless of which planet it was originally.
    The blue-dashed line in the first three panels represents a ``rescaled" blue line, obtained by replacing the value of $\tilde q$ in the solid blue line with $q/a_1$. Note that these rescaled lines are nearly identical to orange lines, indicating that the gravitational interaction between the two planets has a negligible effect on the one-planet ejection fraction. 
    The light-red lines and brown lines (labeled ``large $V_\infty$" and "small $V_\infty$") show the result (for the outer planet only) based on the simulations with $V_{\infty}/v_{2i}=1$ and $V_{\infty}/v_{2i}=0.1$ (where $v_{2i}=\sqrt{GM/a_2}$) for the flyby trajectory (see Section~\ref{sec: additional runs_v}), while the other lines are based on the fiducial simulations (see Section~\ref{sec: method}) with $e=1.1$ for the flyby trajectory.
    {\bf Bottom right:} For BPs, in addition to the result of our fiducial runs that are based on $20^4$ samplings of angles (blue line), we also show the result of an expanded set of simulations based on $30^4$ samplings of angles (orange line).}
    \label{Figs: Fractions of Outcomes}
\end{figure*}

As noted before, we typically sample ${\tilde q} = q/a_2$ values on a uniform grid from $0.1$ to $2$ to determine various branching fractions. 
We also consider $\tilde q = 0.02-0.05$ separately to test the behavior of the branching fractions when $\tilde q$ is very small. 
The first three panels in Fig.~\ref{Figs: Fractions of Outcomes} show the fractions of free-floating planets (FFPs), captured planets (CPs) and survivor planets (SPs), respectively. 
In each case, we show the results for both the inner and outer planets. 

Since the gravitational interactions between the planets are significantly weaker than those between the planets and the stars, the two planets respond to the disturbance from the flyby independently to a good approximation.
Indeed, by using $q/a_1$ as the horizontal axis to rescale the inner-planet results in the three figures, we recover nearly the same results as the outer planet.

For each $\tilde q$, the branching fractions for FFPs, CPs and SPs add up to unity (e.g. the blue/orange lines in the first three panels in Fig.~\ref{Figs: Fractions of Outcomes} add up to unity), indicating that our classification of different outcomes is complete (note that free-floating binary planets constitute a very small fraction of FFPs; see below). 
Our result for the SP fraction agrees with that of \citet{Rodet2021ApJ}, who studied the effect of stellar flybys on single-planet systems in detail.

We note that the fraction of captured planets (the top right panel in Fig.~\ref{Figs: Fractions of Outcomes}) is a non-monotonic function of $\tilde q$: it has a sharp peak at $\tilde q\sim 0.05$, and two broad peaks at $\tilde q\sim 0.8$ and $\tilde q\sim 1.6$.
For $\tilde q\lesssim 0.3$, the capture fraction increases rapidly then decreases with decreasing $\tilde q$. On the other hand, the FFP fraction (the top left panel in Fig.~\ref{Figs: Fractions of Outcomes}) reaches a plateau for $\tilde q\lesssim 0.3$ and then sharply increases for $\tilde q\lesssim 0.05$. 
This complementary behavior for the FFP and CP fractions leads to a smooth behavior for the SP fraction shown in the bottom right panel in Fig.~\ref{Figs: Fractions of Outcomes}.

The bottom right panel in Fig.~\ref{Figs: Fractions of Outcomes} shows the fraction of free-floating binary planets (BPs). 
We see that this fraction is so small that our simulation result exhibits significant fluctuations.
To address this issue, we carry out expanded simulations, sampling $30 \times 30 \times 30 \times 30$ angles (for $i$, $\omega$, $\lambda_1$, and $\lambda_2$) to obtain larger simulation examples. 
The results for these expanded simulations and for our fiducial runs (based on $20 \times 20 \times 20 \times 20$ angles) are shown in the bottom right panel in Fig.~\ref{Figs: Fractions of Outcomes} to provide a rough idea of the impact of data fluctuations.
Interestingly, the BP fraction exhibits a sharp peak at $\tilde q\simeq 0.1$ and a broad peak at $\tilde q\sim 1$, and becomes negligible for $\tilde q\gtrsim 1.5$.
However, even at the sharp peak, the BP fraction is less than $0.25\%$; for most values of $\tilde q$, the fraction is less than $0.05\%$.
\footnote{Our BP fraction is much smaller than the value reported by \citet{Wang2024arXiv02}. 
A possible reason for their large values is that they misidentified the different flyby outcomes (see Section~\ref{sec: method_criteria});
for example, BPs require not only Eq.~(\ref{Eq:inertial}) be satisfied, but also $r_{12}\ll r_{\rm Hill}$. 
In addition, the numerical setup of \citet{Wang2024arXiv02} involves 5 angles to specify the initial conditions;
in fact, only 4 are needed ($i, \omega, \lambda_1, \lambda_2$).}

\subsection{Property of Free-Floating Planets (FFPs)}
\label{sec: Numerical results_FFP}

For free-floating planets, we are interested in the velocity $v_{\rm free}$ relative to the center-of-mass of the star-flyby system. 
At the end of each simulation, when an FFP reaches the velocity ${\bf v}_{\rm end}$, we compute its ``free-floating" velocity as 
\begin{equation}
v_{\rm free}=\sqrt{\left({\bf v}_{\rm end}-{\bf V}_{\rm cm}\right)^2-\frac{2GM_1}{r_1}-\frac{2GM_2}{r_2}},
\label{eq:vinf}
\end{equation}
where $r_1,~r_2$ are the distances between the FFP and the two stars, and 
${\bf V}_{\rm cm}=(M_1{\bf V}_1+M_2{\bf V}_2)/M_{12}=({\bf V}_1+{\bf V}_2)/2$
is the center-of-mass velocity of the star-flyby system (relative to the simulation reference frame).

\begin{figure}[htbp]
\centering
\includegraphics[width=1.02\columnwidth]{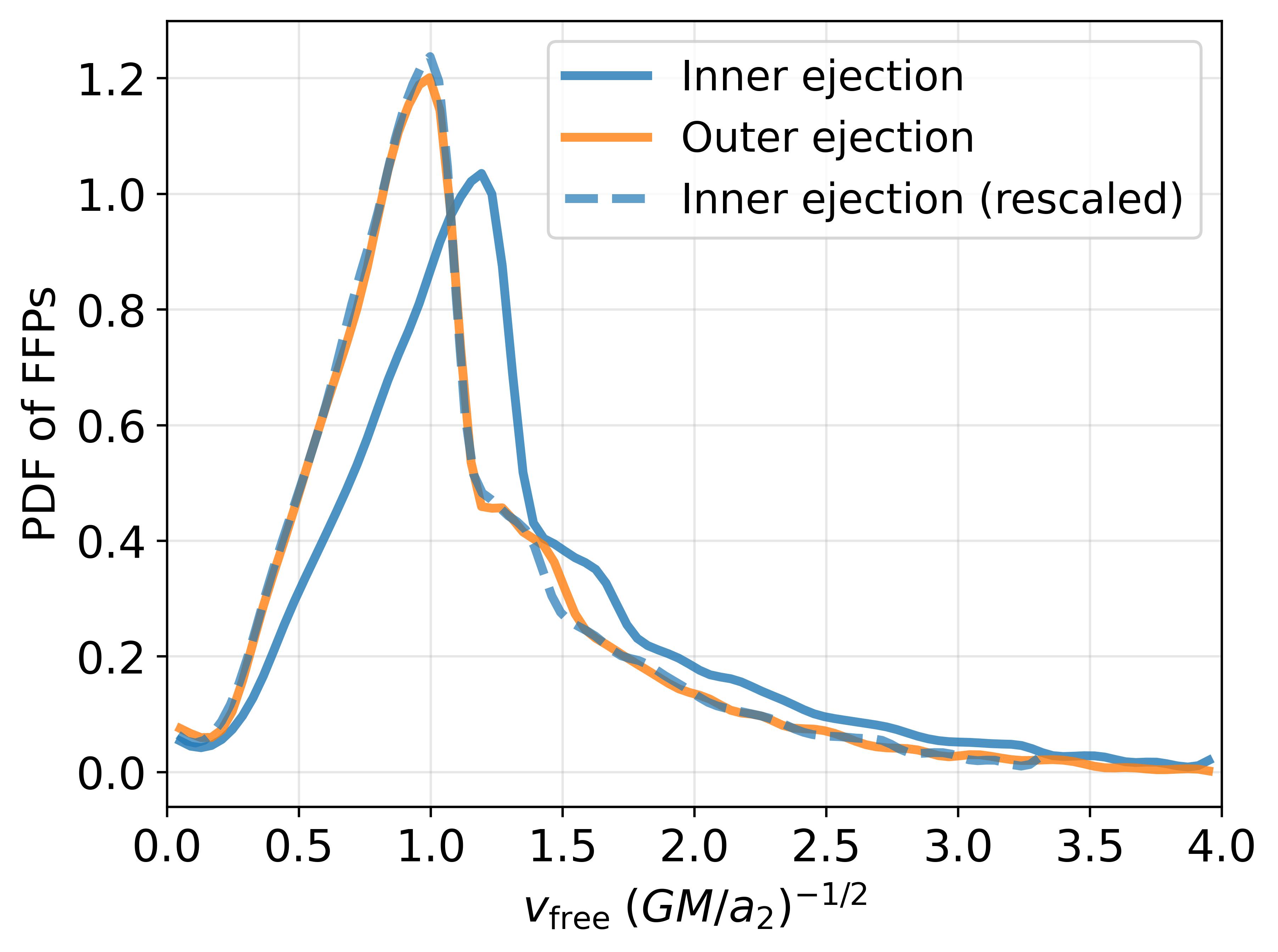}
\caption{
The probability density distribution (PDF) of the free-floating velocity $v_{\rm free}$ of FFPs produced by stellar flybys (sampling over all $\tilde q$ and angles) in the fiducial simulation. 
The blue and orange solid lines represent the results for the inner and outer planets, respectively. 
The blue dashed line shows the rescaled inner planet result after using $\sqrt{GM/a_1}$ as the velocity unit; this dashed line is nearly identical to the solid orange line. 
}
\label{v_infty distribution for free-floating planets}
\end{figure}

We analyze all FFPs generated in our simulations and show the probability density functions (PDFs) for the ``free-floating" velocities of both planets in Fig.~\ref{v_infty distribution for free-floating planets}.
Since the two planets respond to the flyby independently of each other to a good approximation, their PDFs are nearly identical when the velocities are scaled by their initial orbital velocities ($v_{1i}=\sqrt{GM/a_1}$ and $v_{2i}=\sqrt{GM/a_2}$).

We see that the distribution of $v_{\rm free}$ is far from being a Gaussian. 
It rises linearly with $v_{\rm free}$, peaking at $v_{\rm free} \simeq v_{2i}$ (for the outer planet; and $v_{1i}$ for the inner planet), and then decreases gradually, with a bump at $v_{\rm free}\simeq 1.4\, v_{2i}$.

\subsection{Property of Single-Surviving-Planet (SSP) Systems}
\label{sec: Numerical results_SSP}

After a stellar flyby, one of the planets in the original two-planet system may be ejected and become a FFP, or be captured by the flyby. This will result in some single-surviving-planet (SSP) systems. 
We first analyze such an SSP according to whether it is the original inner planet or the outer planet. 
Note that these two systems are formed in different ways (inner planet ejection and outer planet ejection), so the scaling discussed above does not apply here.
Then, considering that we cannot know, in real observations, whether an SSP is originally an inner or outer planet, we also analyze the orbital parameters of all SSP systems, regardless of whether the SSP is originally the inner planet or the outer planet.

\begin{figure*}[htbp]
    \centering
    \begin{minipage}[b]{0.49\linewidth}
        \centering
        \includegraphics[width=\linewidth]{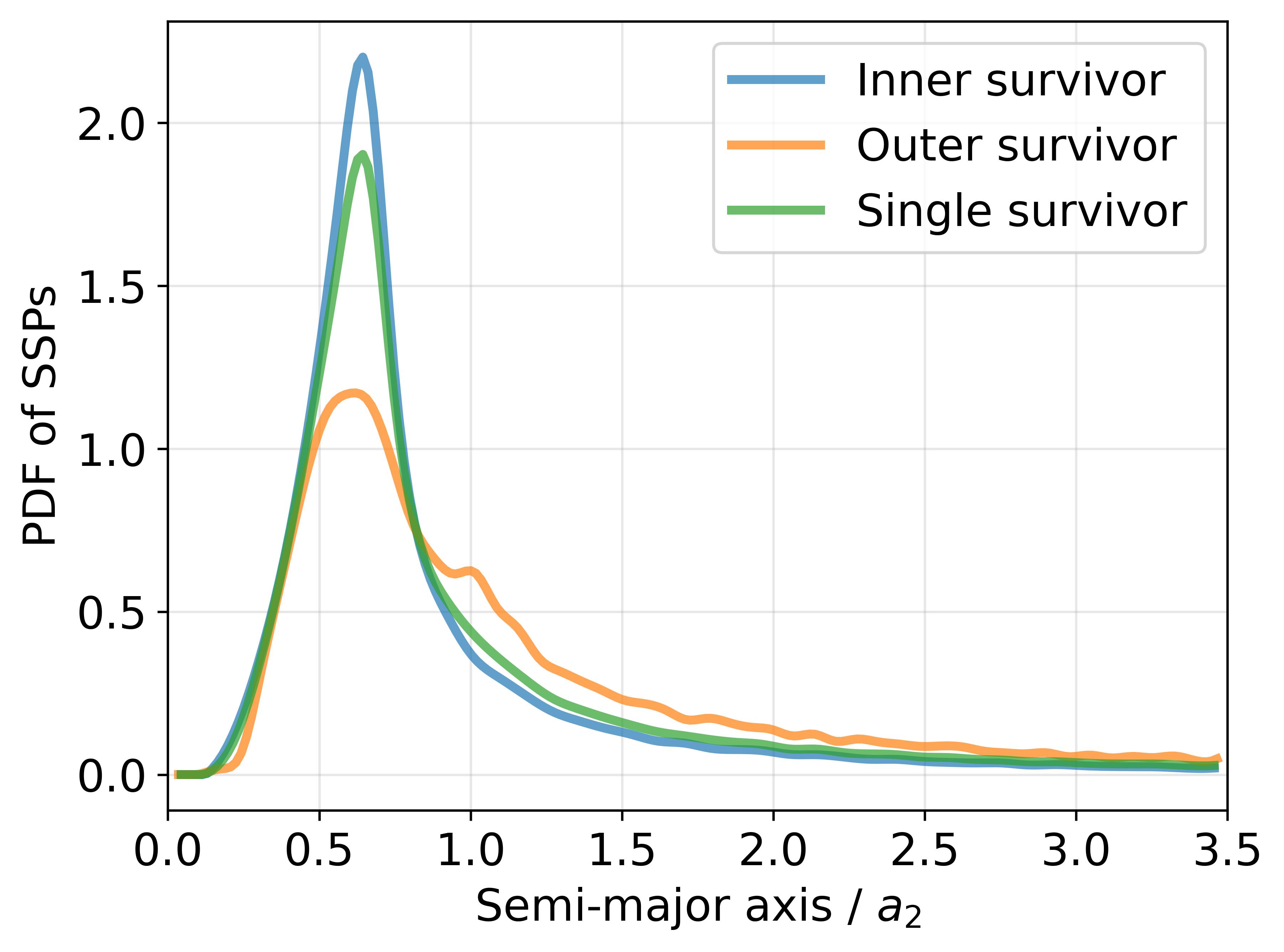}
    \end{minipage}
    \hfill
    \begin{minipage}[b]{0.49\linewidth}
        \centering
        \includegraphics[width=\linewidth]{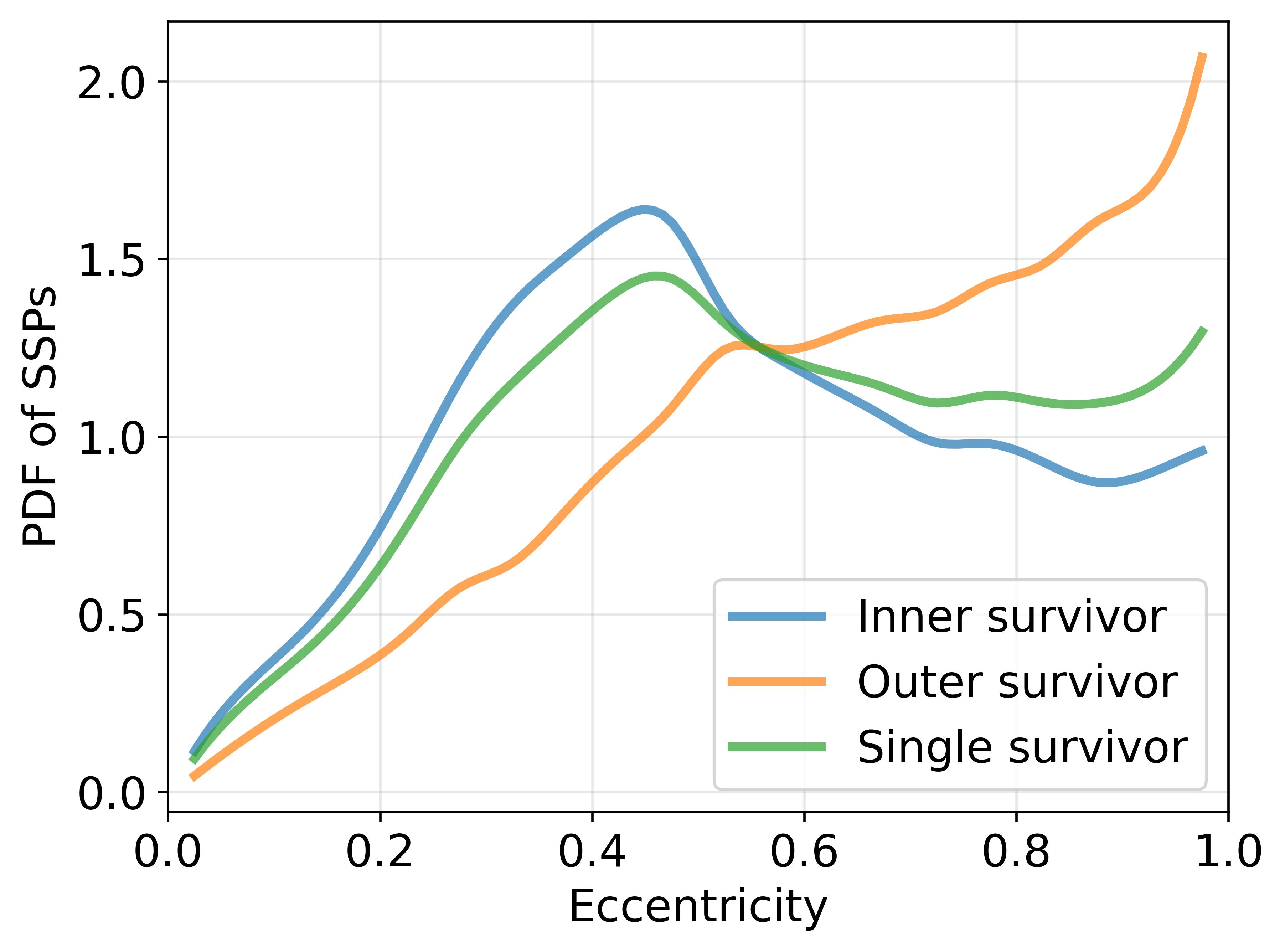}
    \end{minipage}

    \begin{minipage}[b]{0.49\linewidth}
        \centering
        \includegraphics[width=\linewidth]{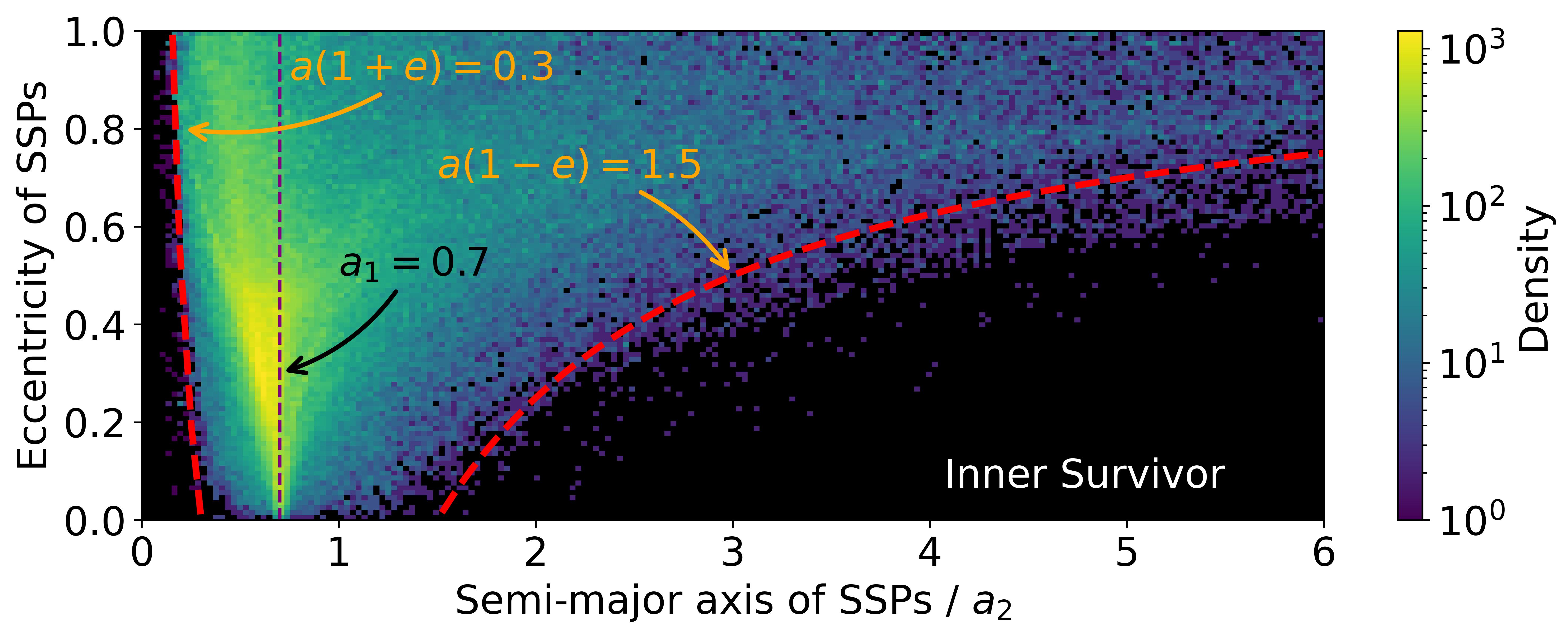}
    \end{minipage}
    \hfill
    \begin{minipage}[b]{0.49\linewidth}
        \centering
        \includegraphics[width=\linewidth]{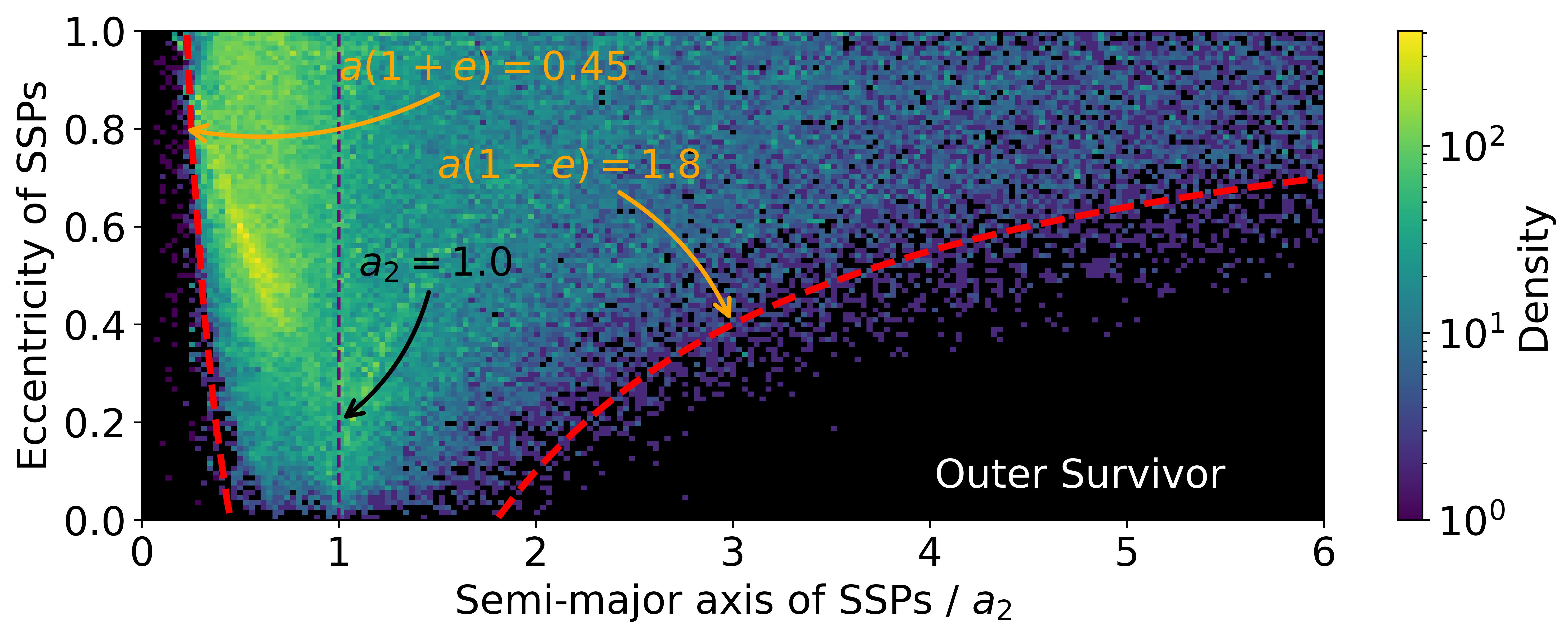}
    \end{minipage}
    
    \caption{
    {\bf Top row:} The probability distributions of the orbital semi-major axis and eccentricity of the single-surviving-planet (SSP) systems. 
    The blue/orange line indicates that the SSP starts out as the inner/outer planet in the original system. 
    The green line refers to all SSPs, regardless of whether the SSP is the inner or outer planet. 
    Note that in both panels, the green line lies between the blue and orange lines, but closer to the blue line, since the inner planet has a higher probability of surviving the flyby (see the bottom left panel of Fig.~\ref{Figs: Fractions of Outcomes}).
    {\bf Bottom row:} The semi-major axis - eccentricity distribution of survivor single-planet (SSP) systems. The middle/bottom panel refers to the inner/outer survivors, respectively. In each panel, the vertical dashed line is the initial semi-major axis of the planet. The red dashed lines (defining the boundaries of the occupied region) correspond to the approximate fitting of the boundaries with the value indicated.}
    \label{Figs: Single planet system orbital parameter}
\end{figure*}

The top row in Fig.~\ref{Figs: Single planet system orbital parameter} shows the semi-major axis and eccentricity distributions of SSPs.
When one planet is ejected by the flyby, the orbital parameters of the SSP can be significantly affected.
From the top left panel in Fig.~\ref{Figs: Single planet system orbital parameter} we see that the inner survivor has its semi-major axis strongly peaked at $0.7a_2$, corresponding to its original value. 
The outer survivor's semi-major axis also peaks at $0.7a_2$, with a broad distribution.  
In general, this is an inward migration of the semi-major axes of the SSPs, consistent with the finding of \citet{Rodet2021ApJ} (see their Fig.~3).

For the eccentricity, we see from the top right panel in Fig.~\ref{Figs: Single planet system orbital parameter} that while many outer survivors have large $e$'s peak (close to unity), the inner survivors have their $e$'s peak around 0.45.
We note that the distributions shown in the top right panel in Fig.~\ref{Figs: Single planet system orbital parameter} are the results of flybys with many $\tilde q$ values.
The eccentricity distribution of SSPs is very different for different $\tilde q$: 
as $\tilde q$ decreases from 2 to 0.1, the peak of eccentricity slowly increases from 0 to 1, and the dispersion of the distribution also gradually increases
[see Appendix \ref{Appendix:q} for a detailed analysis; see also 
Fig.~3 in \citep{Rodet2021ApJ}].

The semi-major axis and eccentricity of SSPs are correlated. 
The middle panel and bottom panel in Fig.~\ref{Figs: Single planet system orbital parameter} show the 2D distribution of SSPs in the $e-a$ plane. We see that SSPs tend to occupy a restricted domain in the $e-a$ plane.
We find that the occupied domain has boundaries approximately set by constant $a(1-e)$ (the pericenter distance) and constant $a(1+e)$ (the apocenter distance). 
Qualitatively, the absence of SSPs in the low-$a$ region can be understood from the fact that it is difficult for the flyby star to ``absorb" a large amount of energy [beyond $GM/(2a_1)$ or $GM/(2a_2)$] from the planet. 
On the other hand, the planet can lose a large fraction of its original binding energy, and thus move to a large-$a$ orbit; but in this process, its eccentricity necessarily becomes large. 
This explains the absence of SSPs in the large-$a$, low-$e$ region in the middle panel and bottom panel in Fig.~\ref{Figs: Single planet system orbital parameter}.

\subsection{Property of Captured Planet (CP) Systems}
\label{sec: Numerical results_CP}

\begin{figure*}[htbp]
    \centering
    \begin{minipage}[b]{0.49\linewidth}
        \centering
        \includegraphics[width=\linewidth]{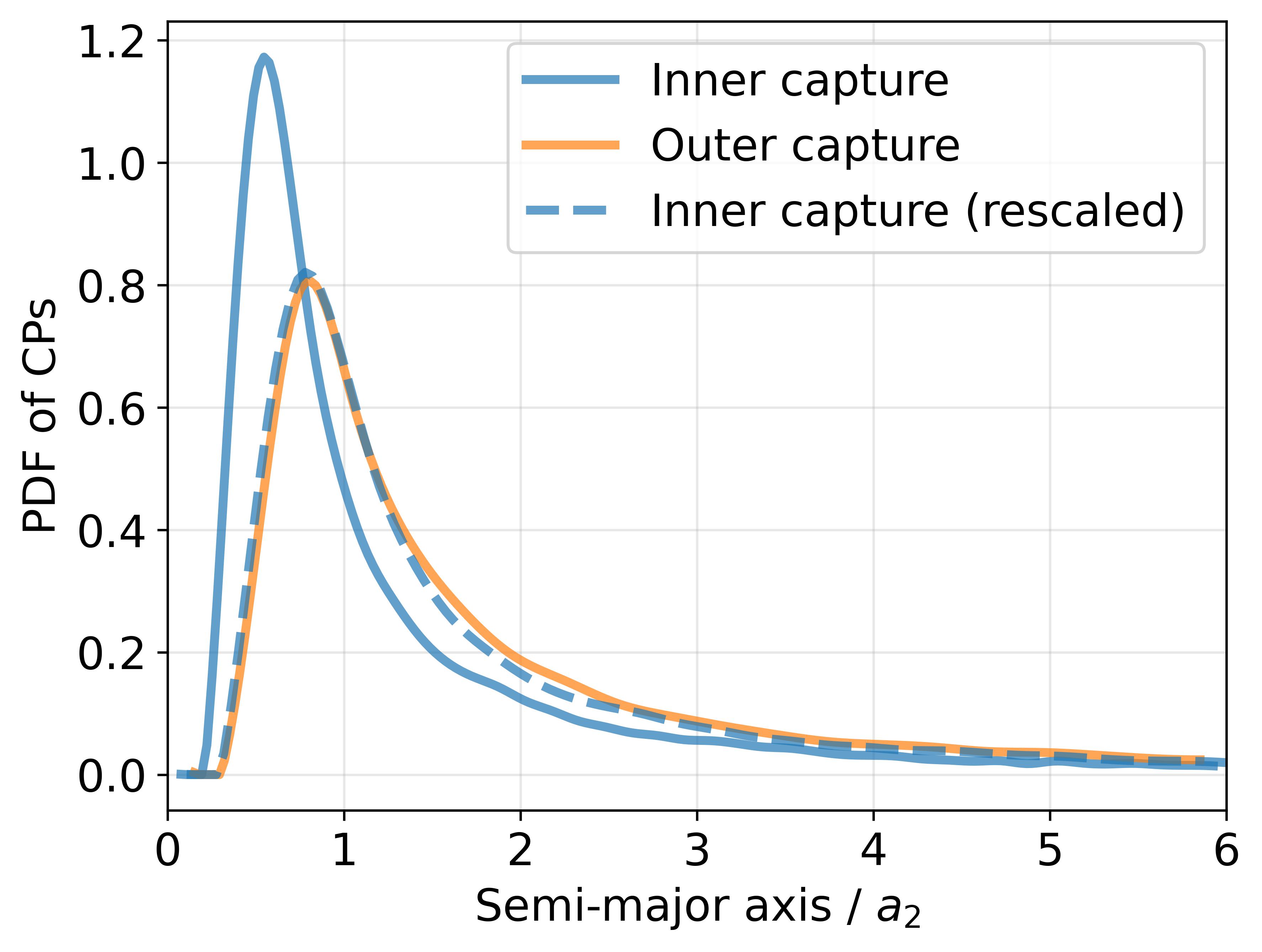}
    \end{minipage}
    \hfill
    \begin{minipage}[b]{0.49\linewidth}
        \centering
        \includegraphics[width=\linewidth]{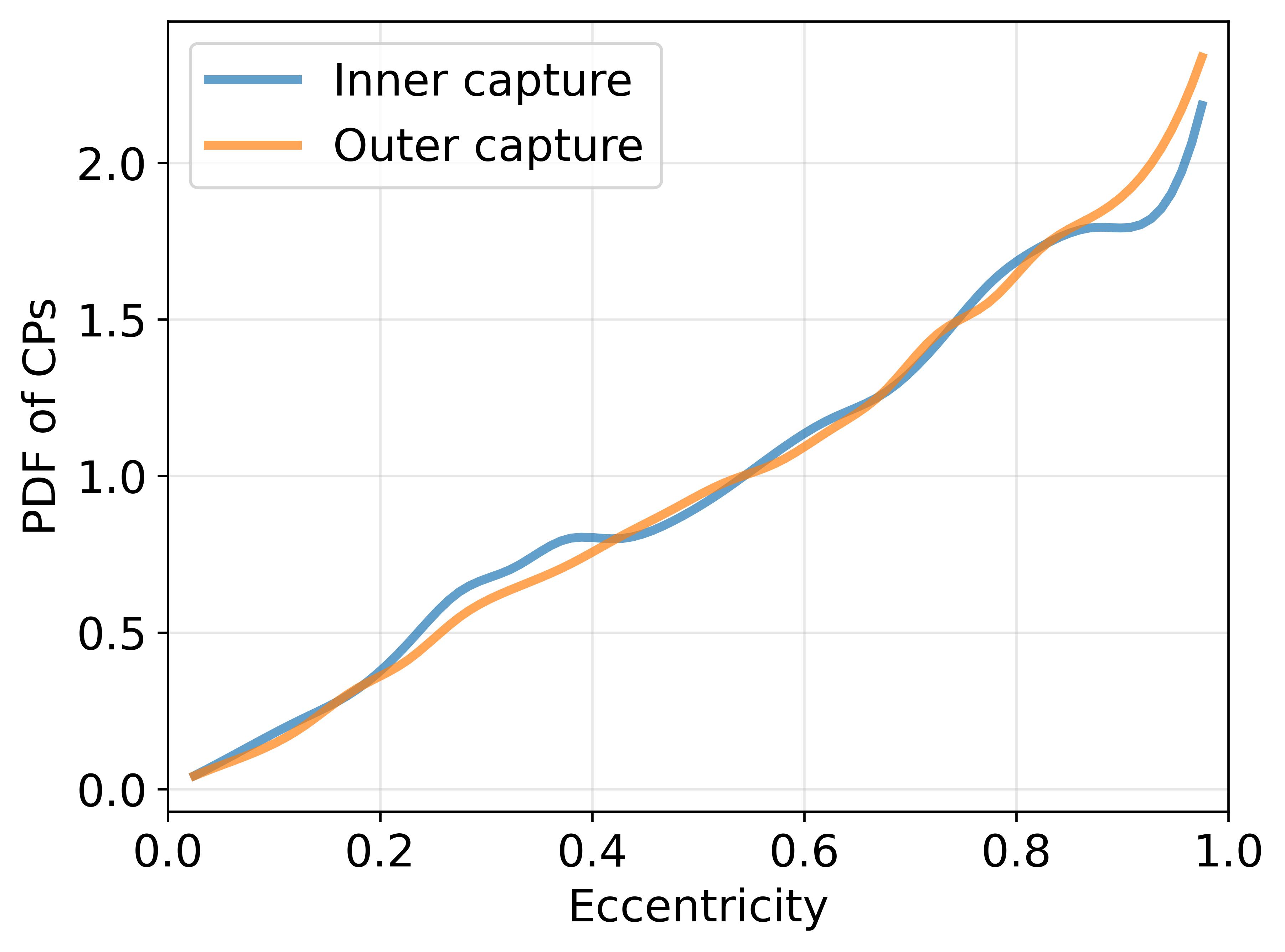}
    \end{minipage}
    
    \begin{minipage}[b]{0.49\linewidth}
        \centering
        \includegraphics[width=\linewidth]{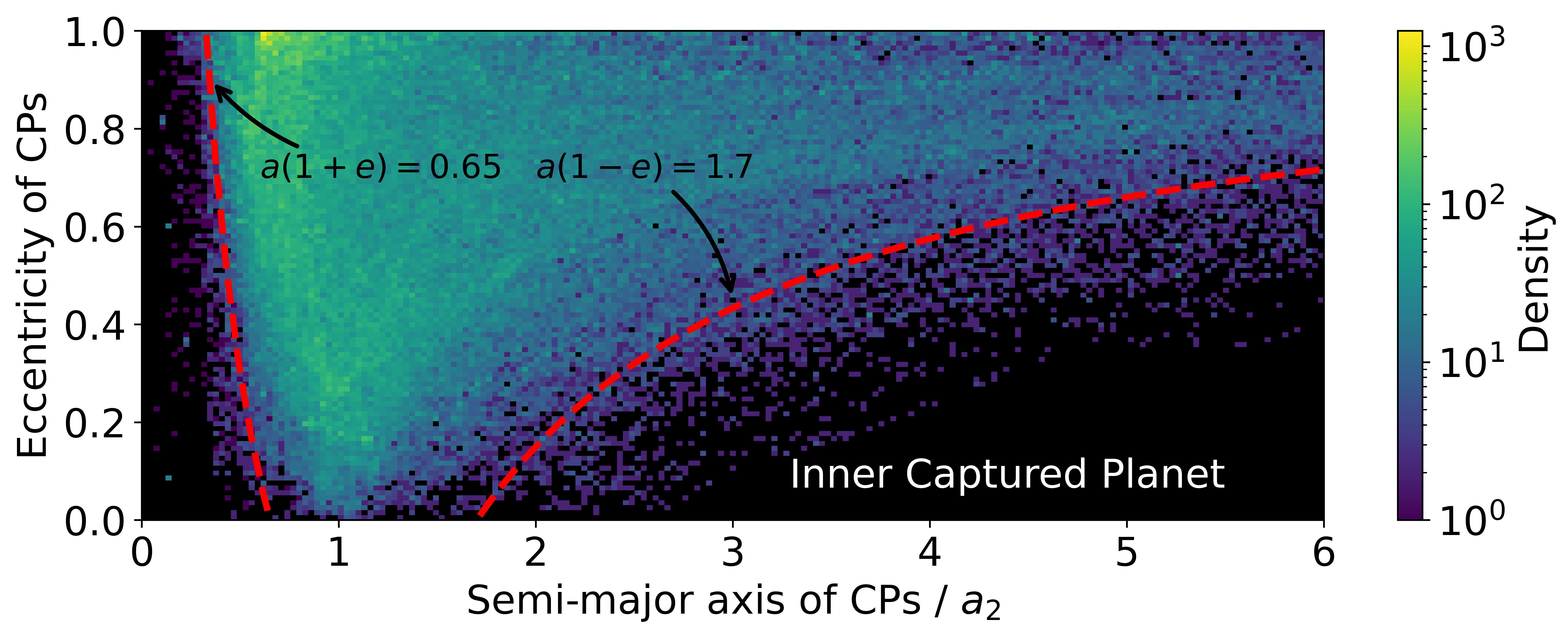}
    \end{minipage}
    \hfill
    \begin{minipage}[b]{0.49\linewidth}
        \centering
        \includegraphics[width=\linewidth]{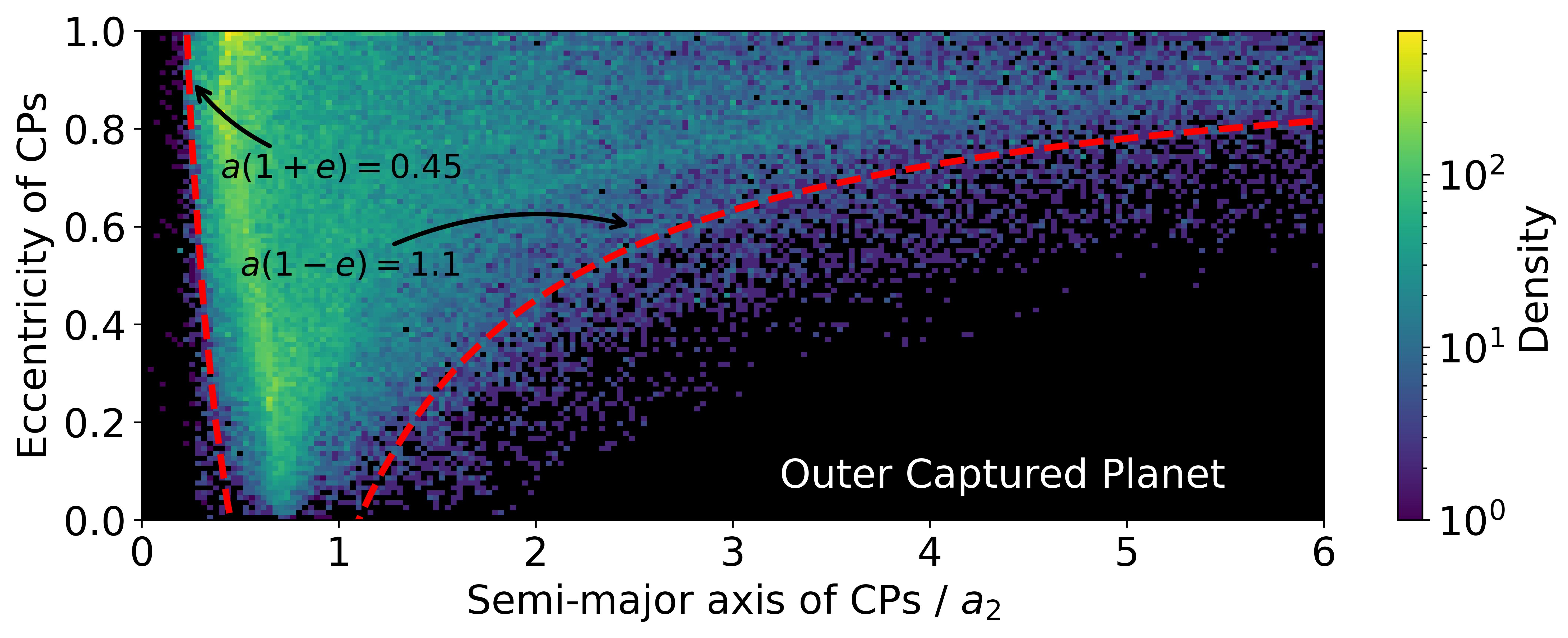}
    \end{minipage}
    \caption{
    Similar to Fig~\ref{Figs: Single planet system orbital parameter}, but for the captured planet (CP) systems. 
    In the upper panels, the orange/blue solid lines correspond to the (original) outer/inner planets that are captured by the flyby star.
    The blue dashed line is the rescaled inner-planet result, using $a/a_1$ as the horizontal coordinate;
    we see that this blue dashed line is nearly identical to the orange line.
    }
    \label{Figs: Captured planet orbital parameter}
\end{figure*}

A planet can be gravitationally captured by the flyby, moving from the original star system to a new star system.

The top row in Fig.~\ref{Figs: Captured planet orbital parameter} shows the semi-major axis and eccentricity distributions of the captured planets (CPs). 
We see that the (original) outer planet, after being captured by the flyby star, exhibits a broad distribution of semi-major axes, with a peak around $0.7a_2$ and a long tail extending to $a\gg a_2$. 
Similarly, the captured inner planet has a semi-major axis distribution that peaks around $0.7a_1$. 
Indeed, when rescaled (using $a/a_1$ and $a/a_2$ for the inner and outer captures, respectively), the two semi-major axis distributions are nearly identical.
The captured planets generally have large eccentricities, with the distribution peaking at $e\sim 1$.

The middle panel and bottom panel in Fig.~\ref{Figs: Captured planet orbital parameter} show the $e-a$ distributions for inner and outer captured planets. 
We see that similar to the case of SSPs (Fig.~\ref{Figs: Single planet system orbital parameter}), the CPs tend to occupy a restricted domain in the $e-a$ plane.

\subsection{Property of Free-Floating Binary Planets (BPs)}
\label{sec: Numerical results_BP}

\begin{figure}[htbp]
        \subfigure
        {
            \begin{minipage}[b]{\linewidth} 
                \centering
                \includegraphics[width=\columnwidth]{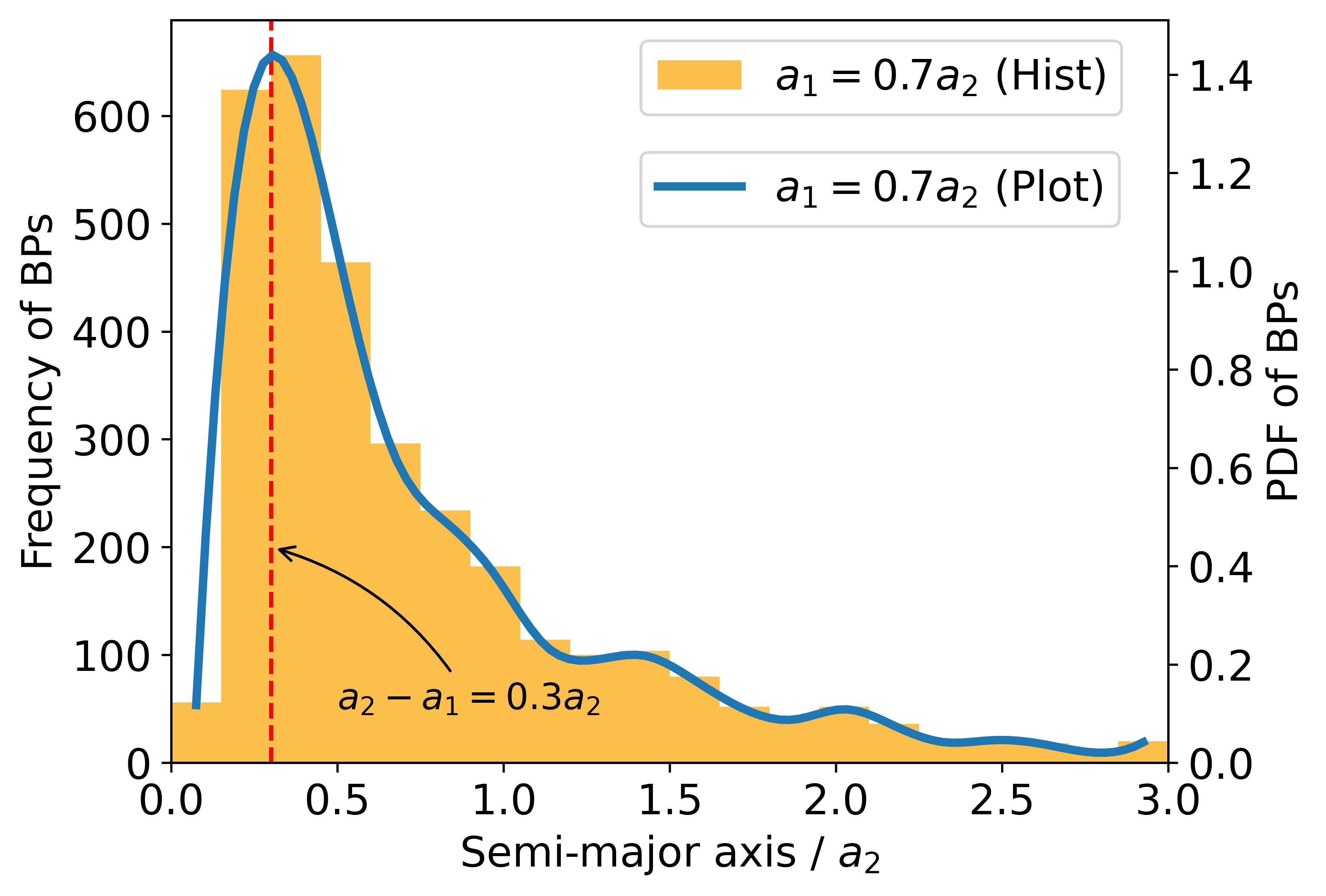}
                \label{JuMBO SMA(a=0.7)}
            \end{minipage}
        }
        \subfigure
        {
            \begin{minipage}[b]{\linewidth}
                \centering
                \includegraphics[width=\columnwidth]{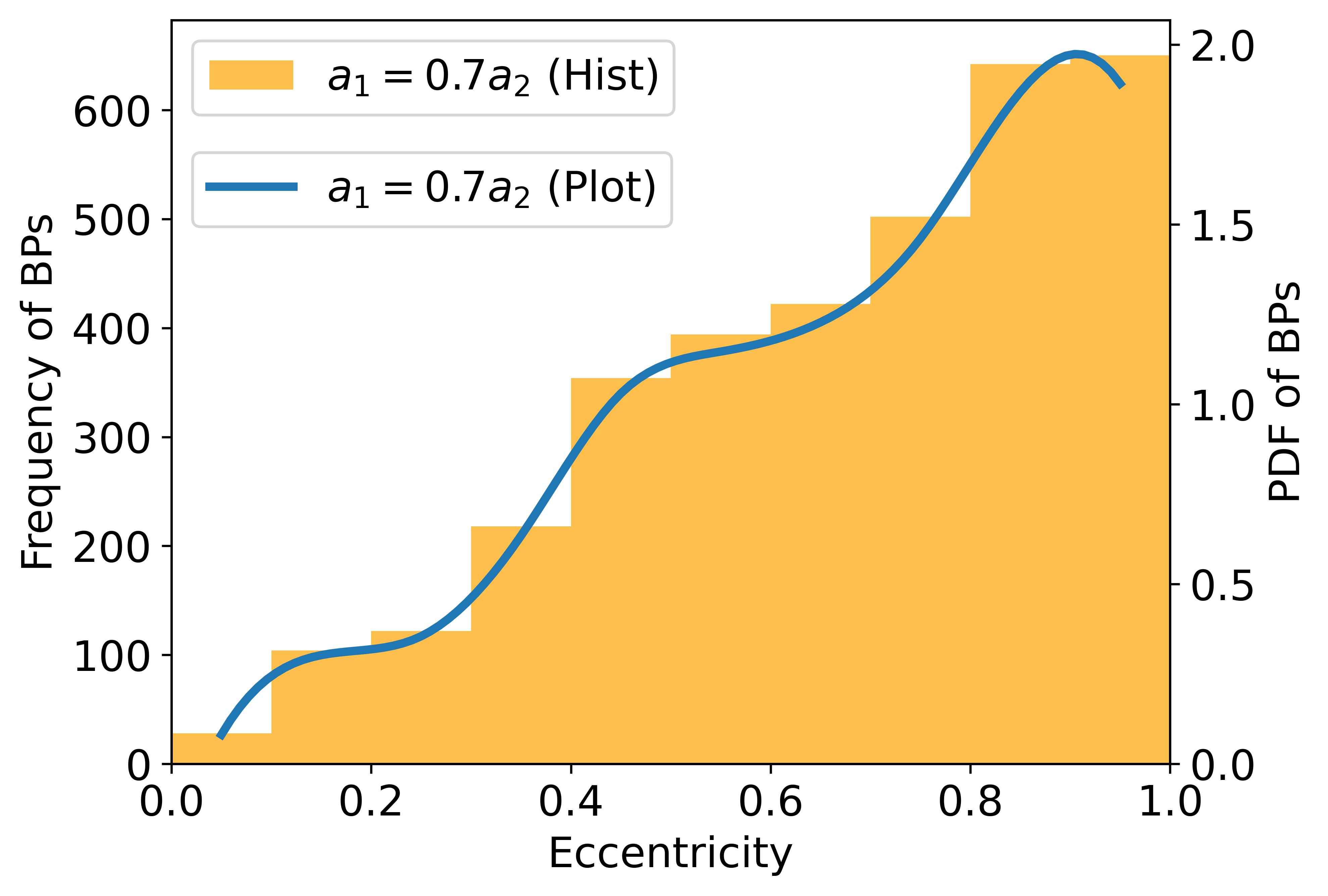}
                \label{JuMBO e(a=0.7)}
            \end{minipage}
        }
        \subfigure
        {
            \begin{minipage}[b]{\linewidth}
                \centering
                \includegraphics[width=\columnwidth]{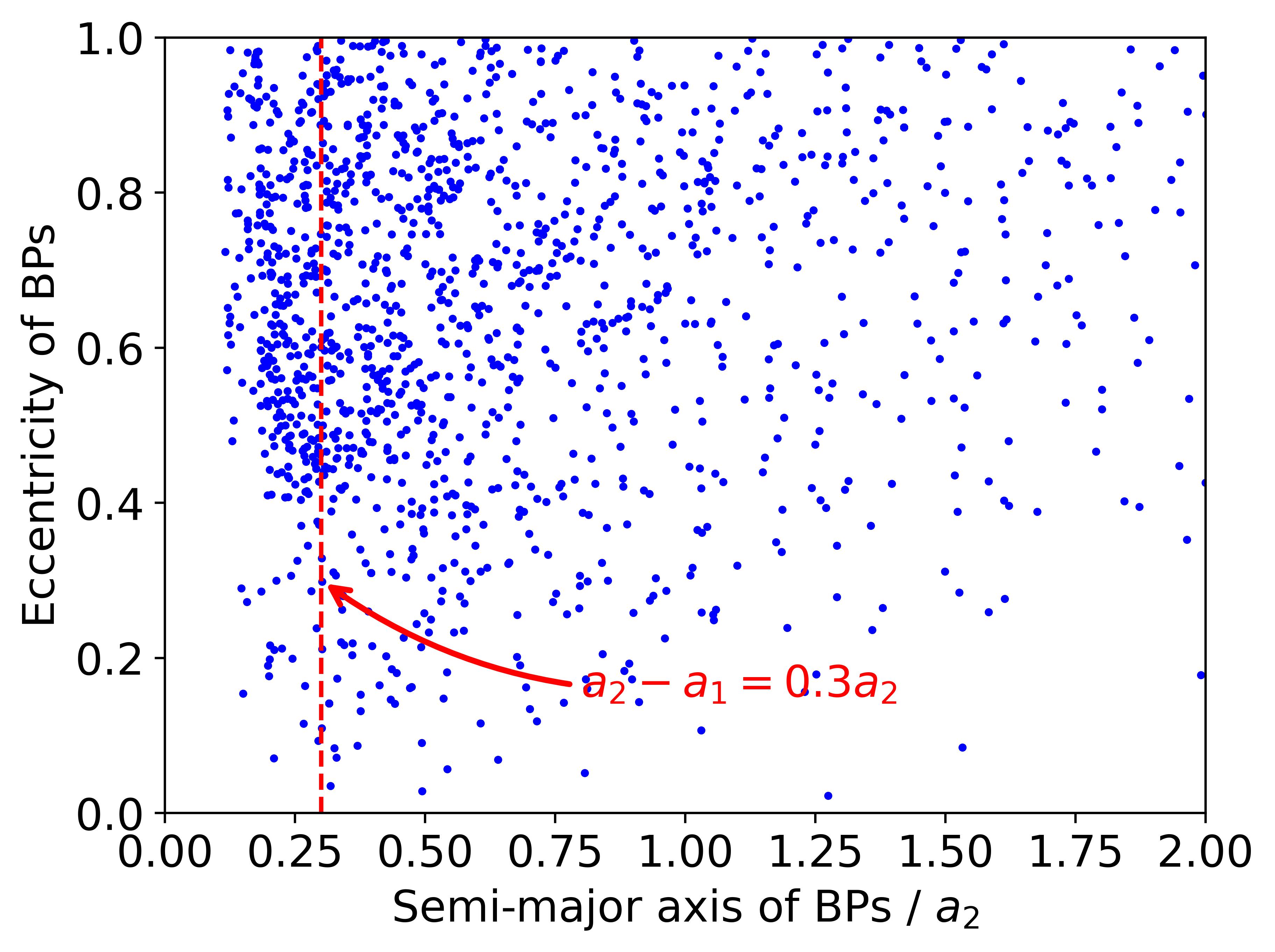}
                \label{JuMBOs' e-a distribution 0.7}
            \end{minipage}
        }
        \caption{
        {\bf Top and middle:} The distribution of the orbital semi-major axis and eccentricity of free-floating binary planets produced by stellar flyby. The initial planets have initial orbits $a_1=0.7a_2$ (as in Fig.~\ref{Figs: Fractions of Outcomes}-\ref{Figs: Captured planet orbital parameter}). The frequency of occurrences in the $30^4$ simulation runs is shown on the left vertical axis, and the PDF is shown on the right vertical axis. It is evident that the distribution of the semi-major axes for BPs peaks at $\Delta a = a_2 - a_1$. 
        {\bf Bottom:} The 2D distribution of BPs in the $a-e$ plane.}
        \label{Figs: JuMBOs' orbital parameter($a_1=0.7a_2$)}
\end{figure}

In Section~\ref{sec: numerical results_branching ratio} (see the bottom right panel of Fig.~\ref{Figs: Fractions of Outcomes}) we show that the fraction of binary planet (BP) formation is rather small (peak $\simeq 0.25\%$ at $\tilde q\simeq 0.1$ and less than 0.05\% for most $\tilde q$'s). 
To properly analyze the orbital property of these BPs, we use the extended simulation runs which sample $30 \times 30 \times 30 \times 30$ angles (for $i$, $\omega$, $\lambda_1$ and $\lambda_2$). 
Fig.~\ref{Figs: JuMBOs' orbital parameter($a_1=0.7a_2$)} show the results.

We see that the semi-major axis distribution of BPs peaks at $a\simeq \Delta a=a_2-a_1$, with a broad tail extending to $a\gg \Delta a$. 
The BPs are also more likely to have large eccentricities. We can see these features more clearly in the 2D distribution shown in the bottom panel of Fig.~\ref{Figs: JuMBOs' orbital parameter($a_1=0.7a_2$)}.

\section{Results: Additional Runs with Different Parameters}
\label{sec: additional runs}

Our fiducial simulation results presented in Section~\ref{sec: numerical results} are for the system parameters $a_1=0.7a_2$, $e=1.1$ and $m_1=m_2=1 M_{\rm J}$. 
In this section, we present some results for different parameters in order to determine how robust the results are and to explore the conditions under which BPs can be produced with a higher probability.

\subsection{Different initial $a_1/a_2$}
\label{sec: additional runs_a}

We expect that BPs can be produced more frequently when the original
systems have smaller $a_2-a_1$. 
To test this, we carry out simulations for systems with $a_1=0.8a_2$ (with the other parameters the same as the fiducial runs of Section \ref{sec: numerical results}).
Note that although this initial orbital configuration violates the Hill stability criterion (Eq.~(\ref{eq:stability})) and is therefore dynamically unstable, the time for the stability to develop is typically much longer than the dynamical time (the initial orbital period of $m_2$) \citep{Pu2021MNRAS}.
This means that the orbits of the two planets remain mostly intact prior to the close encounter with the flyby. 
Thus it is useful to use the $a_1=0.8a_2$ configuration to explore how the BP formation fraction depends on $\Delta a$. 
We also carry out simulations for systems with $a_1=0.6a_2$ to verify our conclusion. 
Since the BP formation fraction with $a_1=0.6a_2$ is very small, we do not carry out the detailed analysis.

\begin{figure}[htbp]
\centering
\includegraphics[width=1.02\columnwidth]{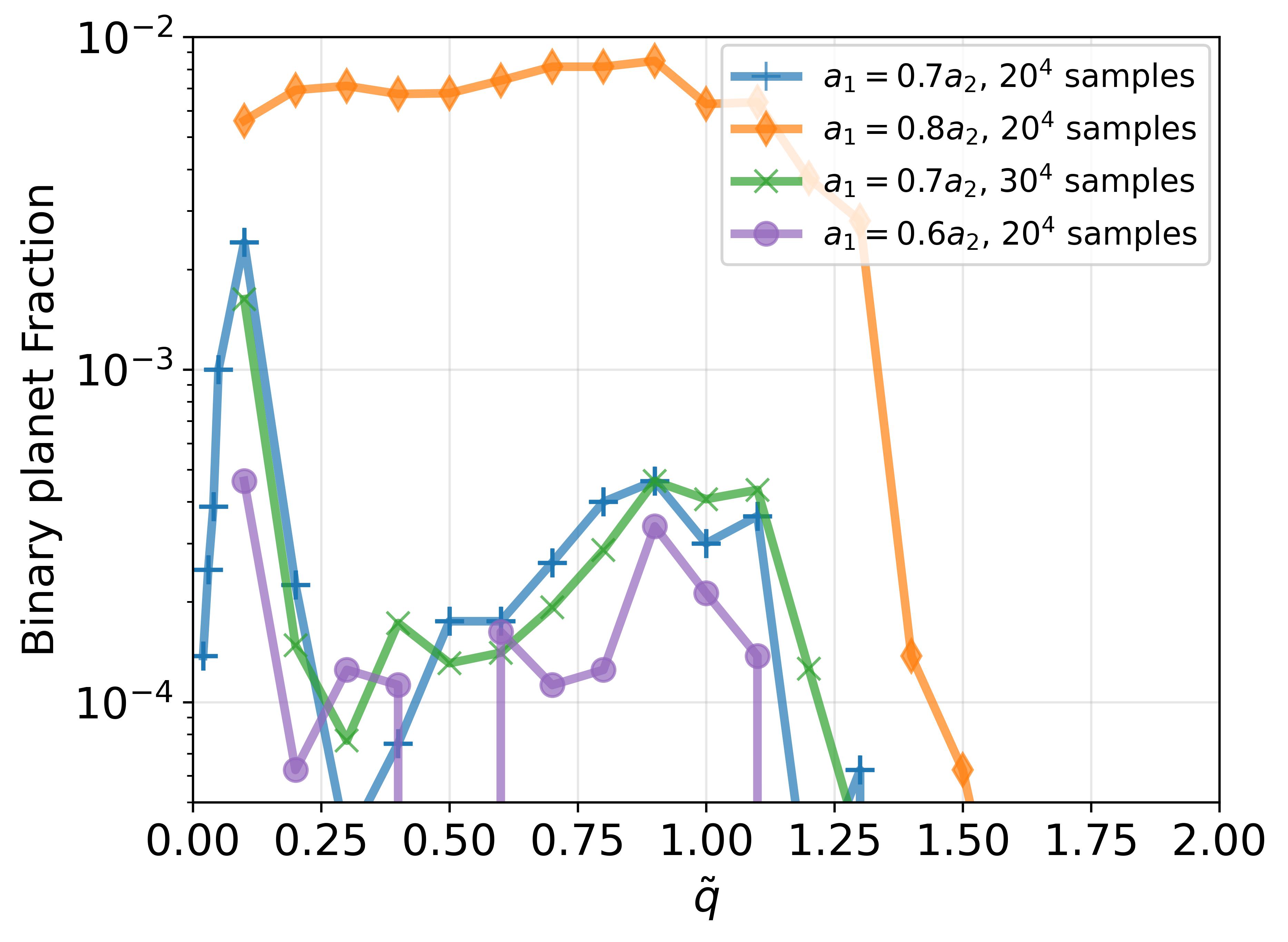}
\caption{
Same as the bottom right panel in Fig.~\ref{Figs: Fractions of Outcomes}, except for the runs with different initial $a_2-a_1$. 
In addition to the results of our fiducial runs with $a_1=0.7a_2$ that are based on $20^4$ samplings (blue) and $30^4$ samplings (green) of angles, we also show the results of an expanded set of simulations with $a_1=0.8a_2$ (orange) and $a_1=0.6a_2$ (purple) based on $20^4$ samplings of angles.
}
\label{JuMBOs0.70.8}
\end{figure}

\begin{figure}[htbp]
        \subfigure
        {
            \begin{minipage}[b]{\linewidth} 
                \centering
                \includegraphics[width=1.02\columnwidth]{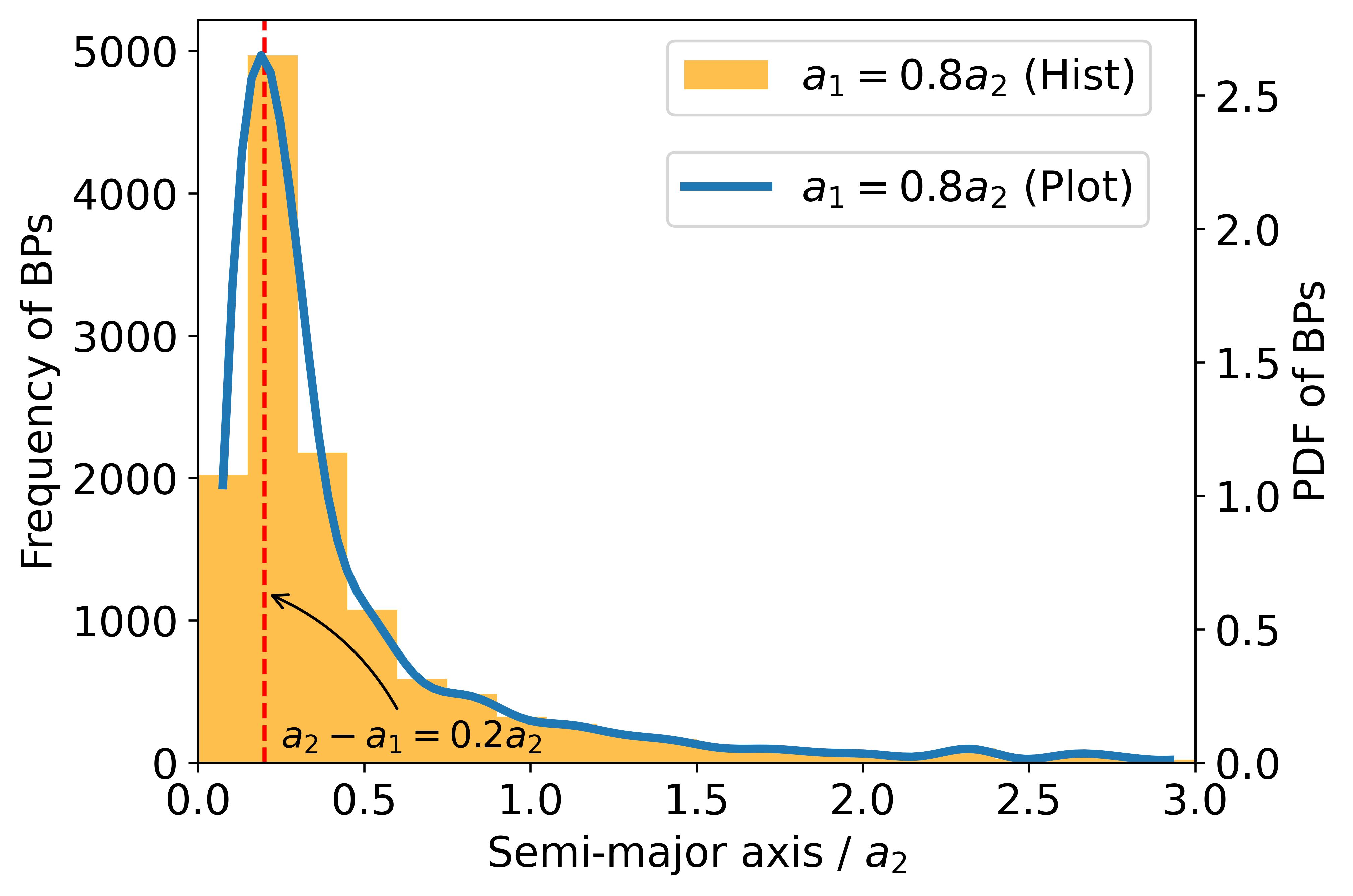}
                \label{JuMBO SMA(a=0.8)}
            \end{minipage}
        }
        \subfigure
        {
            \begin{minipage}[b]{\linewidth}
                \centering
                \includegraphics[width=1.02\columnwidth]{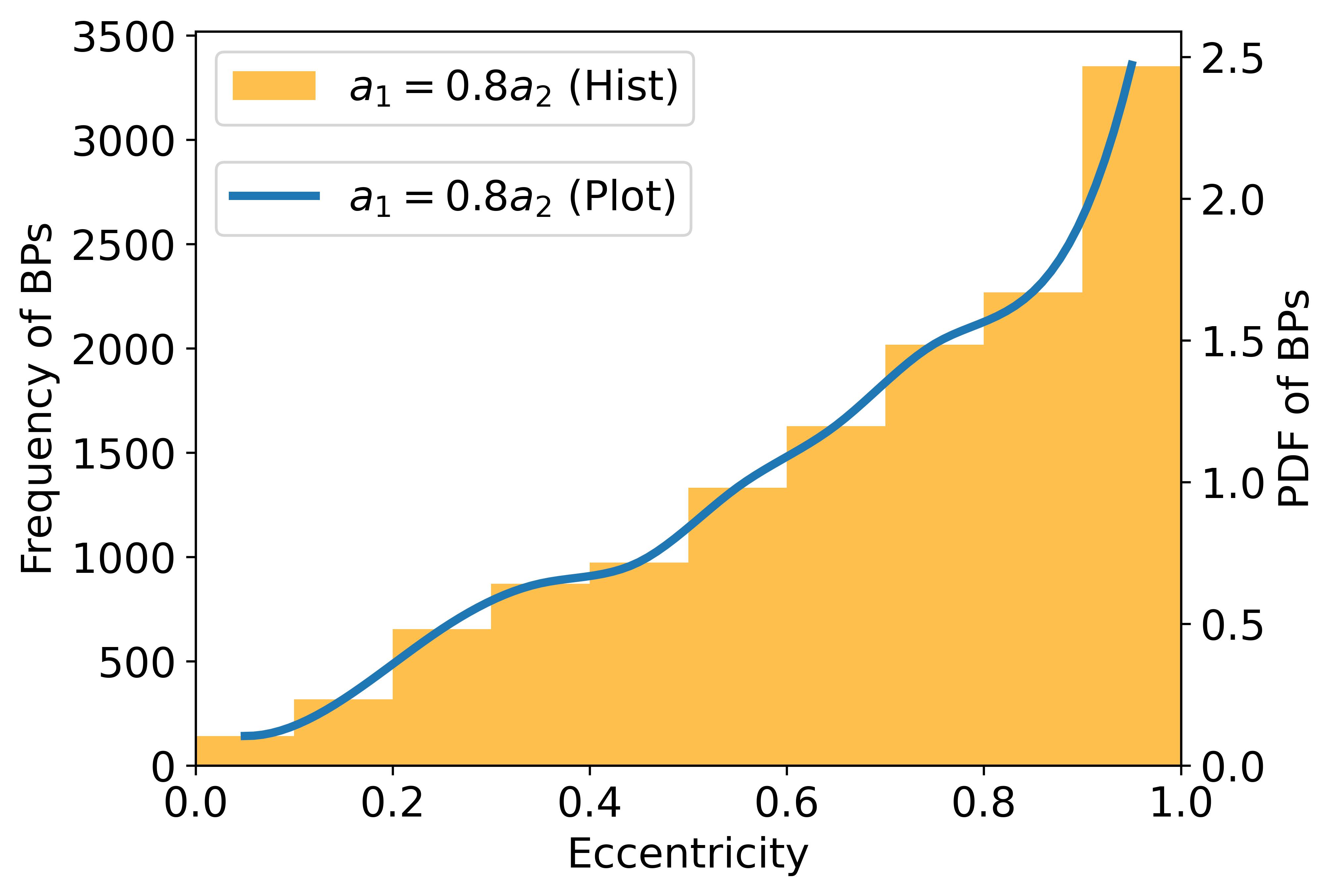}
                \label{JuMBO e(a=0.8)}
            \end{minipage}
        }
        \subfigure
        {
            \begin{minipage}[b]{\linewidth}
                \centering
                \includegraphics[width=1.02\columnwidth]{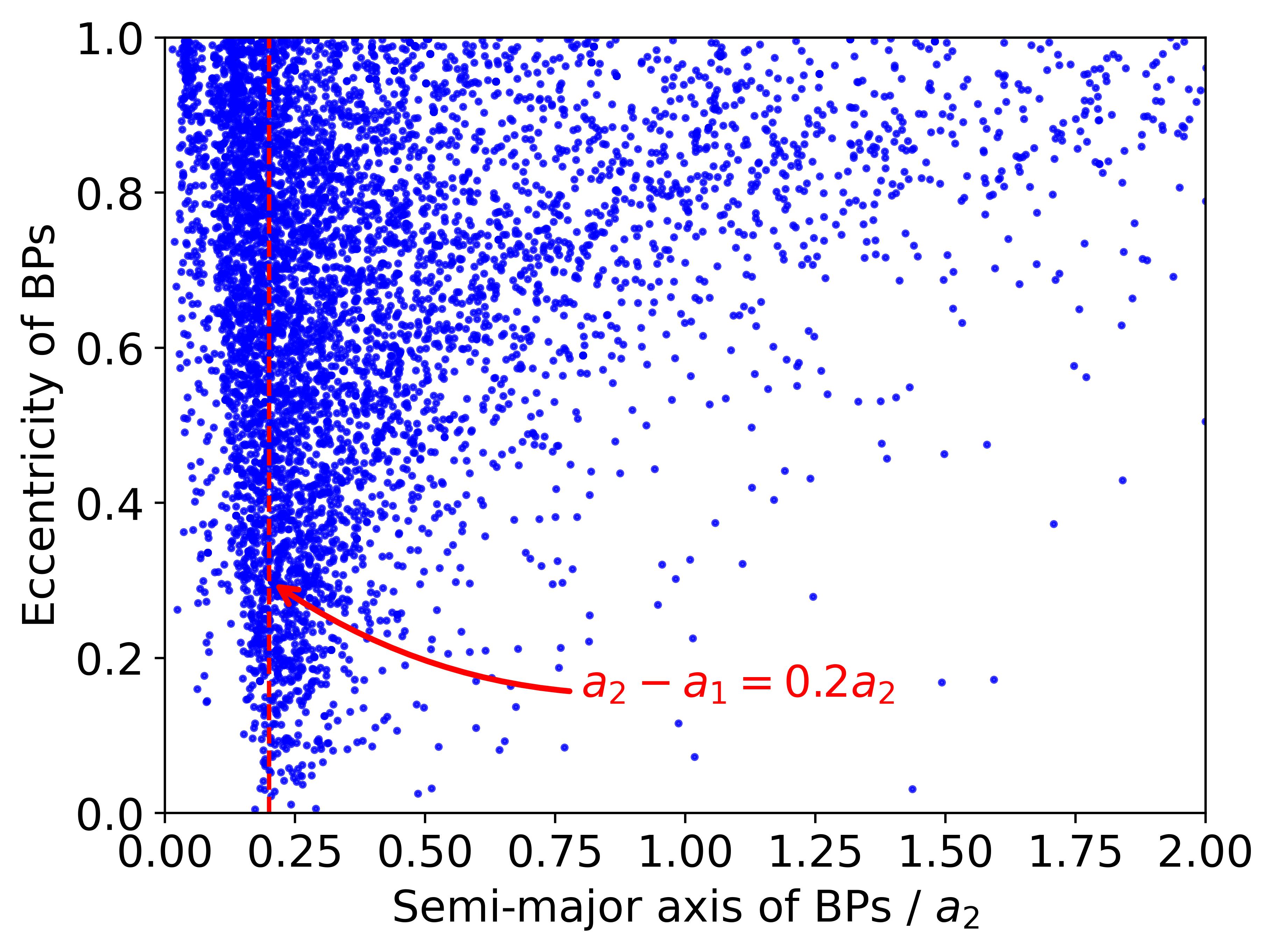}
                \label{JuMBO e-a(a=0.8)}
            \end{minipage}
        }
        \caption{
        Similar to Fig.~\ref{Figs: JuMBOs' orbital parameter($a_1=0.7a_2$)}, except for the case of $a_1=0.8a_2$, and the simulation runs sample $20^4$ angles.
        }
        \label{JuMBOs' orbital parameter($a_1=0.8a_2$)}
\end{figure}

Fig.~\ref{JuMBOs0.70.8} shows that the fraction of BP formation for $a_1=0.8a_2$ is indeed much larger than for $a_1=0.7a_2$ (see the bottom right panel in Fig.~\ref{Figs: Fractions of Outcomes}), typically by a factor of about 10.
(Because of the large number of BPs, our simulation runs adopt $20 \times 20 \times 20 \times 20$ sampling for the four angles.)
Fig.~\ref{JuMBOs' orbital parameter($a_1=0.8a_2$)} shows the semi-major axis and eccentricity distributions of BPs for $a_1=0.8a_2$. 
These are qualitatively similar to the $a_1=0.7a_2$ case (see Fig.~\ref{Figs: JuMBOs' orbital parameter($a_1=0.7a_2$)}). 
In particular, the semi-major axis distribution peaks at $a_2-a_1=0.2a_2$.

\subsection{Different planet masses}
\label{sec: additional runs_m}

Since the formation of free-floating BPs relies on planet-planet interactions, we expect that BPs can be produced more frequently for larger planet masses.
To test this, we carry out simulations for systems with different planet-to-star mass ratio combinations $m/M$ ranging from $0.001$ to $0.004$. 
Note that in order to satisfy the stability condition (Eq.~(\ref{eq:stability})) for different masses, we set $a_1$ to $0.6a_2$ instead of $0.7a_2$ in our fiducial setup.

\begin{figure}[htbp]
\centering
\includegraphics[width=1.02\columnwidth]{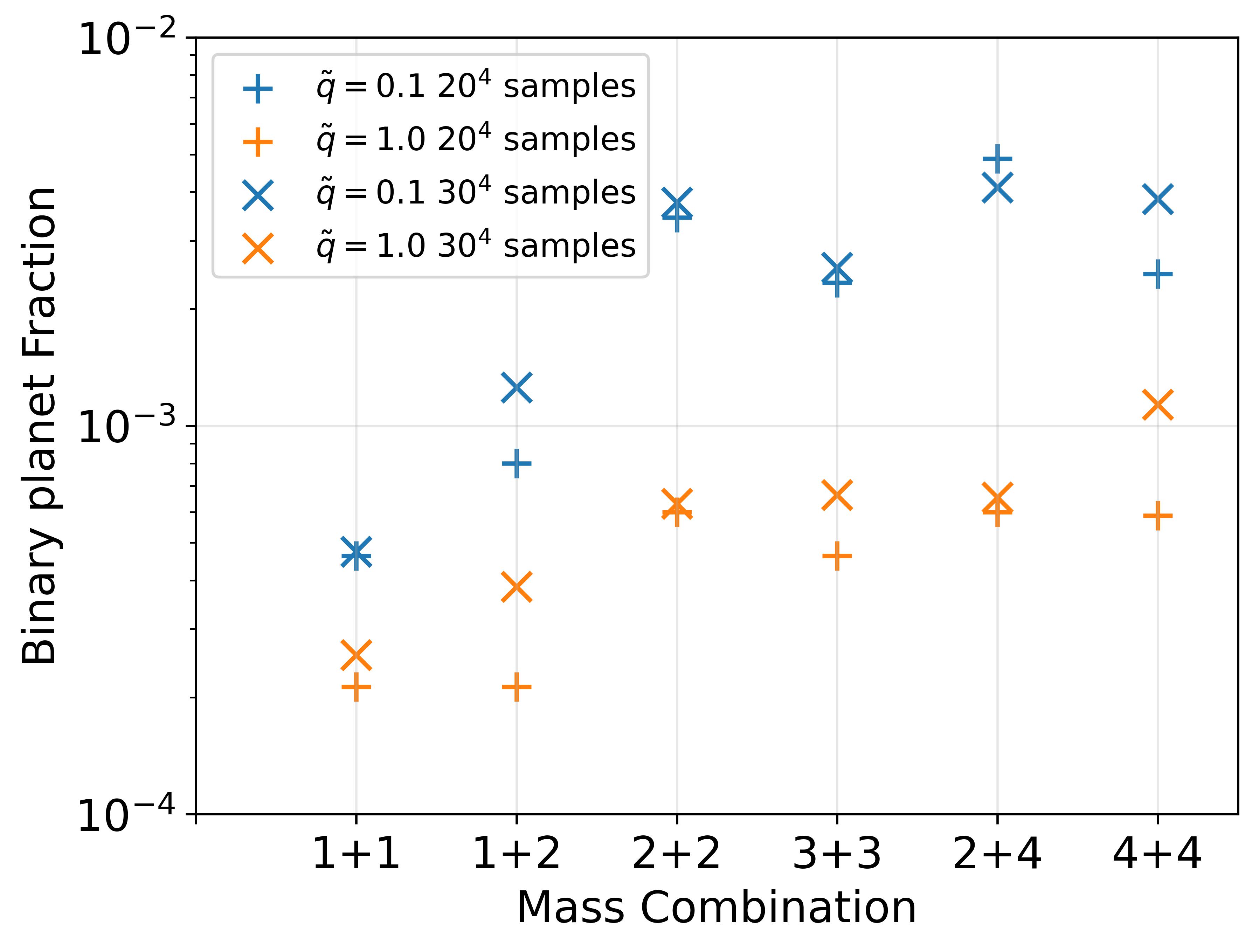}
\caption{
The fractions of forming BPs with different planet mass combinations for the $e=1.1$, $a_1=0.6a_2$ runs. 
The blue markers in the figure represent the results for $\tilde q=0.1$, and the orange markers represent the results for $\tilde q=1$. 
The ``+" and ``$\times$" represent $20^4$ and $30^4$ samplings of angles, respectively. The ticks on the x-axis represent the masses of the two planets (in units of $M_{\rm J}$). 
The vertical distance between different types of markers of the same color serves as an indicator of numerical fluctuations.
}
\label{Differentm}
\end{figure}

From the bottom right panel in Fig.~\ref{Figs: Fractions of Outcomes} we know that the BP fraction exhibits a sharp peak at $\tilde q\simeq 0.1$ and a broad peak at $\tilde q\sim 1$. 
So we focus on these two $\tilde q$ values and show the result in Fig.~\ref{Differentm}.
We see that the fraction of BP formation generally increases with increasing $m/M$, although the dependence may not be exactly monotonic. 
In all cases, we find that the BP formation fraction is less than $0.5\%$ (for $a_1=0.6a_2$).

\subsection{Different flyby velocities: $V_\infty/v_{2i}$}
\label{sec: additional runs_v}

All results presented in Section~\ref{sec: numerical results} (with the setup described in Section~\ref{sec: method}) are based on the assumption of a slightly hyperbolic orbit for the flyby, with $e=1.1$.
Such a flyby trajectory has the advantage that some of the results for the inner and outer planets satisfy a simple scaling relation
(see Fig.~\ref{Figs: Fractions of Outcomes}, \ref{v_infty distribution for free-floating planets} and \ref{Figs: Captured planet orbital parameter}).

For a given pericenter distance $q$, the flyby has a velocity at infinity $V_\infty=\sqrt{GM(e-1)/q}$. 
For $q\sim a_2$ and $e=1.1$, this implies $V_\infty\ll v_{2i}\equiv\sqrt{GM/a_2}$. 
For planetary systems with $a_2\sim 100$~au, we find $v_{2i}\simeq 3$~km/s, comparable to the typical $V_\infty$ in star clusters, corresponding to an appreciably hyperbolic orbit for the flyby.
In order to evaluate how such a hyperbolic flyby orbit affects the results of Section~\ref{sec: numerical results}, we carry out a suite of flyby simulations with $V_\infty=v_{2i}$.

The light red lines in Fig.~\ref{Figs: Fractions of Outcomes} represent the results for the outer planets in the $V_\infty=v_{2i}$ flybys. 
Comparing to the corresponding $e=1.1$ results (the orange lines), we see that the fraction of planet capture decreases significantly, and the fraction of the survivor planets increases correspondingly. 

\begin{figure}[htbp]
        \subfigure
        {
            \begin{minipage}[b]{\linewidth} 
                \centering
                \includegraphics[width=1.02\columnwidth]{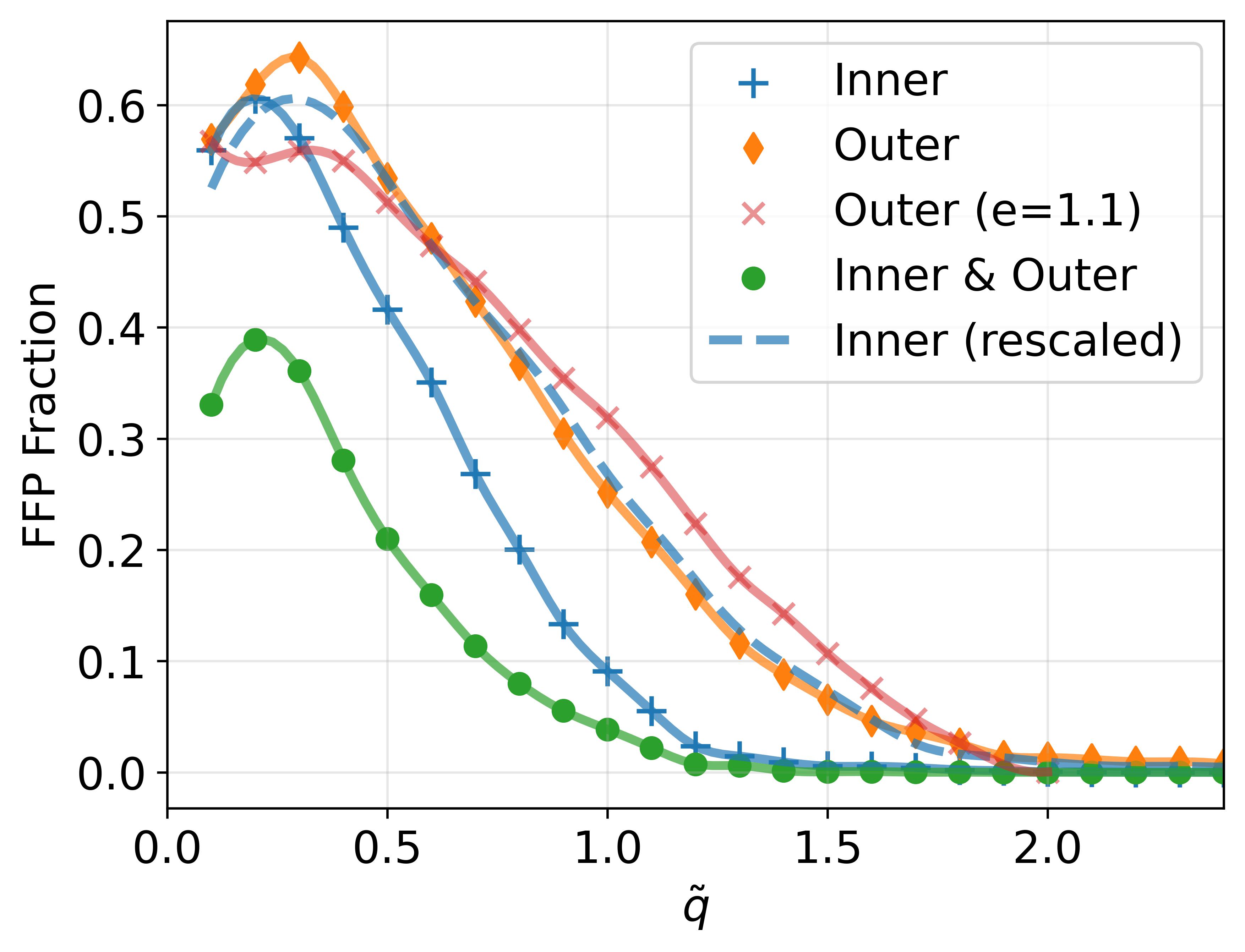}
                \label{Lv The fraction of forming free-floating planets}
            \end{minipage}
        }
        \subfigure
        {
            \begin{minipage}[b]{\linewidth}
                \centering
                \includegraphics[width=1.02\columnwidth]{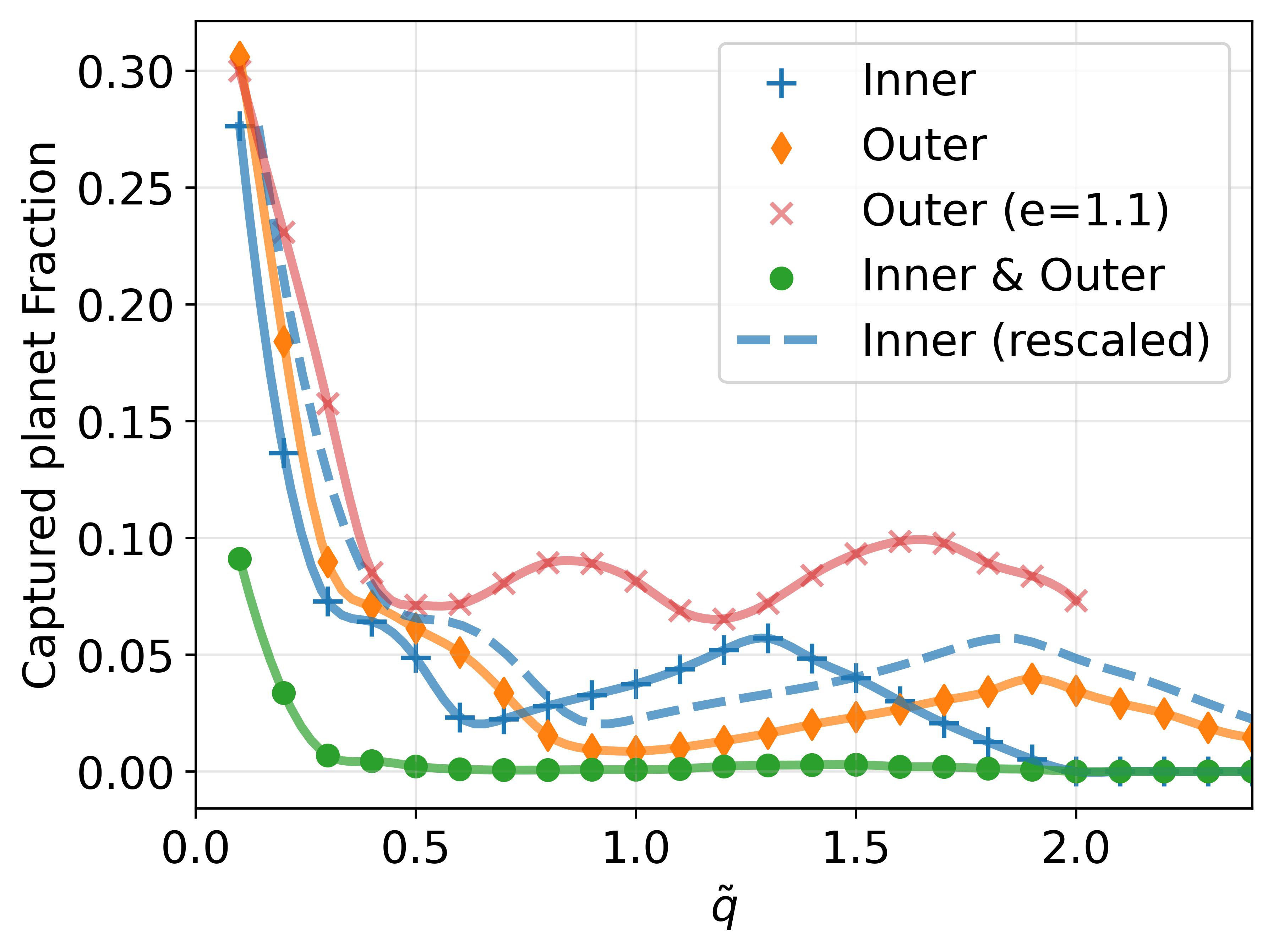}
                \label{Lv The fraction of planet stolen}
            \end{minipage}
        }
        \subfigure
        {
            \begin{minipage}[b]{\linewidth}
                \centering
                \includegraphics[width=1.02\columnwidth]{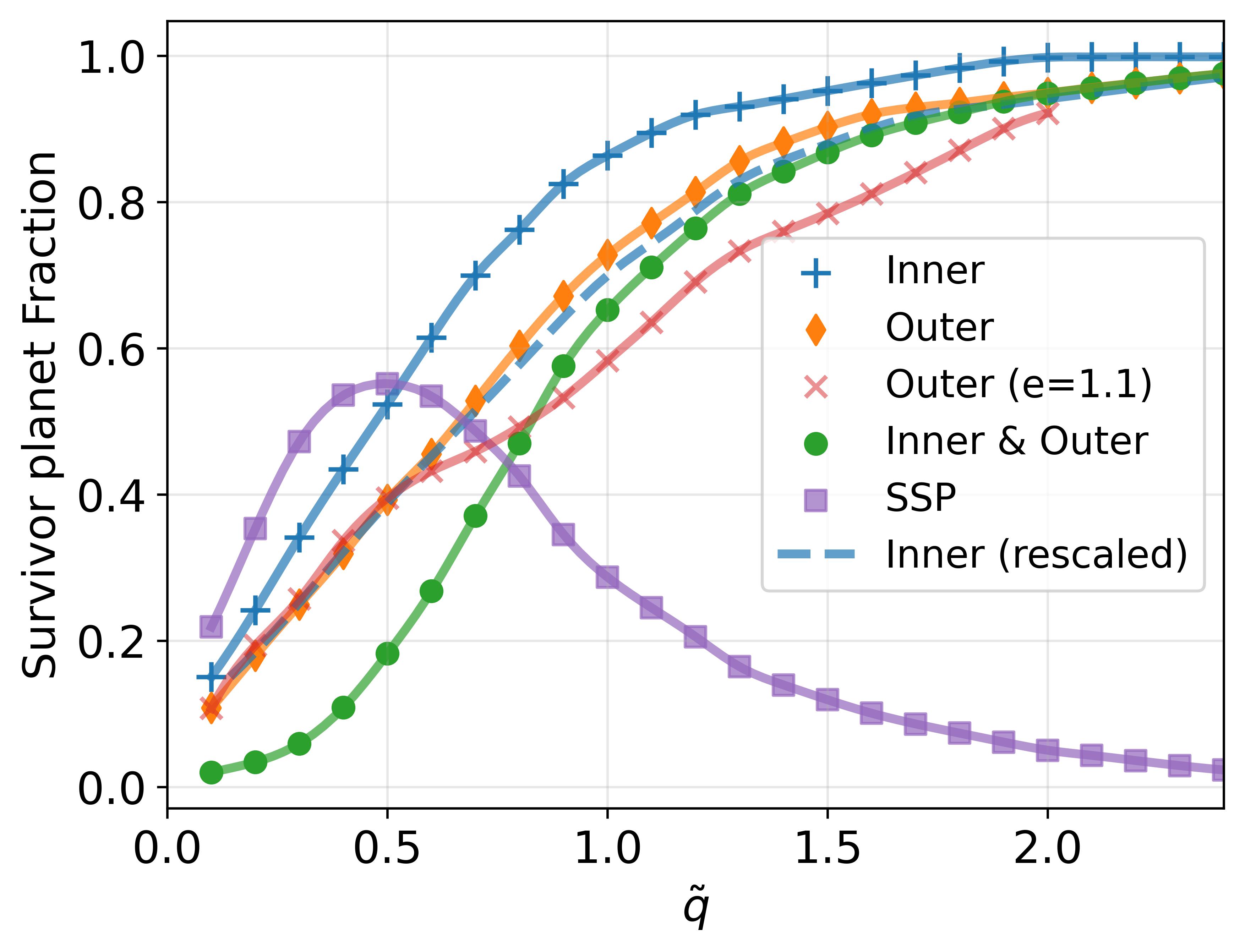}
                \label{Lv The fraction of remaining bound}
            \end{minipage}
        }
        \caption{
        Same as the first three panels in Fig.~\ref{Figs: Fractions of Outcomes}, except for the runs with $V_\infty=v_{2i}\equiv\sqrt{GM/a_2}$. Note that here the light-red lines (labeled ``$e=1.1$") show the result (for the outer planet only) based on the simulations with $e=1.1$ for the flyby trajectory (fiducial runs, see Section~\ref{sec: numerical results}).}
        
        \label{Figs: Lv fraction of outcomes}
\end{figure}

\begin{figure}[htbp]
\centering
\includegraphics[width=1.02\columnwidth]{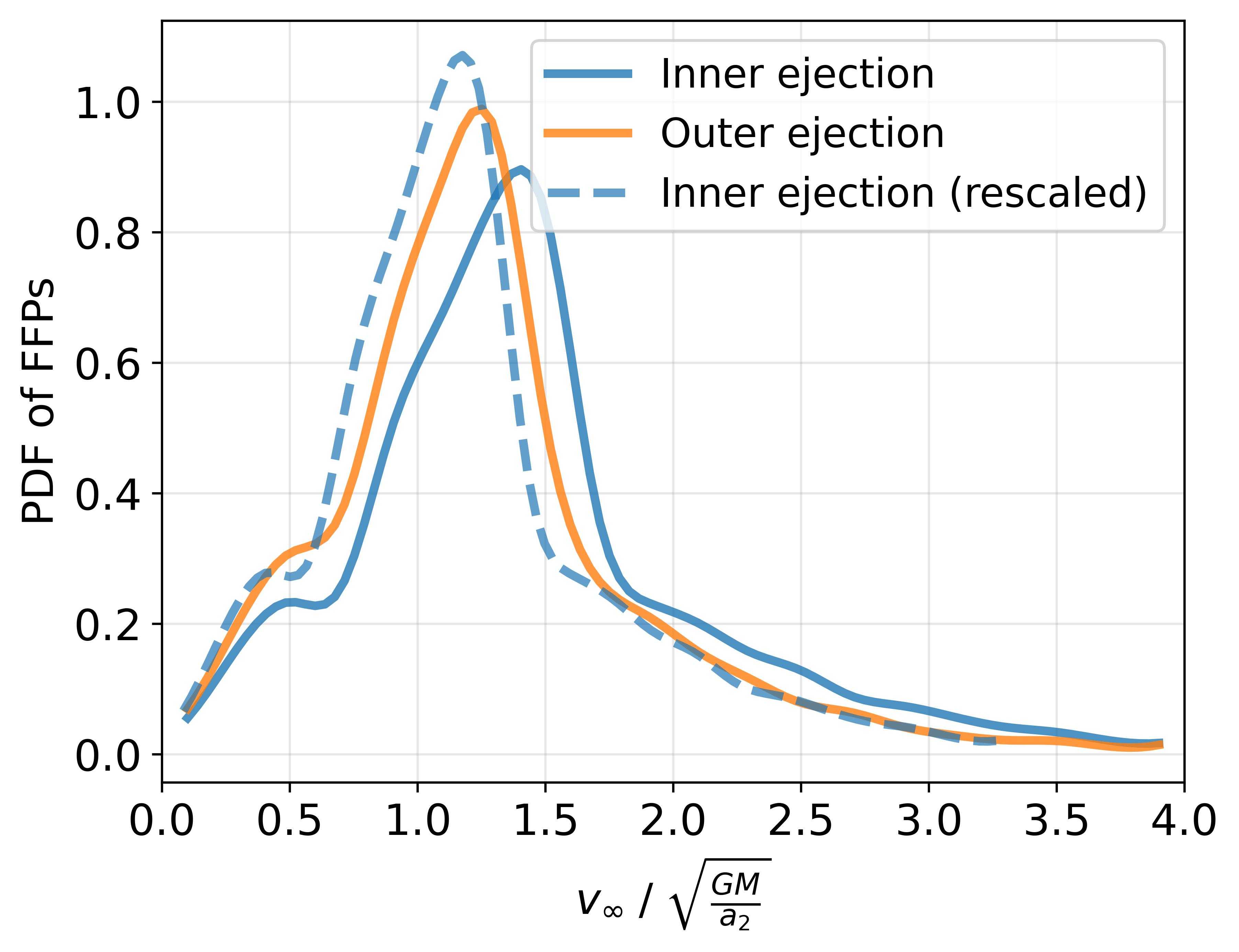}
\caption{
Same as Fig.~\ref{v_infty distribution for free-floating planets}, except for the runs with $V_\infty=v_{2i}\equiv\sqrt{GM/a_2}$.
}
\label{Lv v_infty distribution for free-floating planets}
\end{figure}

\begin{figure*}[htbp]
    \centering
    \begin{minipage}[b]{0.49\linewidth}
        \centering
        \includegraphics[width=\linewidth]{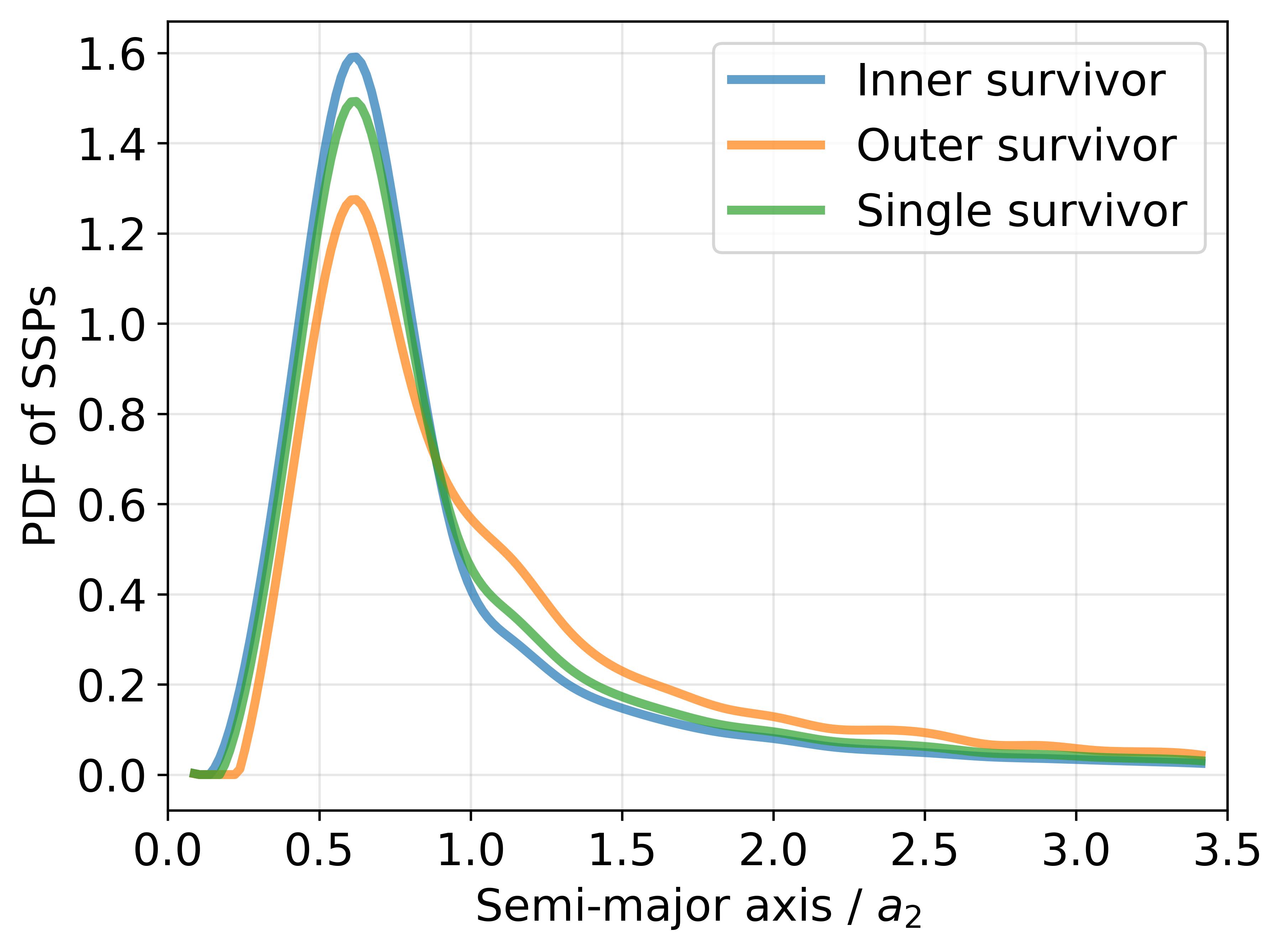}
    \end{minipage}
    \hfill
    \begin{minipage}[b]{0.49\linewidth}
        \centering
        \includegraphics[width=\linewidth]{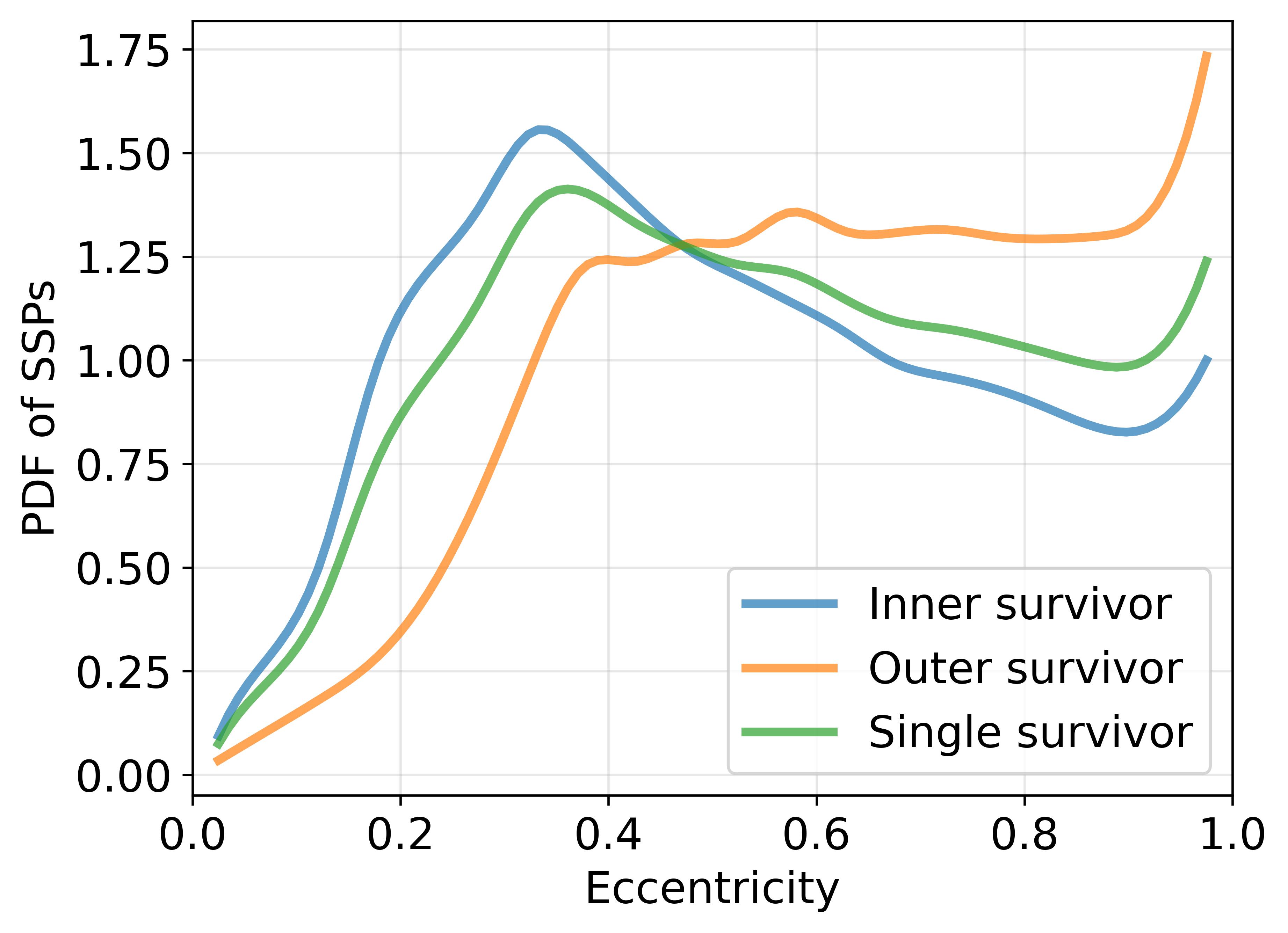}
    \end{minipage}

    \begin{minipage}[b]{0.49\linewidth}
        \centering
        \includegraphics[width=\linewidth]{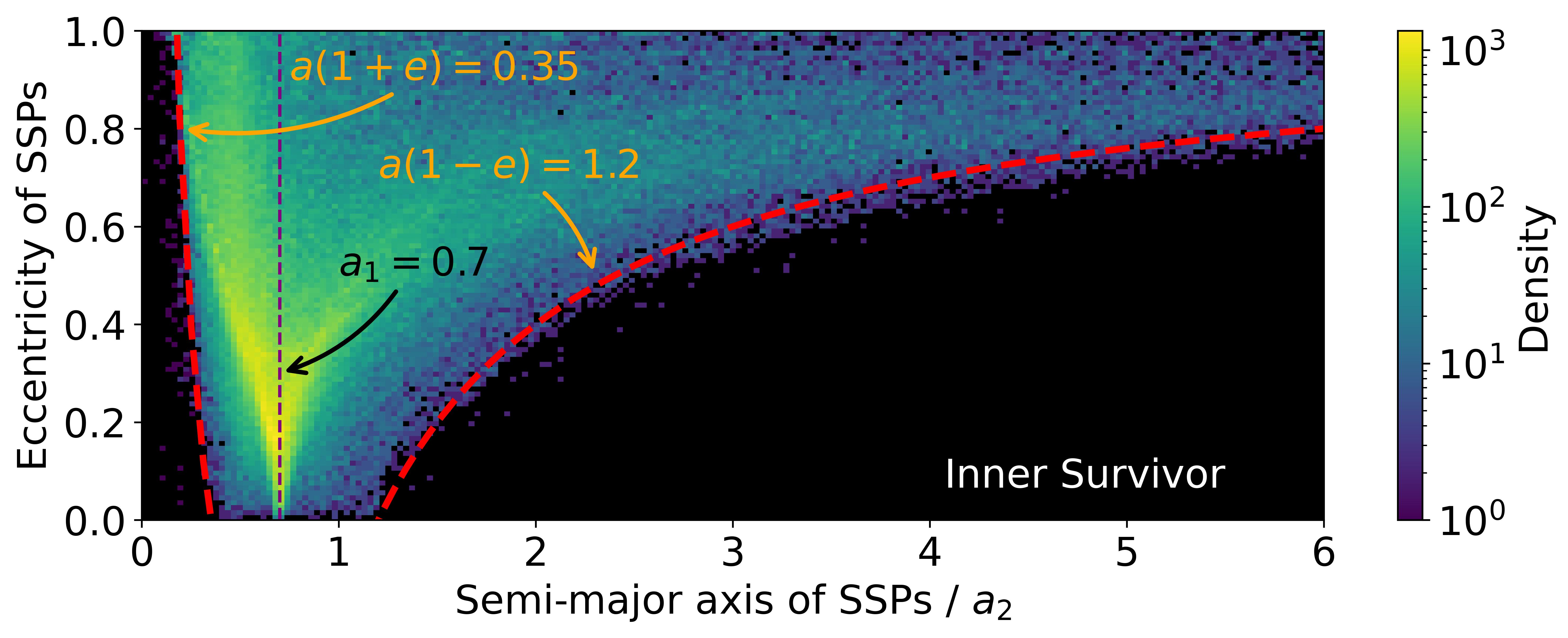}
    \end{minipage}
    \hfill
    \begin{minipage}[b]{0.49\linewidth}
        \centering
        \includegraphics[width=\linewidth]{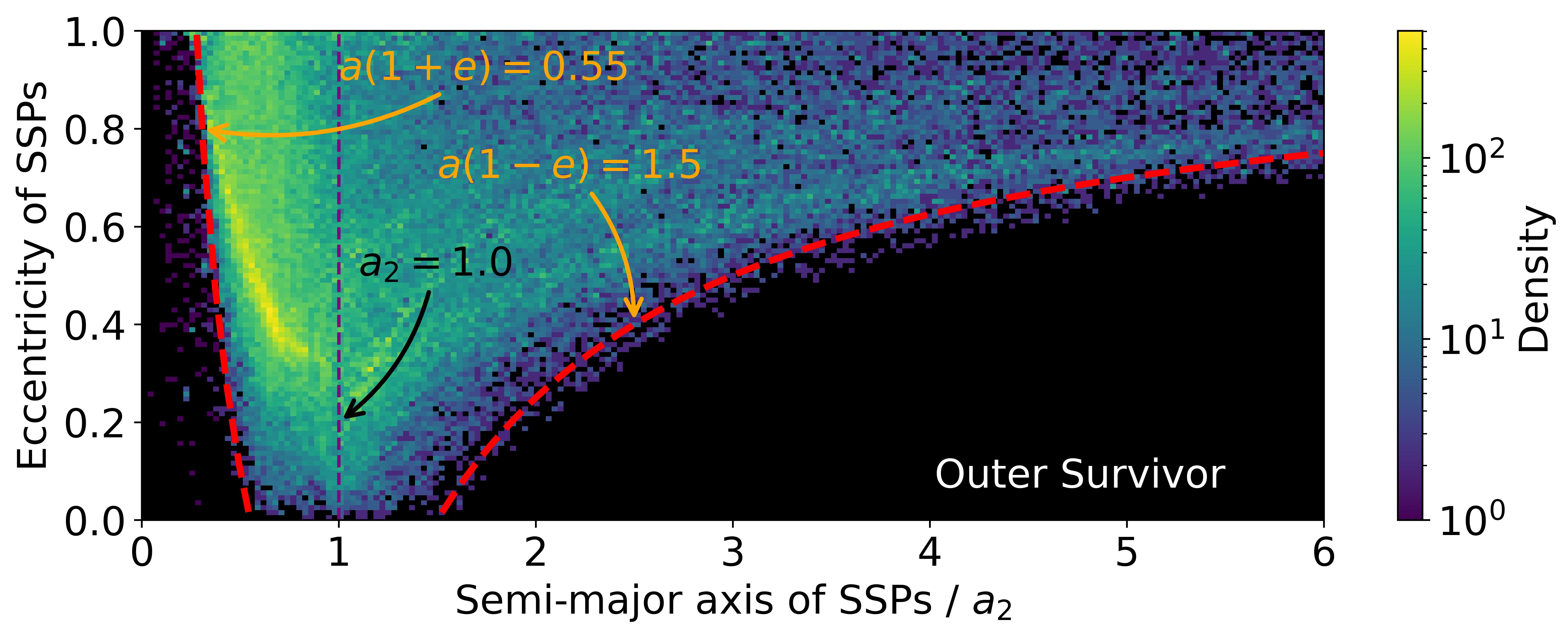}
    \end{minipage}
    \caption{
    Same as Fig.~\ref{Figs: Single planet system orbital parameter}, except for the runs with $V_\infty=v_{2i}\equiv\sqrt{GM/a_2}$.}
    \label{Figs: Lv Single planet system orbital parameter}
\end{figure*}

\begin{figure*}[htbp]
    \centering
    \begin{minipage}[b]{0.49\linewidth}
        \centering
        \includegraphics[width=\linewidth]{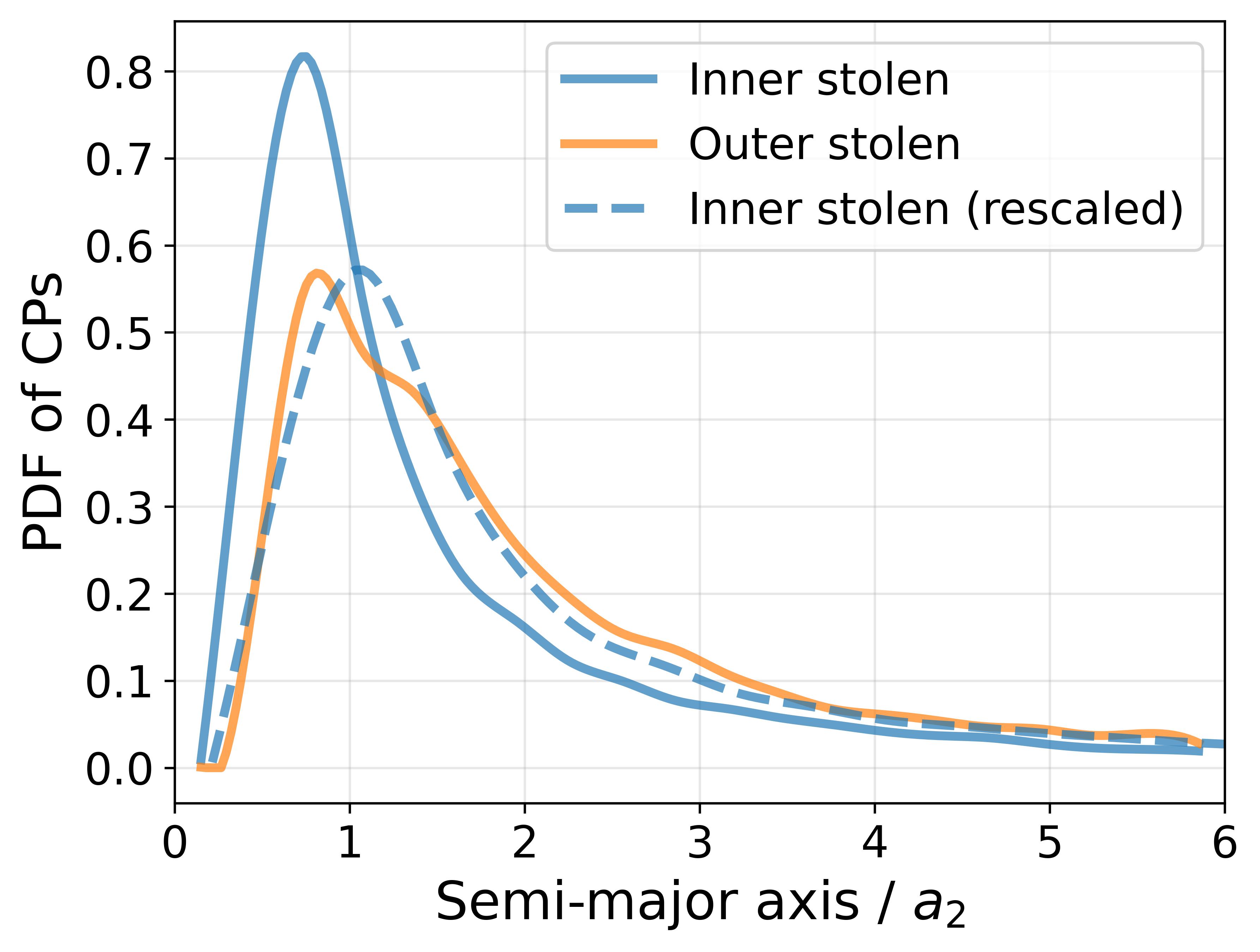}
    \end{minipage}
    \hfill
    \begin{minipage}[b]{0.49\linewidth}
        \centering
        \includegraphics[width=\linewidth]{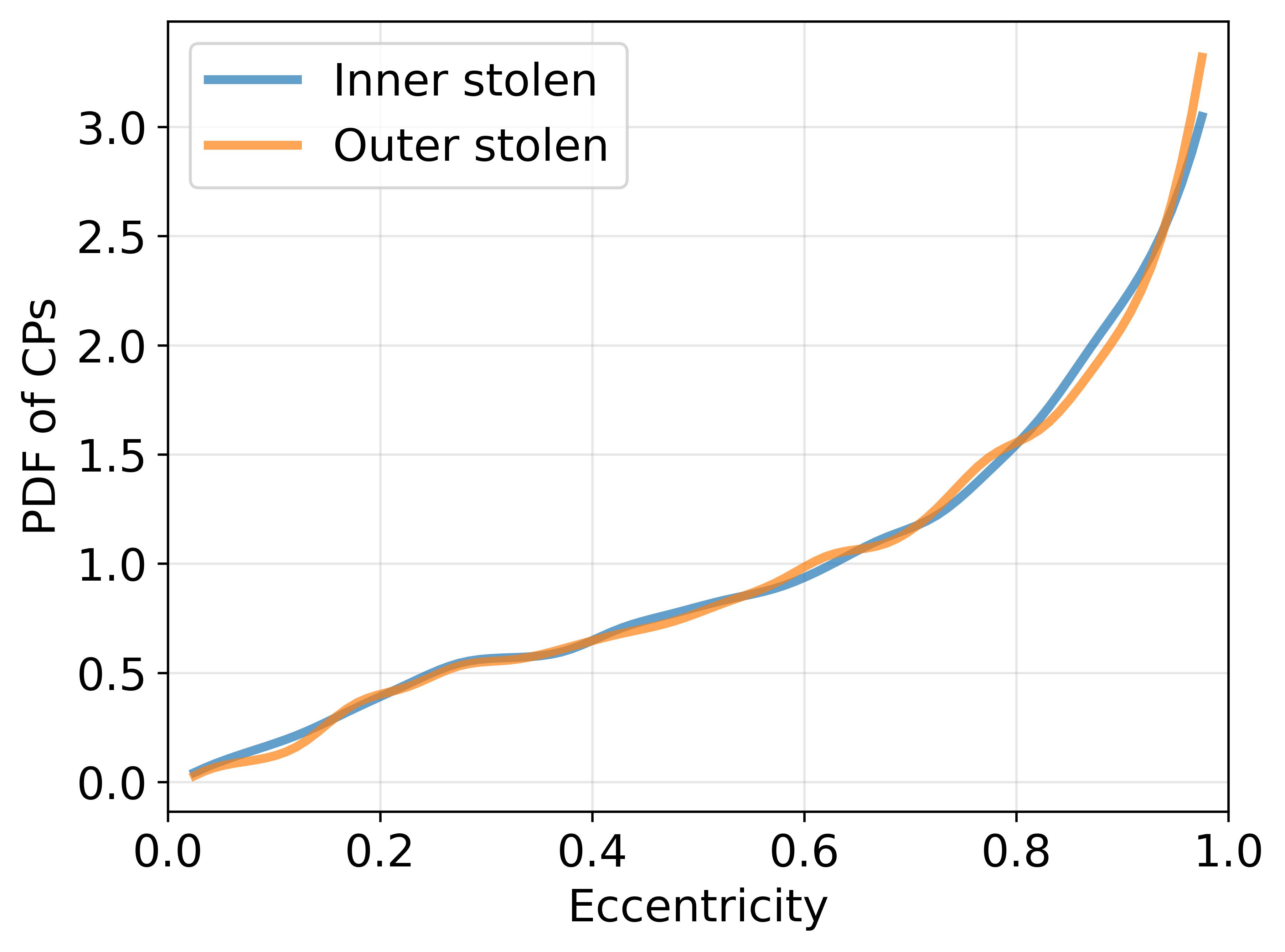}
    \end{minipage}
    
    \begin{minipage}[b]{0.49\linewidth}
        \centering
        \includegraphics[width=\linewidth]{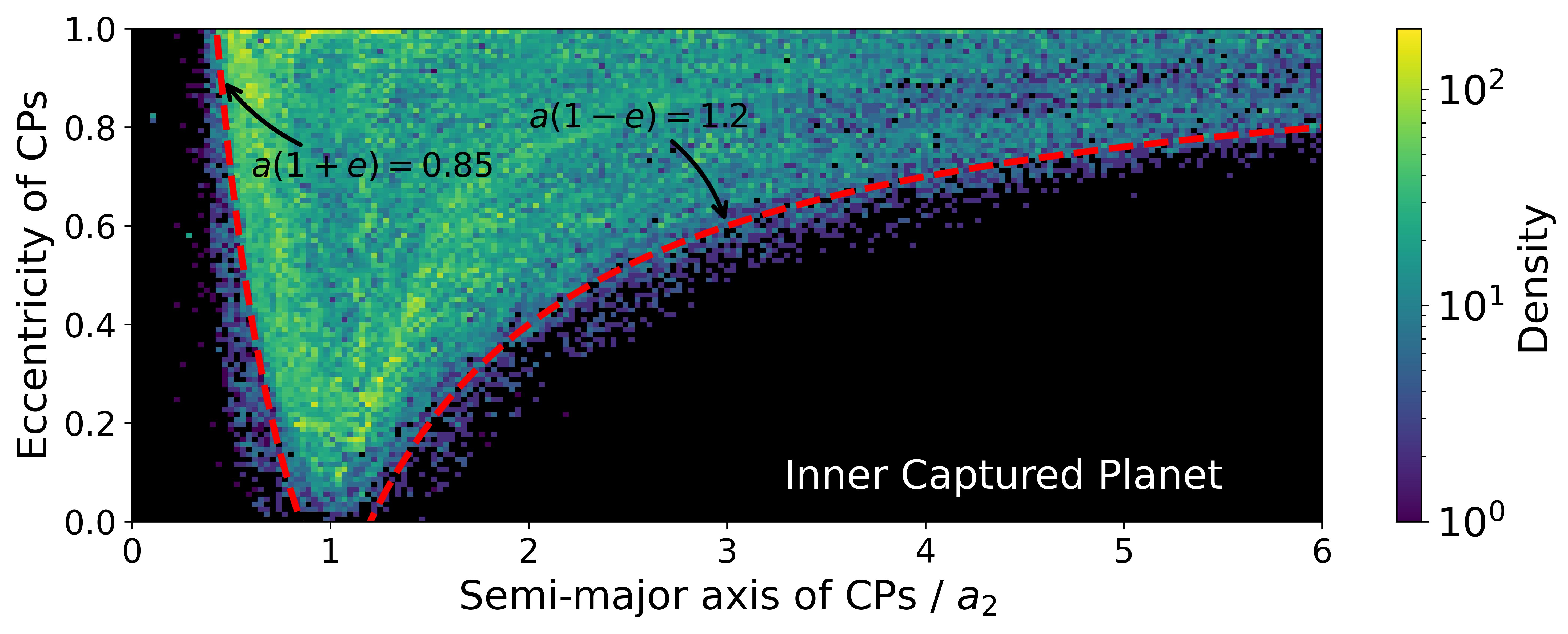}
    \end{minipage}
    \hfill
    \begin{minipage}[b]{0.49\linewidth}
        \centering
        \includegraphics[width=\linewidth]{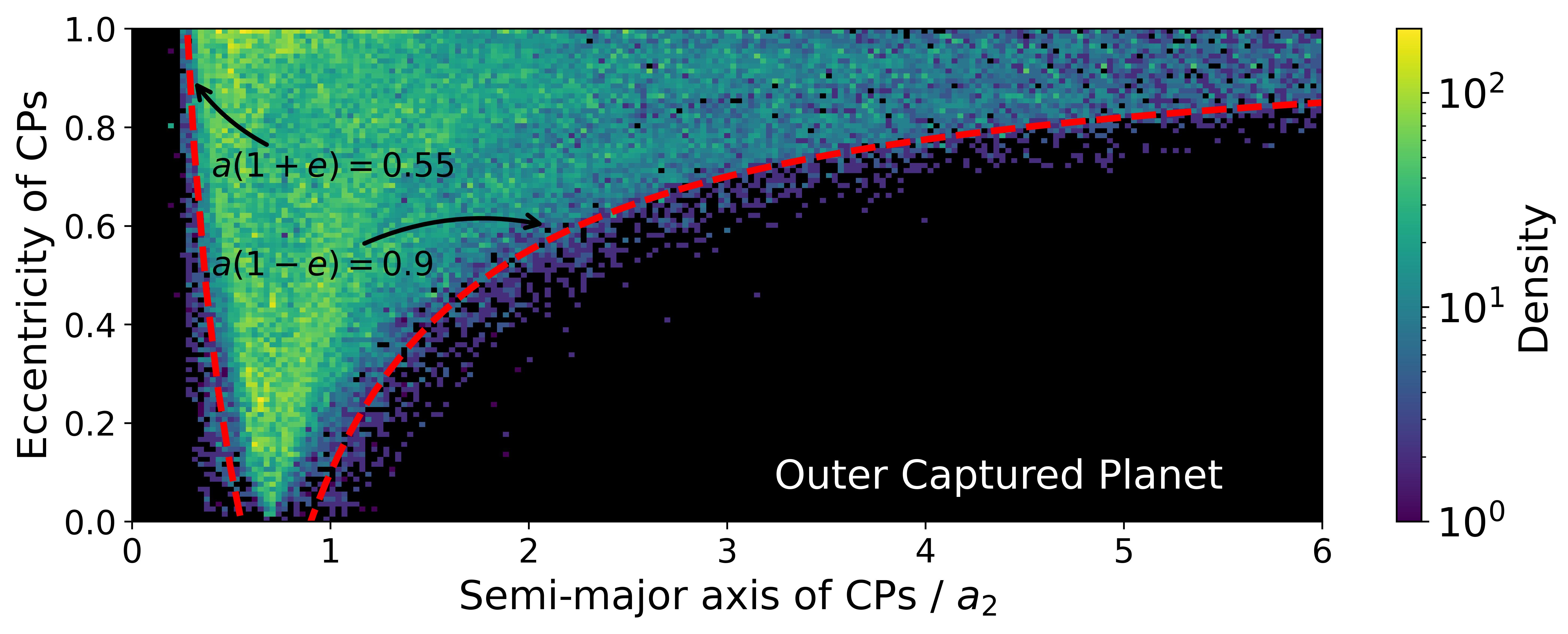}
    \end{minipage}
    
    \caption{
    Same as Fig.~\ref{Figs: Captured planet orbital parameter}, except for the runs with $V_\infty=v_{2i}\equiv\sqrt{GM/a_2}$.}
    \label{Figs: Largev Captured planet orbital parameter}
\end{figure*}

\begin{figure}[htbp]
        \subfigure
        {
            \begin{minipage}[b]{\linewidth} 
                \centering
                \includegraphics[width=\columnwidth]{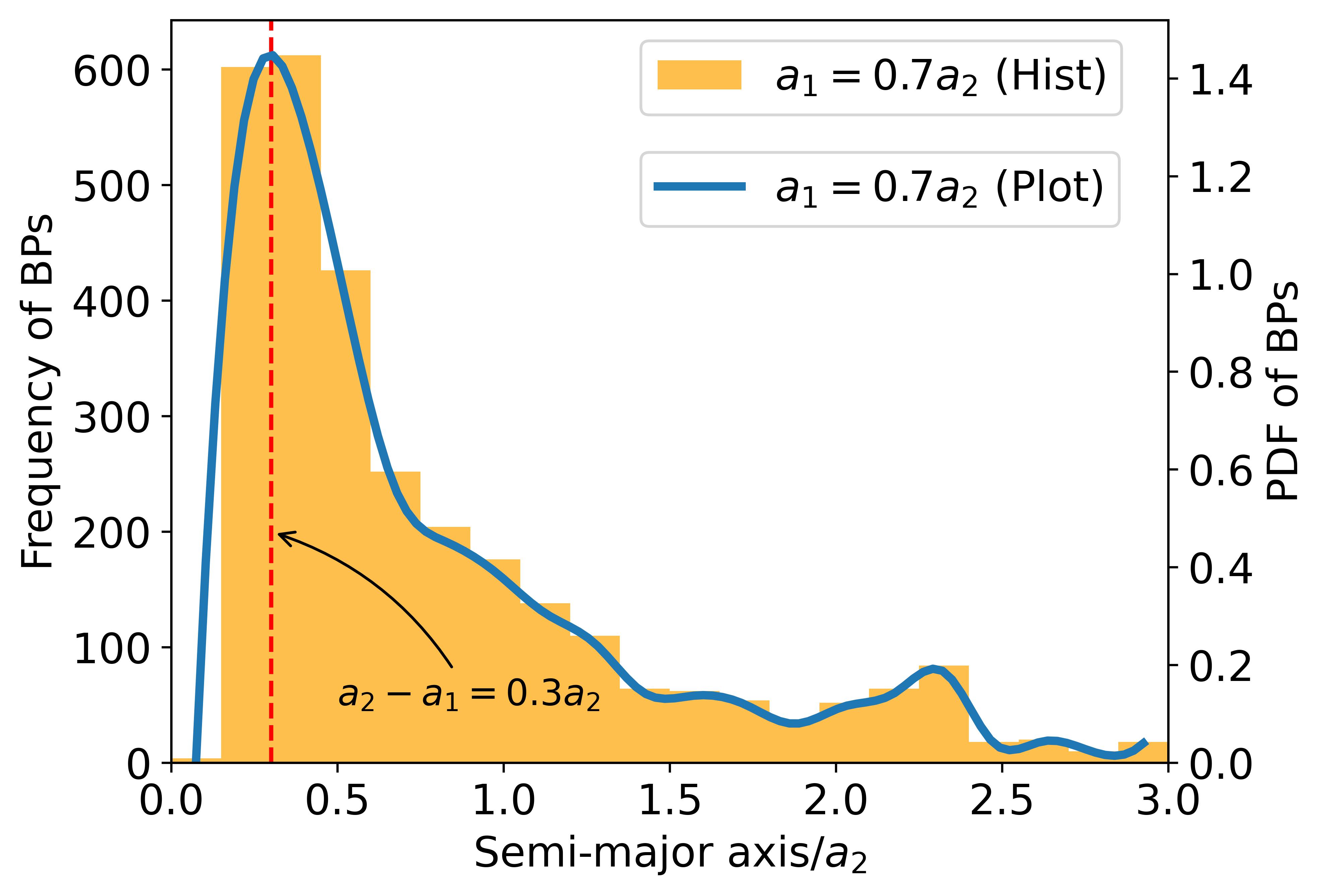}
                \label{Lv JuMBO SMA(a=0.7)}
            \end{minipage}
        }
        \subfigure
        {
            \begin{minipage}[b]{\linewidth}
                \centering
                \includegraphics[width=\columnwidth]{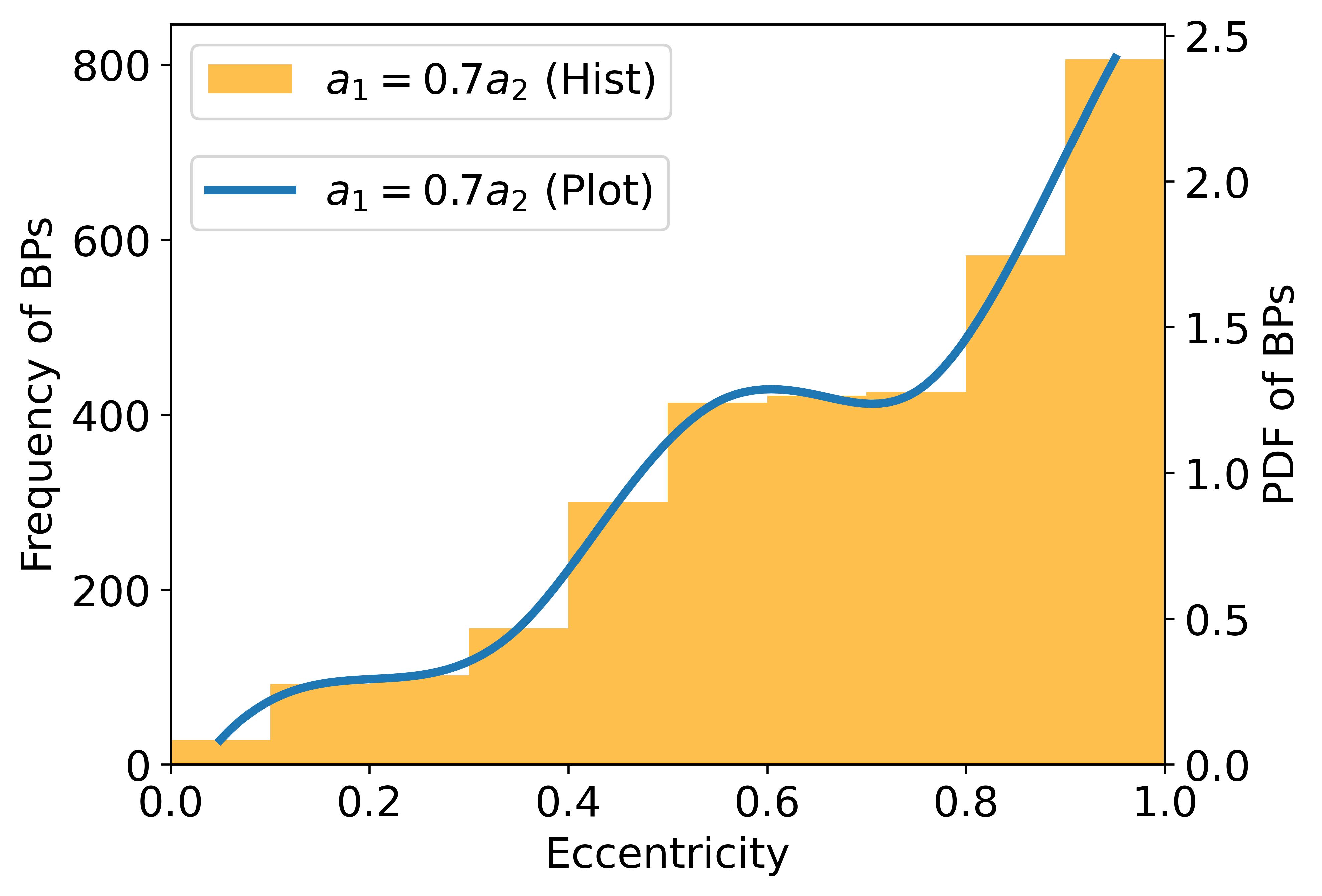}
                \label{Lv JuMBO e(a=0.7)}
            \end{minipage}
        }
        \subfigure
        {
            \begin{minipage}[b]{\linewidth}
                \centering
                \includegraphics[width=\columnwidth]{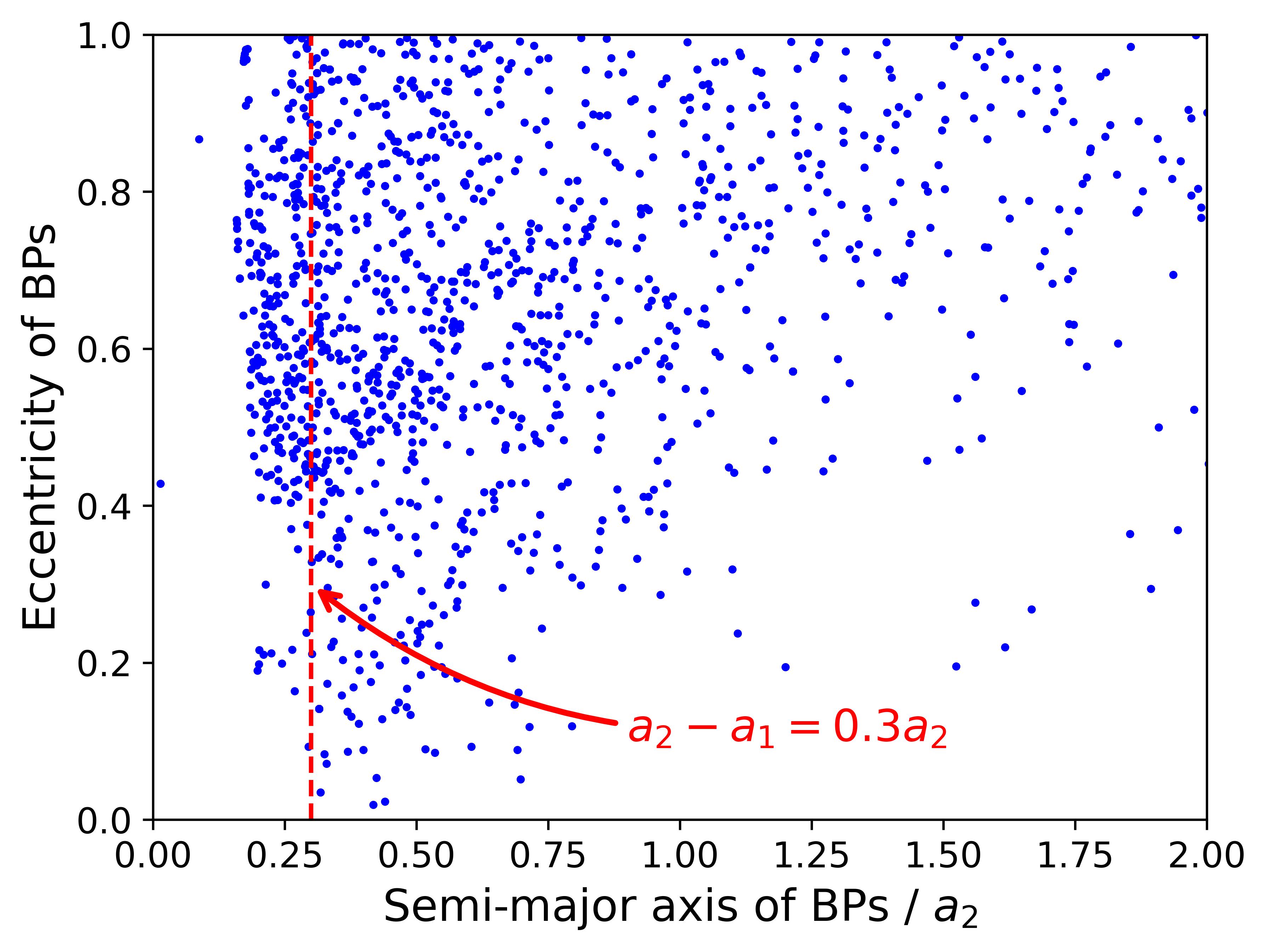}
                \label{Lv JuMBOs' e-a distribution 0.7}
            \end{minipage}
        }
        \caption{
        Same as Fig.~\ref{Figs: JuMBOs' orbital parameter($a_1=0.7a_2$)}, except for the runs with $V_\infty=v_{2i}\equiv\sqrt{GM/a_2}$.}
        \label{Figs: Largev JuMBOs' orbital parameter($a_1=0.7a_2$)}
\end{figure}

Note that the orbital velocities ($v_{1i}=\sqrt{GM/a_1}$ and $v_{2i}$) of the inner planet and the outer planet are different, so the scaling relations we see for the $e=1.1$ runs do not apply for the $V_\infty=v_{\rm orb}(a_2)=v_{2i}$ runs. 
However, the rescaled inner-planet results (with $\tilde q$ replaced by $q/a_1$ in Fig.~\ref{Figs: Lv fraction of outcomes}, and $v_{\rm free}/v_{2i}$ by $v_{\rm free}/v_{1i}$ in Fig.~\ref{Lv v_infty distribution for free-floating planets}) should correspond to the single-planet simulations with $V_\infty=\sqrt{7/10}\,v_{\rm orb}(a_1)$.
Therefore, we can compare the blue dotted line with the yellow solid line in Fig.~\ref{Figs: Lv fraction of outcomes}, \ref{Lv v_infty distribution for free-floating planets} to evaluate the effect of the different $V_\infty/v_{\rm orb}$ value on the results. 
Combining this perspective with the results shown in Section~\ref{sec: numerical results}, we can conclude that as $V_\infty/v_{\rm orb}$ increases, the fraction of free-floating planets does not change much (see the top panel of Fig.~\ref{Figs: Lv fraction of outcomes}),  but the fraction of capture planets decreases significantly (see the middle panel of Fig.~\ref{Figs: Lv fraction of outcomes}), and the fraction of survivor planets also increases significantly. 
At the same time, as can be seen from Fig.~\ref{Lv v_infty distribution for free-floating planets}, the velocities of the 
FFPs tend to increase as $V_\infty/v_{\rm orb}$ increases. 
The other PDFs (for SSPs, CPs and BPs) for the large-$V_\infty$ runs 
(Figs.~\ref{Figs: Lv Single planet system orbital parameter}-\ref{Figs: Largev JuMBOs' orbital parameter($a_1=0.7a_2$)}) are somewhat different from those for the $e=1.1$ runs.

In addition, for completeness, we have analyzed the situation where $V_\infty$ is smaller than $v_{\rm orb}$. We choose 
$V_\infty/v_{\rm orb}=0.1$ for this examination, and the results are presented in Appendix~\ref{Appendix:smallV}.

Of particular interest is whether different values of $V_\infty/V_{\rm orbit}$ can strongly influence the fraction of forming BPs. Our result for $V_\infty/v_{\rm orb}=0.1$ is shown as the brown line in the lower right panel of Fig.~\ref{Figs: Fractions of Outcomes}, and we can see that the different velocities do not significantly affect the fraction of forming BPs.

\section{Production Rates of FFP, CP, SSP and BP in Star Clusters}
\label{sec: production rates}

The results of Sections \ref{sec: numerical results}-\ref{sec: additional runs} can be applied to star clusters to estimate the production rates of various exotic planets/systems due to stellar flybys. 
Consider a typical two-planet system (like the one studied in the previous sections) embedded in a star cluster with stellar density $n_\star$, and velocity distribution $f(V_\infty$), such that 
$\int f(V_\infty)\, dV_\infty =1$.
For a given two-planet system, the rate of producing type-$\alpha$ systems (where $\alpha=$ FFP, SSP, CP and BP) is given by
\begin{equation}
\mathcal{R}_\alpha=n_\star \bigl\langle \sigma_\alpha V_\infty\bigr\rangle\equiv\int_0^\infty {\rm d}V_\infty\,n_\star f(V_\infty)V_\infty  \sigma_\alpha(V_\infty),
\label{eq:rate}
\end{equation}
where $\sigma_\alpha$ is the cross-section for producing type-$\alpha$ systems,
\begin{equation}
\sigma_\alpha(V_\infty)=
\int_0^{\infty} \pi {\rm d} (b^2)(q,V_\infty)\,p_\alpha(\tilde q, V_\infty).
\label{Ri equation}
\end{equation}
Here $p_\alpha(\tilde q, V_\infty)$ is the fraction of producing $\alpha$-type systems, as calculated in Sections \ref{sec: numerical results}-\ref{sec: additional runs_v}. 
For example, for $\alpha=$FFP, the relevant figure is the upper left panel of Fig.~\ref{Figs: Fractions of Outcomes} or the upper panel of Fig.~\ref{Figs: Lv fraction of outcomes}.
The impact parameter $b(q,V_\infty)$ associated with an hyperbolic trajectory of periastron $q$ and velocity at infinity $V_\infty$, is 
\begin{equation}
b^2=q^2\Bigl(1+\frac{2GM_{12}}{qV_\infty^2}\Bigr).
\end{equation}
Thus the cross-section can be written as
\begin{equation}
\sigma_\alpha(V_\infty)=
2\pi {GM_{12} a_2\over V_\infty^2}
\left( P_{\alpha 1}+{M_1\over M_{12}}\tilde V_\infty^2 P_{\alpha 2}\right),
\label{10}
\end{equation}
where $\tilde V_\infty^2\equiv V_\infty^2/v_{2i}^2$, with $v_{2i}=\sqrt{GM_1/a_2}$ (recall that $M_1$ is the initial host star mass) and
\begin{eqnarray}
&& P_{\alpha 1}\equiv  \int_0^{\infty}
{\rm d} \tilde q\, p_\alpha(\tilde q,V_\infty),\label{eq:P1}\\
&& P_{\alpha 2}\equiv  \int_0^{\infty}
{\rm d} \tilde q\, \tilde q\, p_\alpha(\tilde q,V_\infty).
\label{eq:P2}
\end{eqnarray}

So far the equations in this section are exact. 
To proceed further, we need to know how $p_\alpha$ depends on $V_\infty$. 
Our analysis in Section~\ref{sec: additional runs_v} suggests that, for the range of parameters of interest, the value of $V_\infty$ has a small/modest effect on $p_\alpha(\tilde q, V_\infty)$. 
We now make the approximation that $p_\alpha(\tilde q, V_\infty)$ is independent of $V_\infty$, which implies $P_{\alpha 1}$ and $P_{\alpha 2}$ are independent of $V_\infty$, and also assume that the velocities of the stars satisfy the Maxwell–Boltzmann distribution with dispersion $\sigma_\star$, i.e.
\begin{equation}
f(V_\infty)=\sqrt{\frac{2}{\pi}}\frac{V_\infty^2}{\sigma_\star^3}\exp\Bigl(-\frac{V_\infty^2}{2\sigma_\star^2}\Bigr).
\end{equation}
Then the rate of producing type-$\alpha$ systems (Eq.~(\ref{eq:rate})) is given by 
\begin{equation}
{\cal R}_\alpha\simeq {\cal R}_{\rm close}
\left[P_{\alpha 1}+\Bigl(\frac{\sigma_\star}{v_{2i}}\Bigr)^2P_{\alpha 2}\right],
\label{14}
\end{equation}
where ${\cal R}_{\rm close}$ is the close encounter rate 
\begin{eqnarray}
\mathcal{R}_{\rm close}=&&\frac{2\sqrt{2\pi}a_2 GM_{12}n_\star}{\sigma_\star}\nonumber\\
\simeq && 10\Bigl(\frac{n_\star}{10^3 {\rm pc}^{-3}}\Bigr)\Bigl(\frac{M_{12}}{2M_\odot}\Bigr)\Bigl(\frac{a_2}{50\, \rm au}\Bigr)\nonumber\\
&&\times \Bigl(\frac{\sigma_\star}{\rm 1\, km \,s^{-1}}\Bigr)^{\!\!-1} {\rm Gyr^{-1}}.
\end{eqnarray}

Note that the equations above have adopted the parameters of planet 2 
(the outer planet) for scaling. 
Similar expressions can be applied to planet 1 (the inner planet). 
In that case, $a_2$ and $v_{2i}$ should be replaced by $a_1$ and $v_{1i}$ in Eqs.~(\ref{10})-(\ref{14}), and $\tilde q$ in $p_\alpha(\tilde q,V_\infty)$ should be replaced by $q/a_1$. 
For our $e=1.1$ runs (Section~\ref{sec: numerical results}), the coefficient $P_{\alpha 1}$ and $P_{\alpha 2}$ should be the same for different planets. 
Such scalings would not apply for the $V_\infty=v_{2i}$ and $V_\infty=0.1v_{2i}$ runs.

Table \ref{The integrated probability} lists the values of $P_{\alpha 1}$ and $P_{\alpha 2}$ for $\alpha=$ FFP, CP, SSP based on the results of our $e=1.1$ (Section~\ref{sec: numerical results}), $V_\infty=v_{2i}$ and $V_\infty=0.1v_{2i}$ (Section~\ref{sec: additional runs_v}) runs.
Note that for FFP and CP, the results of $e=1.1$ runs are independent of the planet mass and the initial $a_1/a_2$ after rescaling. 
However, for the $V_\infty=v_{2i}$ and $V_\infty=0.1v_{2i}$ (Section~\ref{sec: additional runs_v}) runs, the results of the inner planet and outer planet are different, so we show them separately. We note that the values of $P_{\rm CP}$ shown in the table appear to differ significantly (by a factor of $\sim 2$) between different runs, while the other quantities ($P_{\rm FFP}$ and $P_{\rm SSP}$) are more ``universal".
\footnote{
Note that our numerical simulations only cover the range $\tilde q\leq 2$ for our fiducial runs (and $\tilde q\leq2.4$ for the $V_\infty=v_{2i}$ runs). 
However, we see from Fig.~\ref{The fraction of planet stolen} (orange line) and Fig.~\ref{Lv The fraction of planet stolen} (orange line) that $p_{\rm CP}$ can still be appreciable beyond this $\tilde q$ range.
We have computed $P_{\rm CP1}$ and $P_{\rm CP2}$ approximately by using a linear extrapolation of $p_{\rm cp}(\tilde q)$ beyond our numerical coverage.
Nevertheless, the difference between the values of $P_{\rm CP}$ for the two runs are real (see Fig.~\ref{The fraction of planet stolen}).}.
We see that all these $P_{\alpha 1}$ and $P_{\alpha 2}$ have values in the range between 0.1 and 0.6. 
Thus, in general, the production rates of FFPs and SSPs are comparable, while the rate for CPs is a factor of a few smaller.

Table \ref{The integrated BP probability} lists the values of $P_{\rm BP1}$ and $P_{\rm BP 2}$ based on our $e=1.1$ (Section~\ref{sec: numerical results}, Section~\ref{sec: additional runs_a} ) and different $V_\infty$ (Section~\ref{sec: additional runs_v}) runs. 
If we restrict to the initially stable systems ($a_1/a_2=0.6$ and 0.7), the values are all less than $0.05\%$. 
Thus the production rate of free-floating BPs is at most $0.1\%$ of the FFP rate.
Even for the unstable systems ($a_1/a_2=0.8$), the rate of BPs is at most $1\%$ of the FFP rate.

\begin{table*}[htbp]
    \centering
    \begin{tabular}{ccccccc}
      \toprule
        & $P_{\rm FFP1}$ & $P_{\rm FFP2}$ & $P_{\rm CP1}$ & $P_{\rm CP2}$ & $P_{\rm SSP1}$ & $P_{\rm SSP2}$\\ 
      \midrule
        Fiducial runs ($e=1.1$)&$ 0.57$  &$ 0.38 $ & $ 0.22$ & $ 0.2$ & $ 0.62$ & $ 0.51$ \\
        $V_\infty=v_{2i}$ runs (outer planet) &$ 0.56$  &$ 0.34 $ & $ 0.12$ & $ 0.08$ & $ 0.56$ & $ 0.44$ \\
        $V_\infty=v_{2i}$ runs (inner planet) &$ 0.54$  &$ 0.34 $ & $ 0.15$ & $ 0.14$ & / & / \\
        $V_\infty=0.1v_{2i}$ runs (outer planet) &$ 0.55$  &$ 0.39 $ & $ 0.26$ & $ 0.24$ & $ 0.63$ & $ 0.53$ \\
        $V_\infty=0.1v_{2i}$ runs (inner planet) &$ 0.55$  &$ 0.39 $ & $ 0.25$ & $ 0.24$ & / & / \\
      \bottomrule
\end{tabular}
    \caption{
    The dimensionless quantities $P_{\alpha 1}$ and $P_{\alpha 2}$ (with $\alpha=$ FFP, CP and SSP) as defined in Eqs.~(\ref{eq:P1})-(\ref{eq:P2}).
    These quantities can be used to compute the formation rates of FFPs, CPs and SSPs in a star cluster [see Eqs.~(\ref{10}) and (\ref{14})]. 
    Note that for FFPs and CPs, there is no difference in these quantities between the inner planet and the outer planet in our fiducial ($e=1.1$) runs (but note that when applying to the inner planet, $a_2,~v_{2i}$ should be replaced by $a_1,~v_{1i}$).
    For SSPs, we do not care if they were originally inner planets or outer planets, so there is only one $P_{\rm SSP1}$ and $P_{\rm SSP2}$ for the $V_\infty=v_{2i=\sqrt{GM/a_2}}$ runs and $V_\infty=0.1v_{2i}$ runs}.
    
    \label{The integrated probability}
\end{table*}

\begin{table}[htbp]
    \centering
    \begin{tabular}{ccc}
      \toprule
       & $P_{\rm BP1}$ & $P_{\rm BP2}$\\ 
      \midrule
        $e=1.1$, $a_1=0.7a_2$ runs & $ 5\times 10^{-4}$& $2.25\times 10^{-4}$\\ 
        $e=1.1$, $a_1=0.6a_2$ runs & $ 1.85\times 10^{-4}$& $1.0\times 10^{-4}$\\ 
        $e=1.1$, $a_1=0.8a_2$ runs & $ 8.5\times 10^{-3}$& $5.6\times 10^{-3}$\\ 
        $V_\infty=v_{2i}$, $a_1=0.7a_2$ runs & $ 5\times 10^{-4}$& $2.0\times 10^{-4}$\\ 
        $V_\infty=0.1v_{2i}$, $a_1=0.7a_2$ runs & $ 3.5\times 10^{-4}$& $2.6\times 10^{-4}$\\ 
      \bottomrule
\end{tabular}
    \caption{
    The dimensionless quantities $P_{\rm BP1}$ and $P_{\rm BP2}$ that can be used to compute the formation rate of free-floating binary planets. 
    The 5 rows are for the runs with different flyby trajectories ($e=1.1$, $V_\infty=v_{2i}$ or $V_\infty=0.1v_{2i}$ ) and initial planet spacings. 
    }
    \label{The integrated BP probability}
\end{table}

Using the equations and tables presented above, we can easily compute/estimate the FFP, CP, BP production rates in any cluster. 
Specifically, the rate of producing type-$\alpha$ systems per unit volume in the cluster is given by Eq.~(\ref{14}) multiplied by the number density ($n_{\rm 2P}$) of two-planet systems (with the required configuration). 
Obviously, the main uncertainty of such an estimate comes from the uncertainty in $n_{\rm 2P}$ of various planet masses and orbital configurations.

\section{Summary and Discussion}
\label{sec: summary and discussion}

We have carried out a systematic study on the effects of stellar flybys on two-planet systems. 
The planets are initially on circular orbits, with modest semi-major axis ratios ($a_1/a_2=0.6-0.8$) and the flyby star (on a weakly hyperbolic trajectory) has the same mass as the initial host star of the planets. 
Our goal is to obtain, through numerical experiments, scale-free results that can be used in various contexts concerning planetary systems in star clusters.
When a two-planet system experiences stellar flybys, one or both planets can be ejected, forming free-floating planets (FFPs), captured planets (CPs) around the flyby star, and (rare) free-floating binary planets (BPs); the remaining single-surviving-planets (SSPs) are heavily disturbed, with their orbital parameters greatly modified.
Statistically, when averaging over the flyby orientations, the outcomes of flybys depend mainly on the dimensionless pericenter distance $\tilde q=q/a_2$ (where $a_2$ is the initial semi-major axis of the outer planet), and weakly on the eccentricity ($e>1$) or the incoming velocity of the flyby;
the formation rate of BPs also depends on the planet masses and the initial semi-major axis ratio $a_1/a_2$. The main results of this paper consist of a series of figures (Sections \ref{sec: numerical results}-\ref{sec: additional runs}) that give
\begin{itemize}
    \item the formation fractions (or branching ratios) of FFPs, CPs, SSPs and BPs as a function of $\tilde q$ (Fig.~\ref{Figs: Fractions of Outcomes}, \ref{JuMBOs0.70.8}, \ref{Differentm}, \ref{Figs: Lv fraction of outcomes});
    \item the velocity distribution of FFPs produced by stellar flybys (Fig.~\ref{v_infty distribution for free-floating planets}, \ref{Lv v_infty distribution for free-floating planets});
    \item the orbital semi-major axis and eccentricity distributions of SSPs
    (Fig.~\ref{Figs: Single planet system orbital parameter}, \ref{Figs: Lv Single planet system orbital parameter}), CPs (Fig.~\ref{Figs: Captured planet orbital parameter}, \ref{Figs: Largev Captured planet orbital parameter}),
    and free-floating BPs (Fig.~\ref{Figs: JuMBOs' orbital parameter($a_1=0.7a_2$)}, \ref{JuMBOs' orbital parameter($a_1=0.8a_2$)}, \ref{Figs: Largev JuMBOs' orbital parameter($a_1=0.7a_2$)}).
\end{itemize}
In addition, in Section \ref{sec: production rates}, we derive the analytical expressions (with numerically derived coefficients; see Tables \ref{The integrated probability}-\ref{The integrated BP probability}) that can be used to compute the formation rates of FFPs, SSPs, CPs and BPs in general cluster environments.

In a star cluster, the production rates of FFPs, SSPs, CPs and BPs due to flybys obviously depend on the stellar density, velocity dispersion and the population of stars with planets (of different architectures), our calculations do reveal several more generic/robust results. 
For example, the production rates of FFPs and SSPs are generally similar, while the rate for CPs is a factor of a few smaller (Section~\ref{sec: production rates}); the FFPs have a broad non-Gaussian velocity distribution, peaking at the initial orbital velocity of the planet; the CPs have a broad semi-major axis distribution, with a peak at 0.7 times the original semi-major axis, and the eccentricity distribution is super-thermal.

On the other hand, the production rate of free-floating BPs by flybys depends strongly on the axis ratio $a_1/a_2$ of the initial systems and on the planet masses. 
For the Jupiter-mass planets explored in this paper, we find that the formation fraction of BPs is always less than $1\%$, even for unrealistically closed-pack systems ($a_1/a_2=0.8$, which is intrinsically unstable); for more realistic systems ($a_1/a_2\lesssim 0.7$), the fraction is always less than $0.2\%$ (bottom right panel of Fig.~\ref{Figs: Fractions of Outcomes} and \ref{Differentm}). 
The fraction remains less than $1\%$ when considering $4M_{\rm J}$ planets. 
Overall, when averaging over all possible flybys, the production rate of BPs is less than $0.1\%$ of that for FFPs (see Section~\ref{sec: production rates}). 

Concerning the recent JWST detection of a large population of JuMBOs in the Trapezium \citep{Pearson2023arXiv01}, our calculations show (unsurprisingly) that these BPs cannot be produced by stellar flybys [cf.  \citet{Portegies2024arXiv01}], contrary to the recent claim by \citet{Wang2024arXiv02}. 
We think that these JuMBOs are most likely formed in a scaled-down version of star formation, i.e. via fragmentation of molecular cloud cores or weakly-bound disks or pseudo-disks in the early stages of star formation.

Future observational facilities, such as CSST, Rubin Observatory and Roman Telescope, will likely detect a large number of FFPs and other exotic planetary systems. 
We hope that the general scale-free results presented in this paper will be useful for future studies to evaluate the effects of stellar flybys on planetary systems.
eir exoplanet in upcoming JWST analyses.

\

We thank Drs. Yihan Wang and Rosalba Perna for useful discussions.

\software{
IAS15 \citep{reboundias15},
Matplotlib \citep{Hunter2007}, 
NumPy \citep{Walt2011},
Rebound \citep{rebound}.
}

\newpage
\appendix

\section{Properties of FFP, SSP, CP and BP for different flyby Pericenter distances}
\label{Appendix:q}
In Section~\ref{sec: Numerical results_FFP}-\ref{sec: Numerical results_CP}, we have presented the semi-major axis (SMA) and eccentricity distributions of FFPs, SSPs, CPs over all dimensionless flyby pericenter distances ($\tilde q=q/a_2$). 
However, the distributions of SMA and eccentricity for different $\tilde q$'s are quite different. 
Here, we present these distributions for select values of $\tilde q$'s.
We choose five $\tilde q$ values less than 1 and five $\tilde q$ values 
greater than 1.

From the analysis presented in the main text, we know that there is no 
difference in the behaviors of the inner planet and the outer planet when proper scalings are applied. Here we focus on the results for the outer planet.

\begin{figure}[htbp]
        \subfigure
        {
            \begin{minipage}{0.5\linewidth} 
                \centering
                \includegraphics[width=1.02\columnwidth]{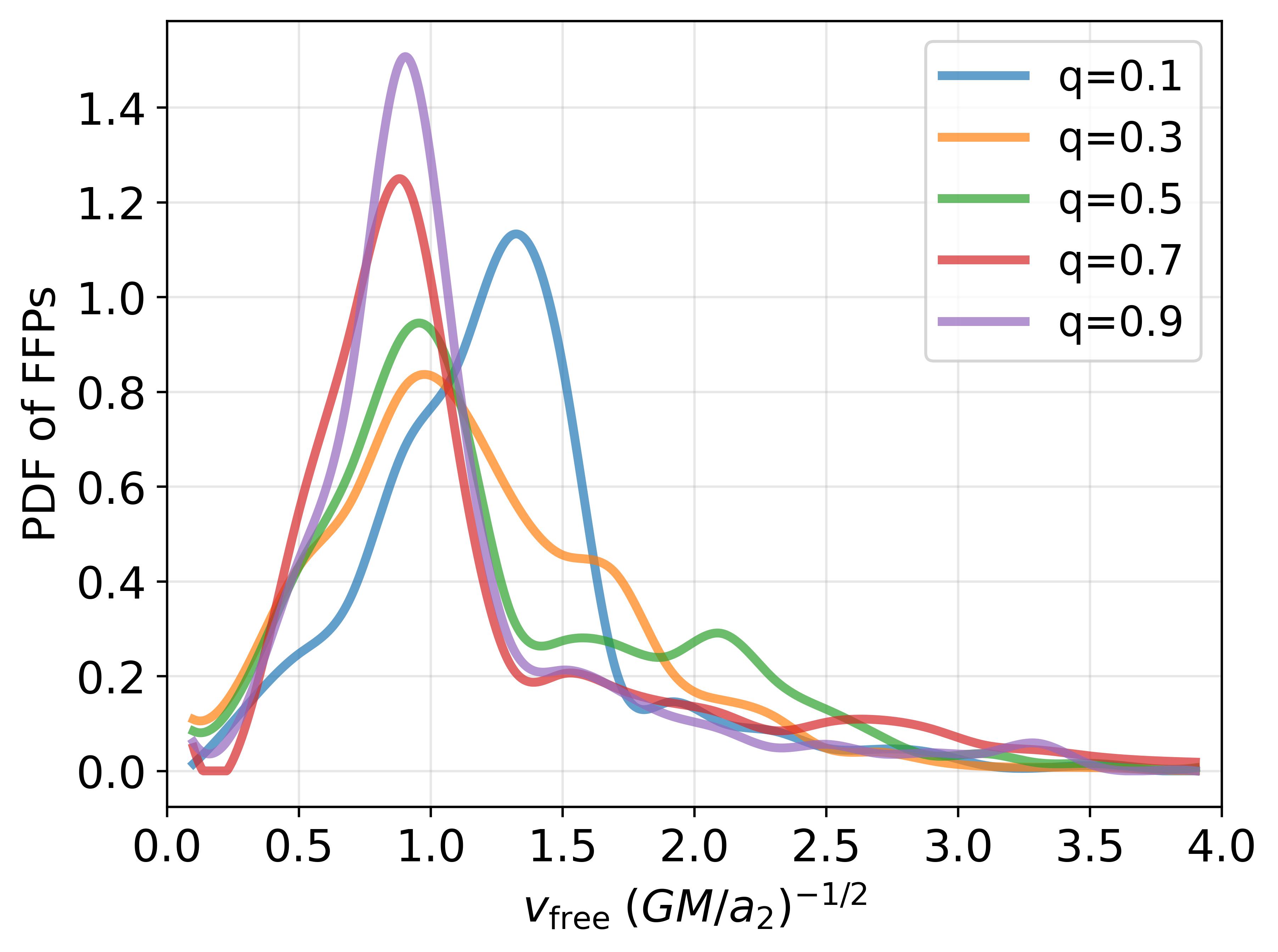}
                \label{app Vinf q1}
            \end{minipage}
        }
        \subfigure
        {
            \begin{minipage}{0.5\linewidth}
                \centering
                \includegraphics[width=1.02\columnwidth]{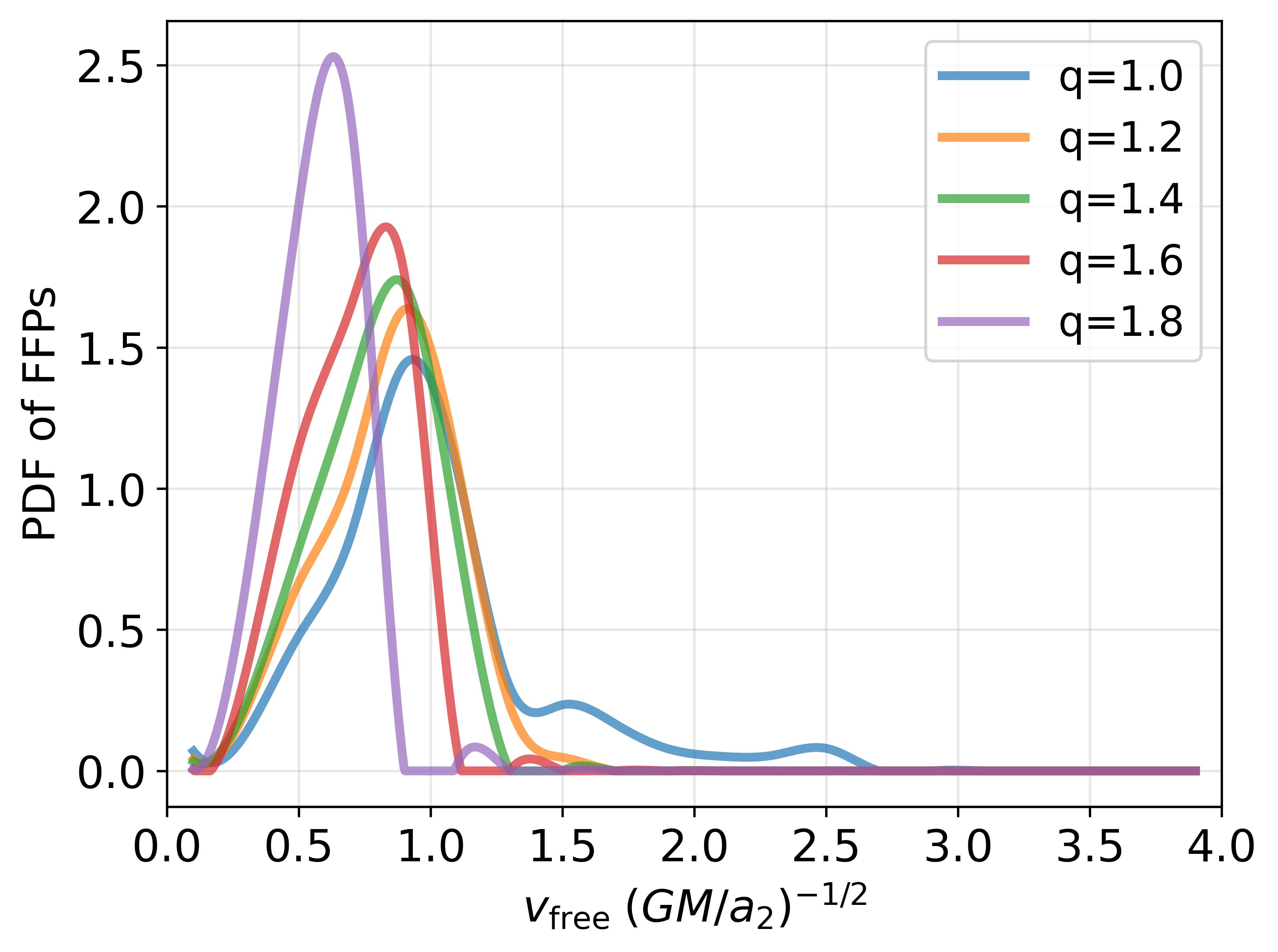}
                \label{app Vinf q2}
            \end{minipage}
        }
        \caption{The probability distributions of free-floating velocity $v_{\rm free}$ of FFPs produced by stellar flybys with different $\tilde q$'s.}
        \label{app Vinf}
\end{figure}
Figure \ref{app Vinf} shows the distributions of the free-floating velocity $v_{\rm free}$ of FFPs.
We see that for ``closer" flybys (those with smaller $\tilde q$), the dispersion of $v_{\rm free}$ is larger, but the position of the velocity peak is basically unchanged.

\begin{figure}[htbp]
        \subfigure
        {
            \begin{minipage}{0.5\linewidth} 
                \centering
                \includegraphics[width=1.02\columnwidth]{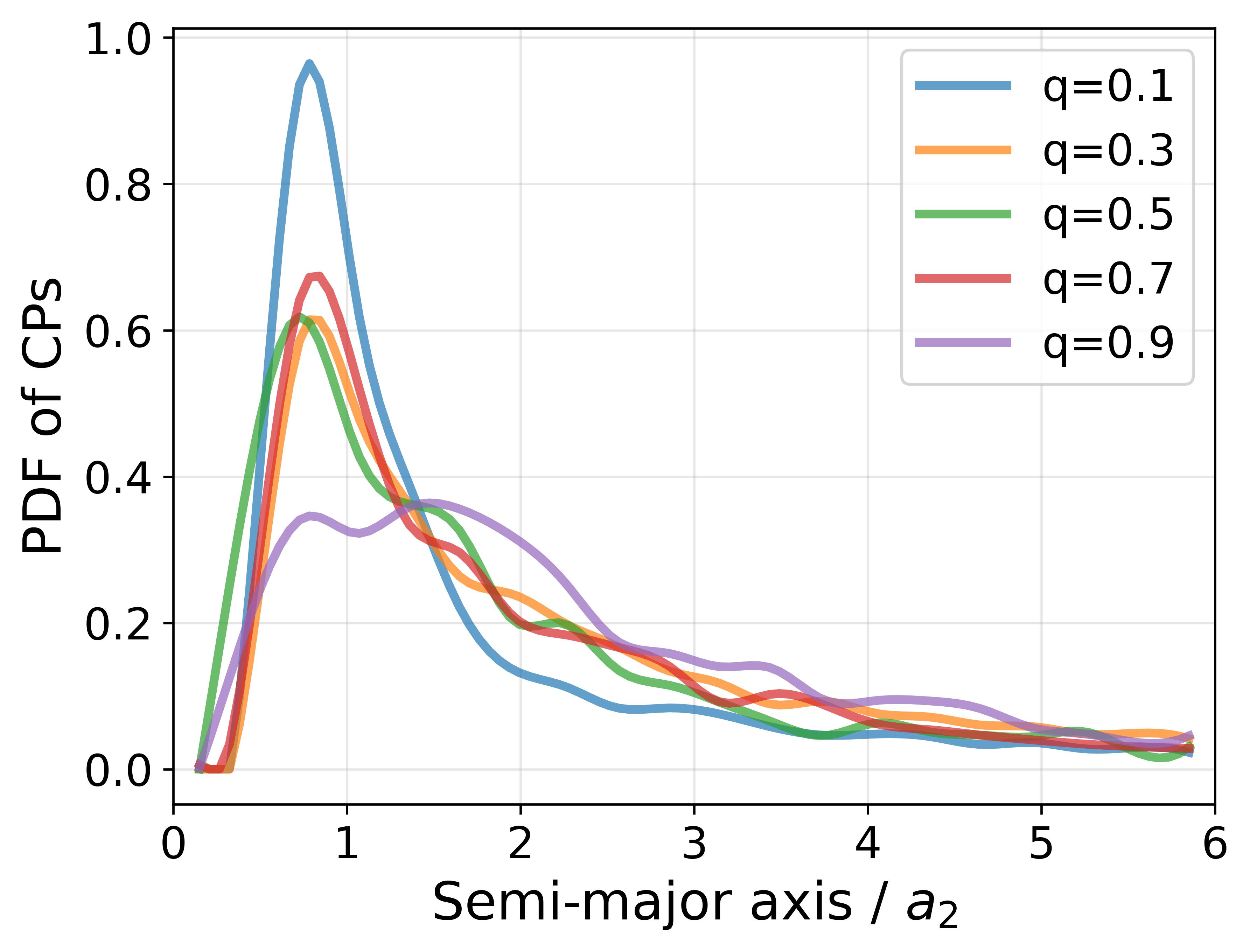}
                \label{app CP SMA q1}
            \end{minipage}
        }
        \subfigure
        {
            \begin{minipage}{0.5\linewidth}
                \centering
                \includegraphics[width=1.02\columnwidth]{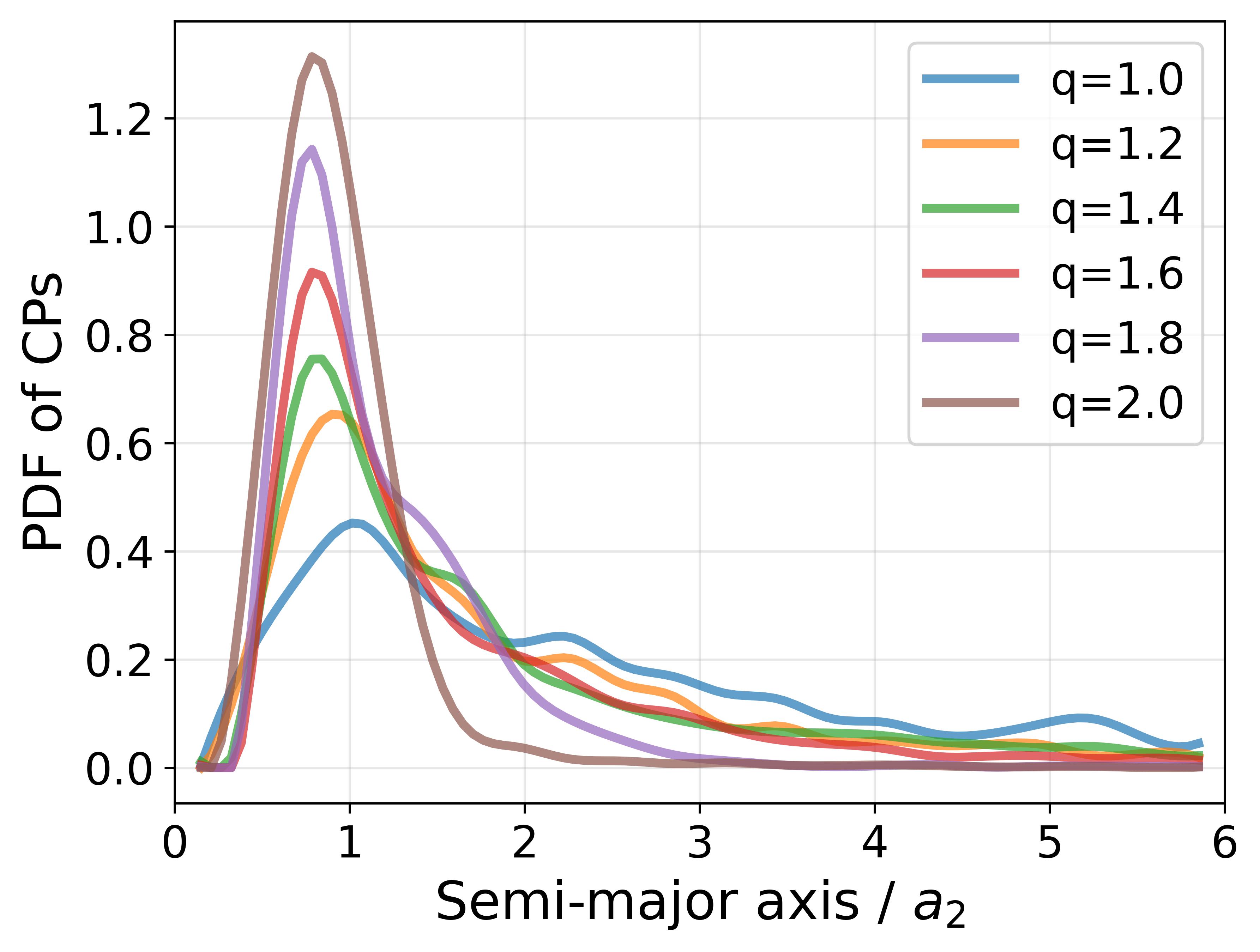}
                \label{app CP SMA q2}
            \end{minipage}
        }
        \caption{The probability distributions of the orbital semi-major axis of CPs with different $\tilde q$'s.}
        \label{app CP SMA}
\end{figure}

\begin{figure}[htbp]
        \subfigure
        {
            \begin{minipage}{0.5\linewidth} 
                \centering
                \includegraphics[width=1.02\columnwidth]{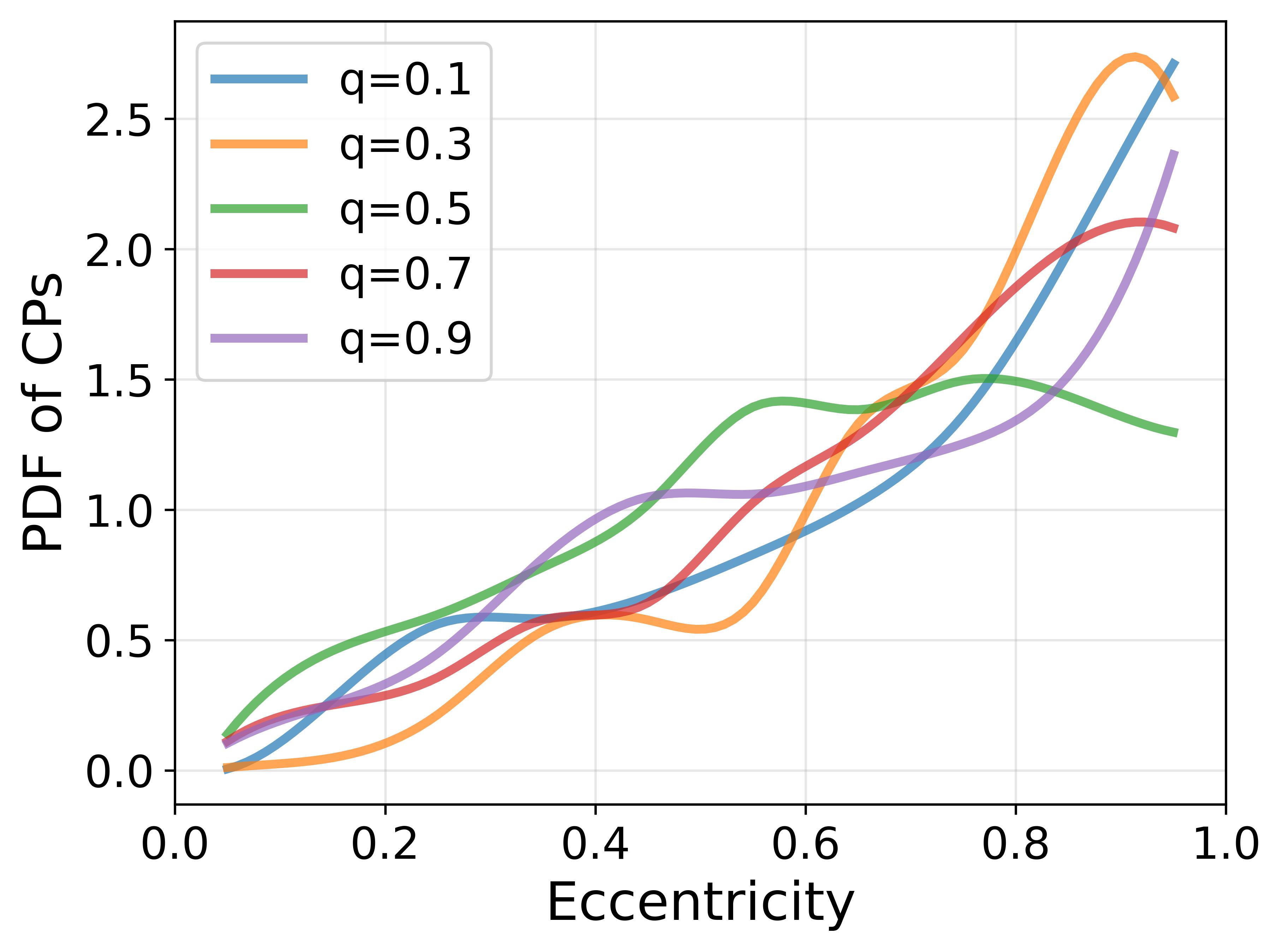}
                \label{app CP e q1}
            \end{minipage}
        }
        \subfigure
        {
            \begin{minipage}{0.5\linewidth}
                \centering
                \includegraphics[width=1.02\columnwidth]{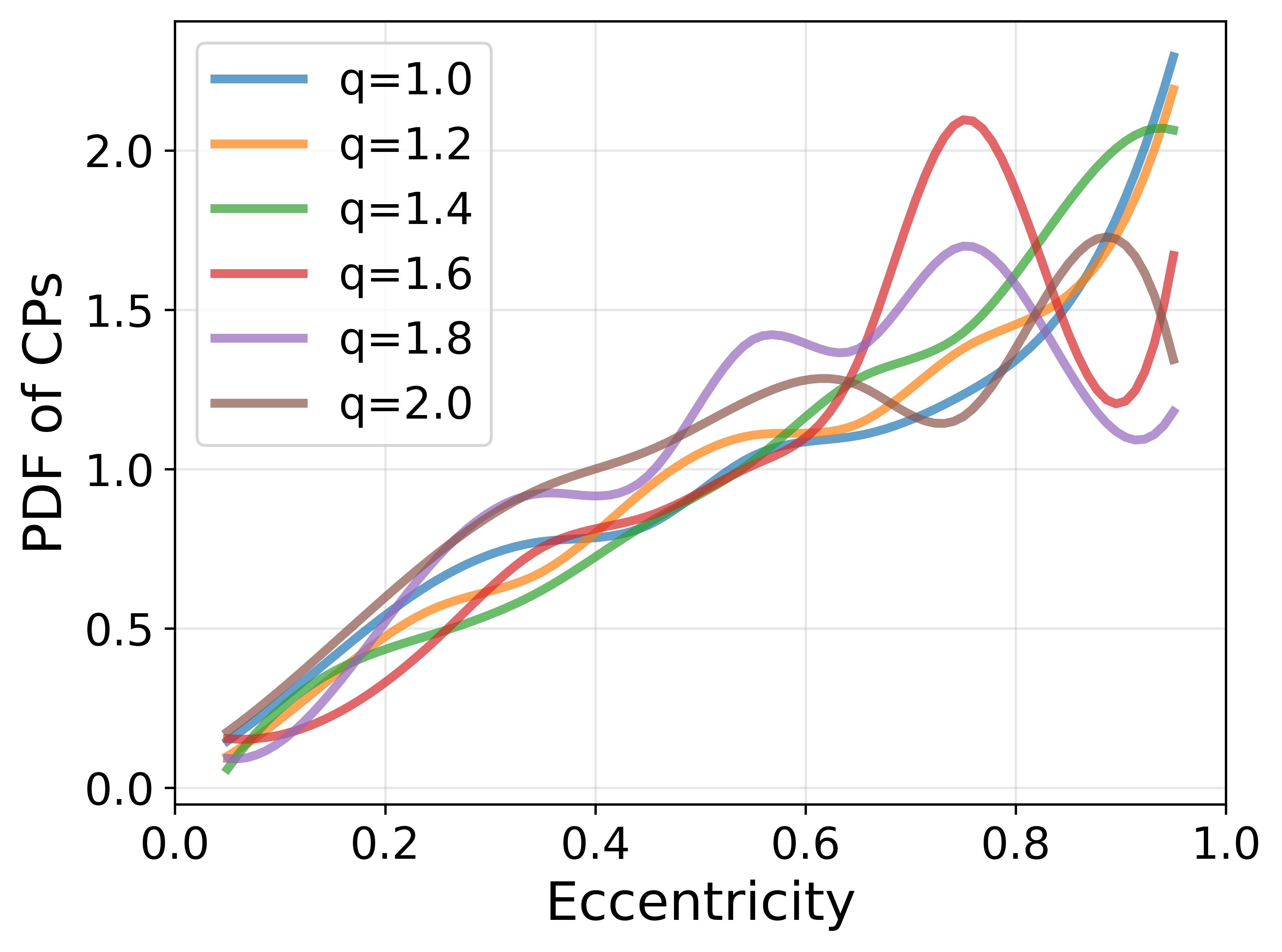}
                \label{app CP e q2}
            \end{minipage}
        }
        \caption{The probability distributions of the orbital eccentricity of CPs  with different $\tilde q$'s.}
        \label{app CP e}
\end{figure}
Fig.~\ref{app CP SMA} and \ref{app CP e} show the orbital parameter distributions of the CPs.
We see that there is no direct correlation between the eccentricity distribution and $\tilde q$, but the dispersion of the SMAs seems to change from small to large and then becomes smaller when $\tilde q$ decreases from $2.0$ to $0.1$, and the SMA dispersion is the largest at $\tilde q= 1.0$.

\begin{figure}[htbp]
        \subfigure
        {
            \begin{minipage}{0.5\linewidth} 
                \centering
                \includegraphics[width=1.02\columnwidth]{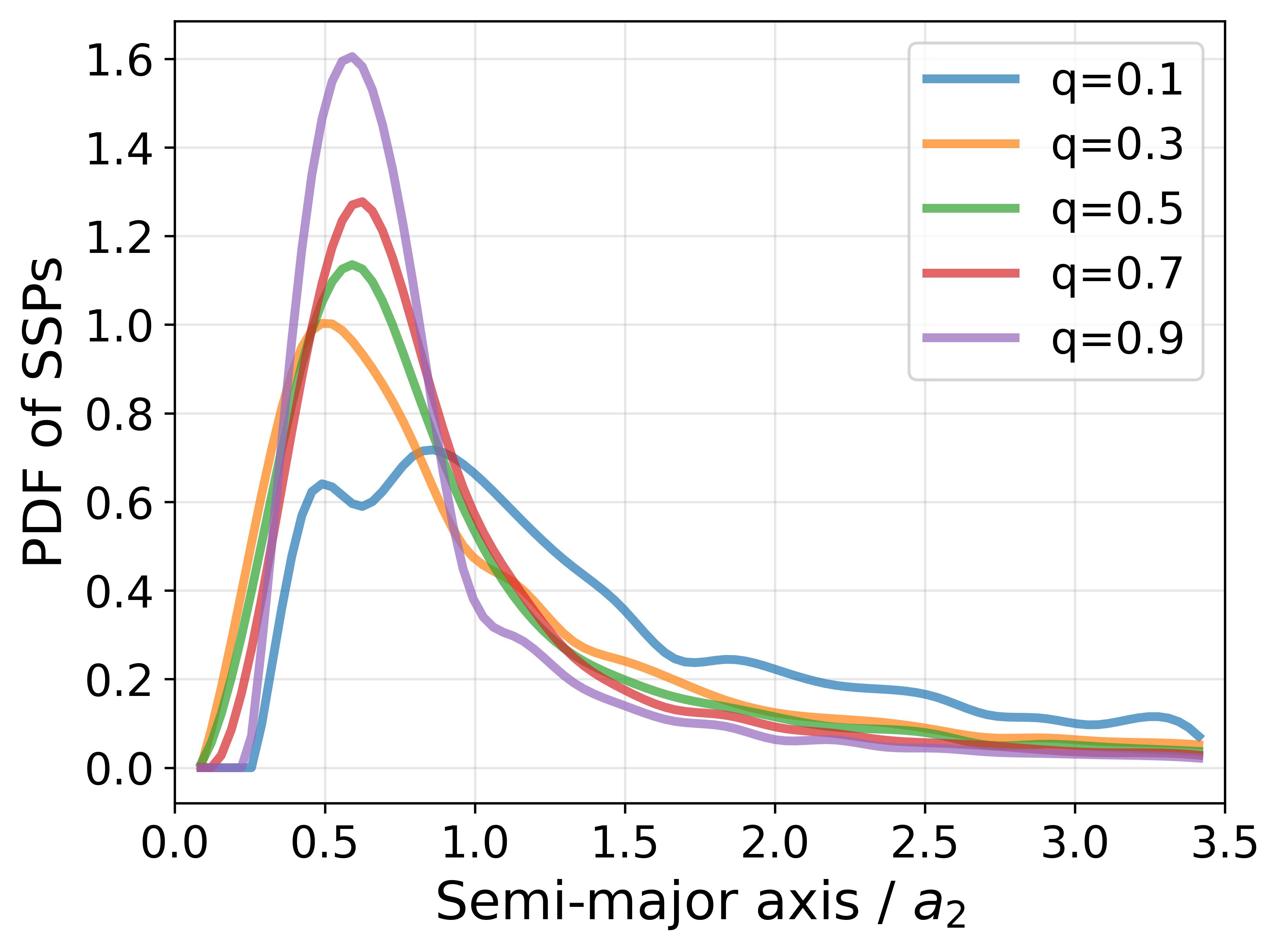}
                \label{app SSP SMA q1}
            \end{minipage}
        }
        \subfigure
        {
            \begin{minipage}{0.5\linewidth}
                \centering
                \includegraphics[width=1.02\columnwidth]{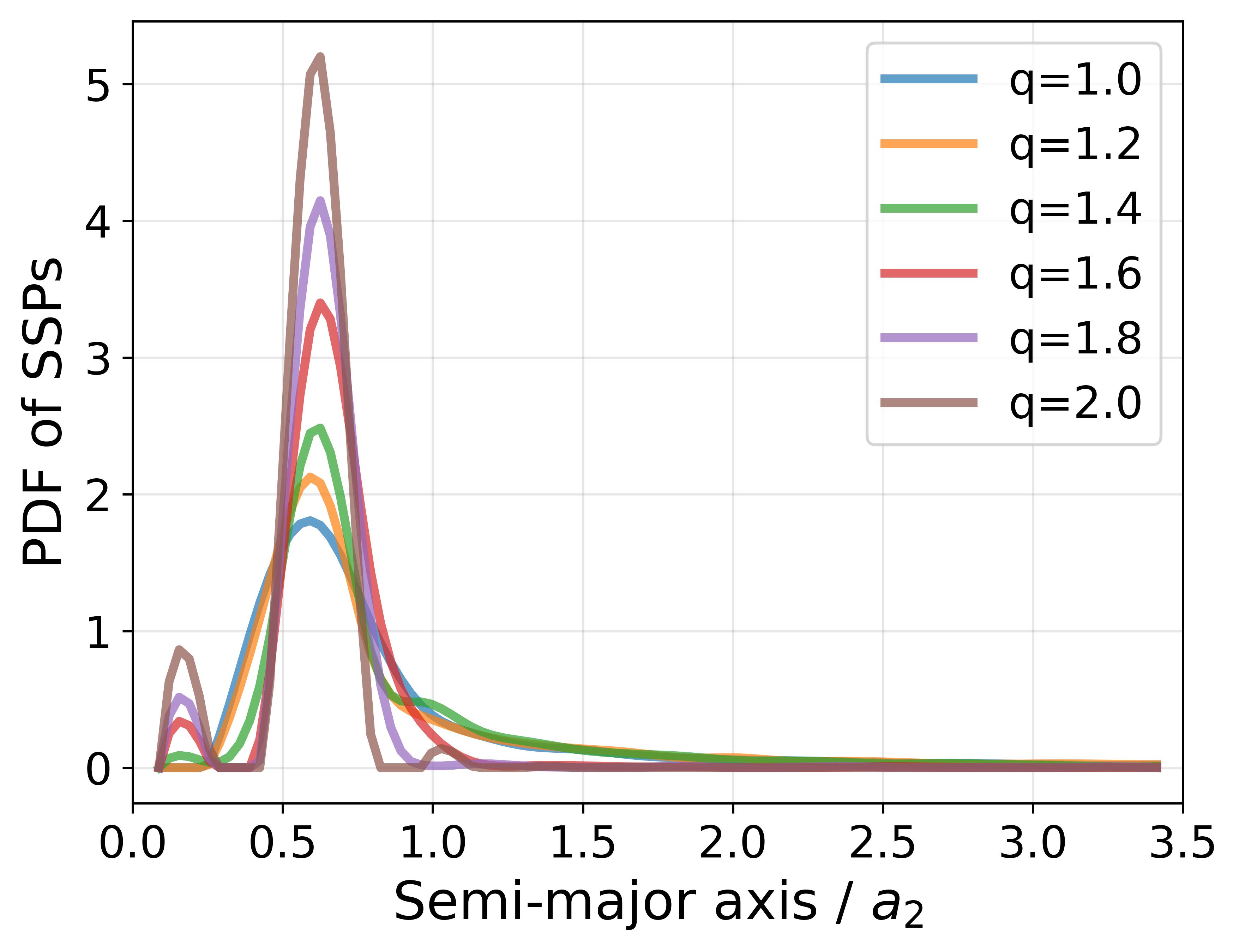}
                \label{app SSP SMA q2}
            \end{minipage}
        }
        \caption{The probability distributions of the orbital semi-major axis of SSPs with different $\tilde q$'s.}
        \label{app SSP SMA}
\end{figure}

\begin{figure}[htbp]
        \subfigure
        {
            \begin{minipage}{0.5\linewidth} 
                \centering
                \includegraphics[width=1.02\columnwidth]{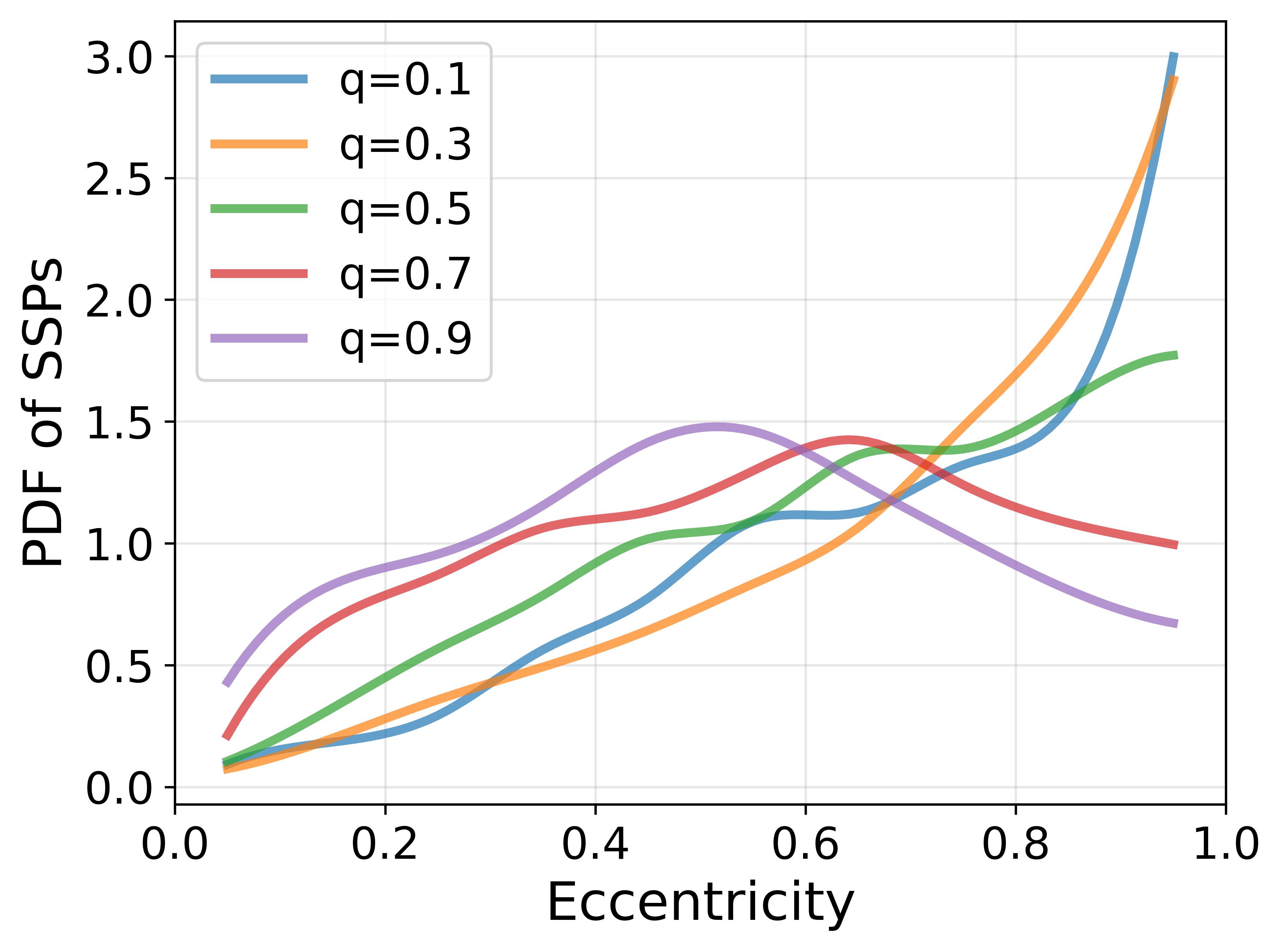}
                \label{app SSP e q1}
            \end{minipage}
        }
        \subfigure
        {
            \begin{minipage}{0.5\linewidth}
                \centering
                \includegraphics[width=1.02\columnwidth]{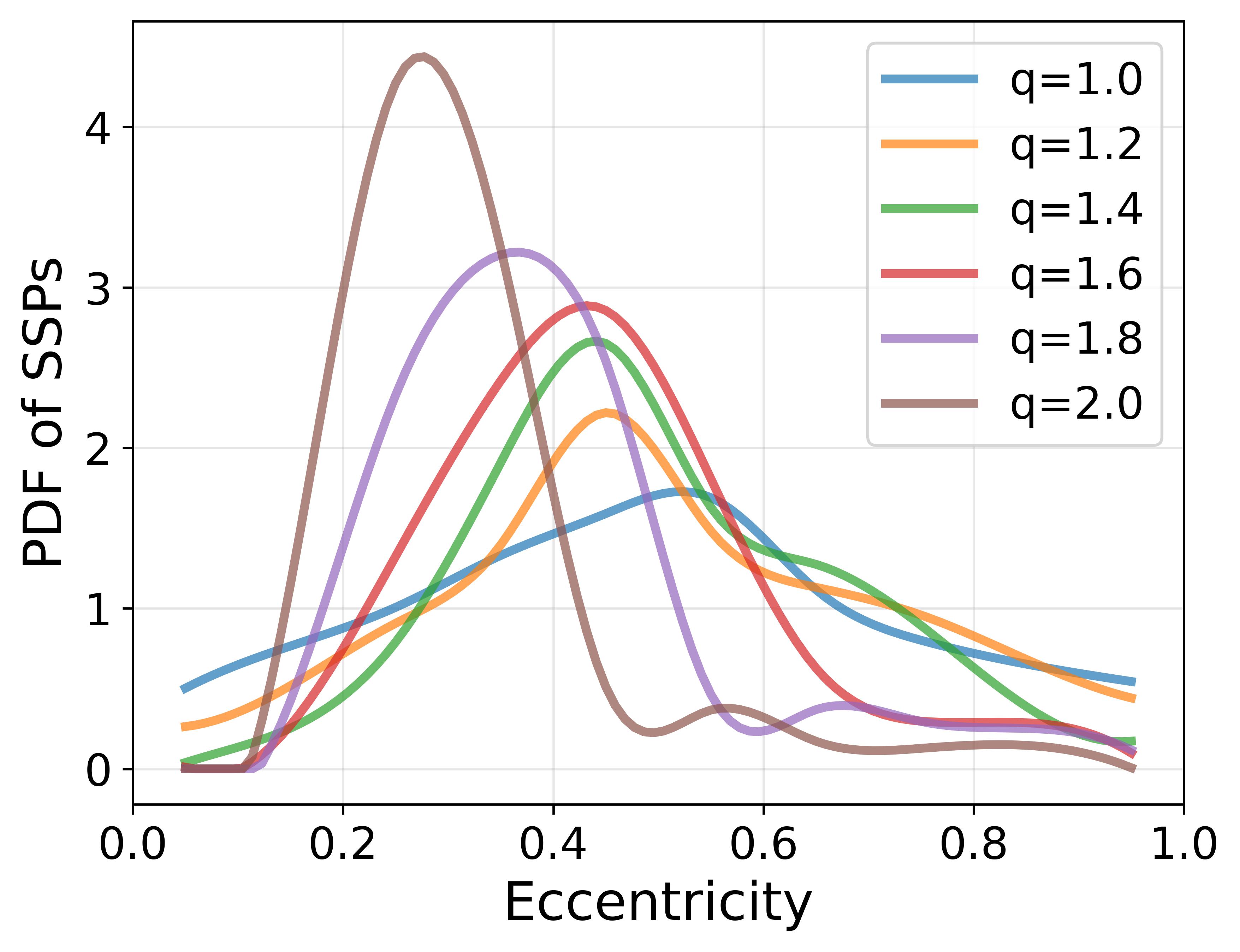}
                \label{app SSP e q2}
            \end{minipage}
        }
        \caption{The probability distributions of the orbital eccentricity of SSPs with different $\tilde q$'s.}
        \label{app SSP e}
\end{figure}

Fig.~\ref{app SSP SMA} and \ref{app SSP e} show the distributions of the orbital parameters of SSPs.
We see that the distribution of the SMAs becomes broader as $\tilde q$ decreases, but the peak of the distribution is still around $0.7a_2$, that is, the initial semi-major axis of the inner planet.
Note that the SSP is more likely to be the inner planet of the original system. 
From Fig.~\ref{app SSP e}, we see that as $\tilde q$ increases, the peak of eccentricity distribution gradually shifts to a smaller value, and the dispersion also becomes smaller.

\section{Results for simulations with small-velocity flyby}
\label{Appendix:smallV}

\begin{figure}[htbp]
    \centering
    \begin{minipage}[b]{0.49\linewidth}
        \centering
        \includegraphics[width=\linewidth]{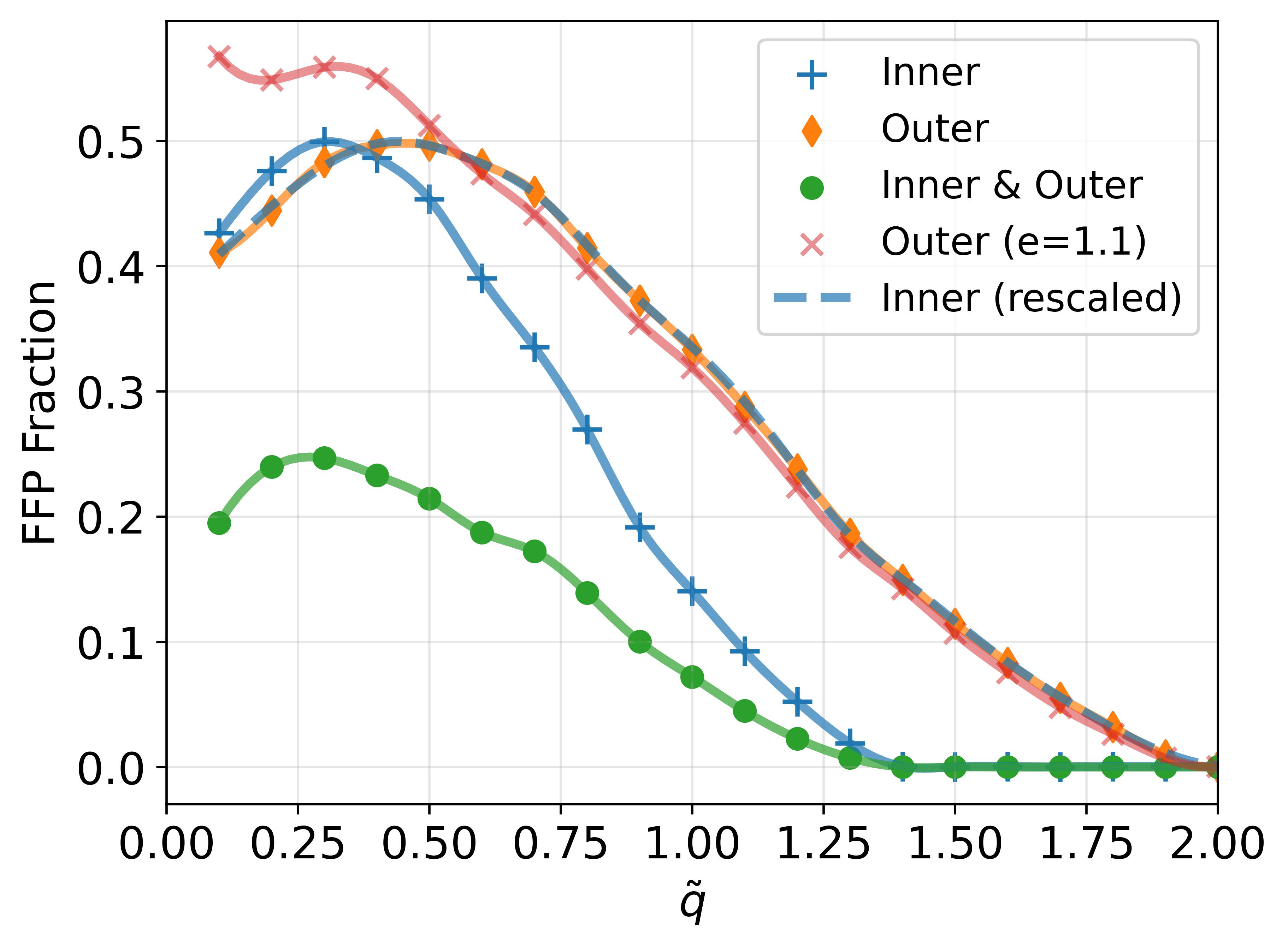}
        \label{fig:Sv_FreeFloating}
    \end{minipage}
    \hfill 
    \begin{minipage}[b]{0.49\linewidth}
        \centering
        \includegraphics[width=\linewidth]{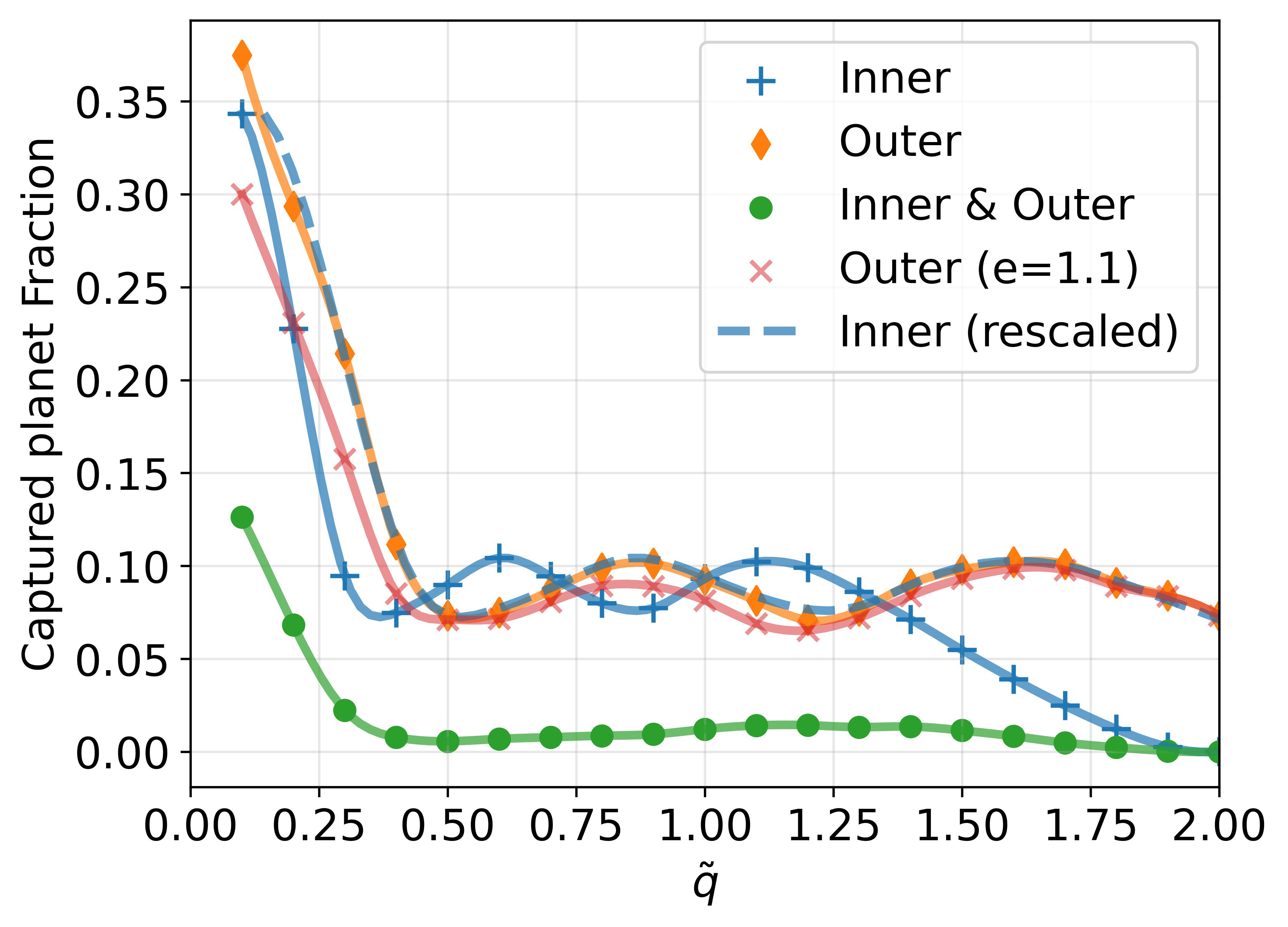}
        \label{fig:Sv_Stolen}
    \end{minipage}
    
    \begin{minipage}[b]{\linewidth}
        \centering
        \includegraphics[width=0.49\linewidth]{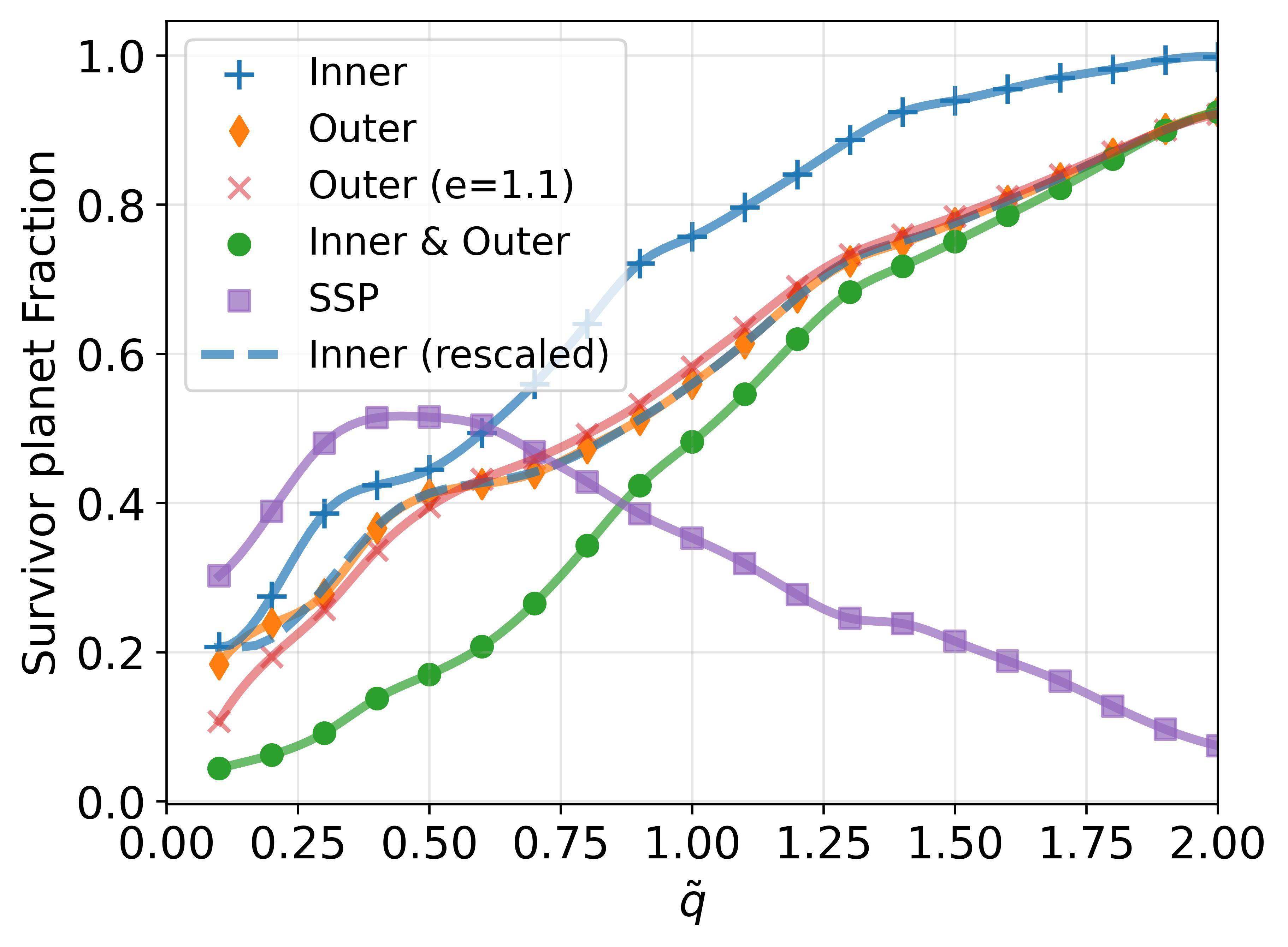}
        \label{fig:Sv_Bound}
    \end{minipage}
    
    \caption{Same as the first three panels in Fig.~\ref{Figs: Fractions of Outcomes}, except for the runs with $V_\infty=0.1 v_{2i}$. Note that here the light-red lines (labeled ``$e=1.1$") show the result (for the outer planet only) based on the simulations with $e=1.1$ for the flyby trajectory (fiducial runs, see Section~\ref{sec: numerical results}).}
    \label{fig:Sv_fraction_outcomes}
\end{figure}

\begin{figure}[htbp]
\centering
\includegraphics[width=0.49\columnwidth]{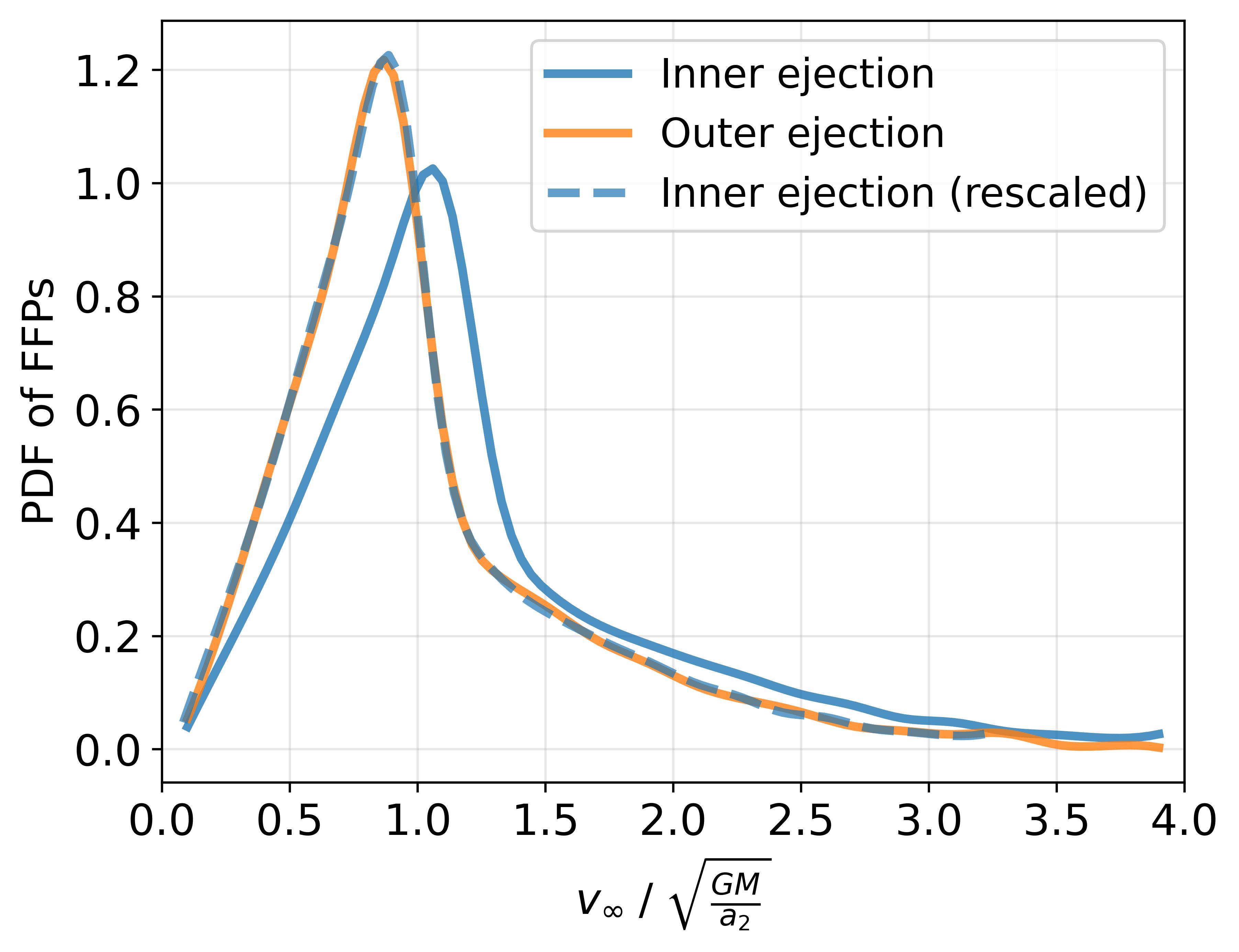}
\caption{
Same as Fig.~\ref{v_infty distribution for free-floating planets}, except for the runs with $V_\infty=0.1 v_{2i}$.
}
\label{Sv v_infty distribution for free-floating planets}
\end{figure}

\begin{figure*}[htbp]
    \centering
    \begin{minipage}[b]{0.49\linewidth}
        \centering
        \includegraphics[width=\linewidth]{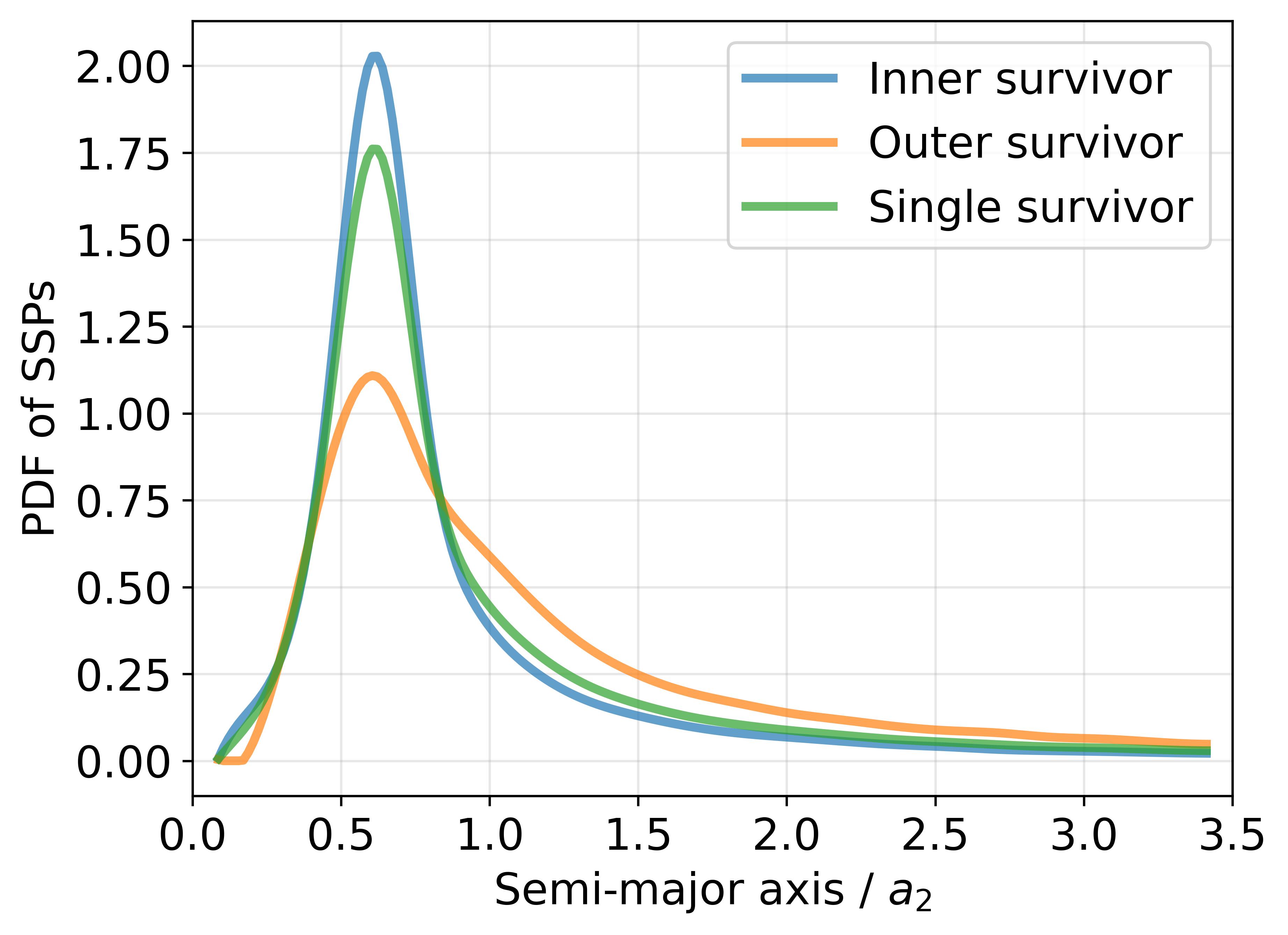}
    \end{minipage}
    \hfill
    \begin{minipage}[b]{0.49\linewidth}
        \centering
        \includegraphics[width=\linewidth]{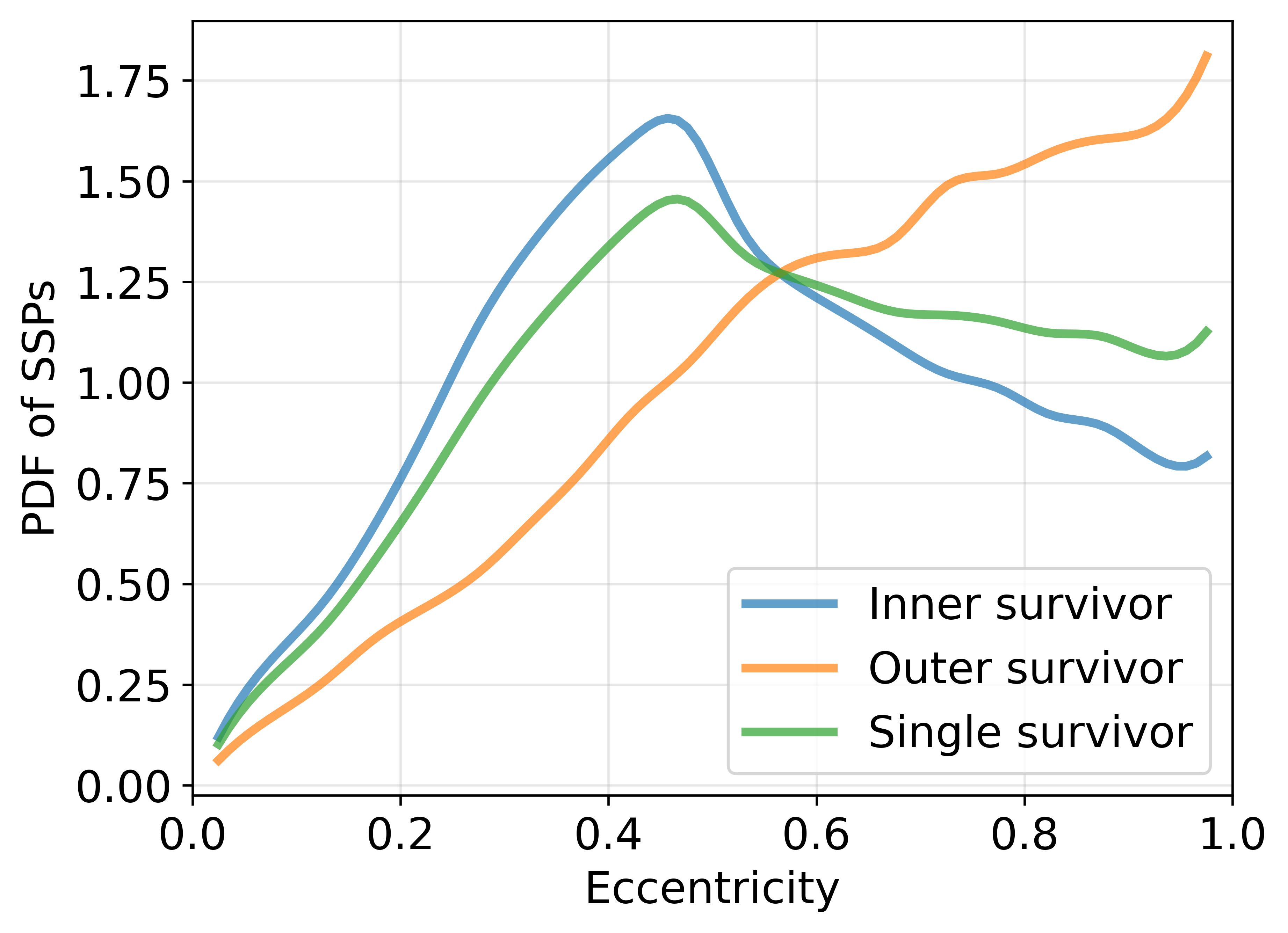}
    \end{minipage}
    
    \begin{minipage}[b]{0.49\linewidth}
        \centering
        \includegraphics[width=\linewidth]{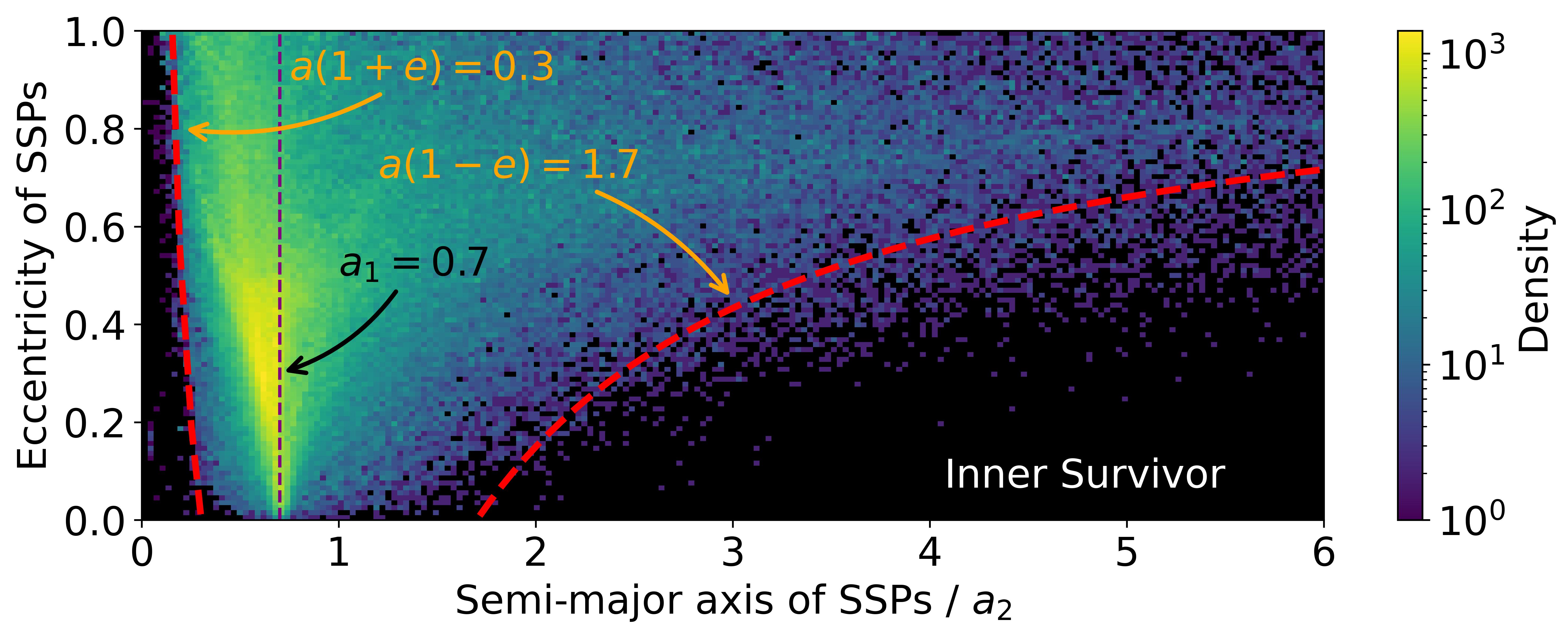}
    \end{minipage}
    \hfill
    \begin{minipage}[b]{0.49\linewidth}
        \centering
        \includegraphics[width=\linewidth]{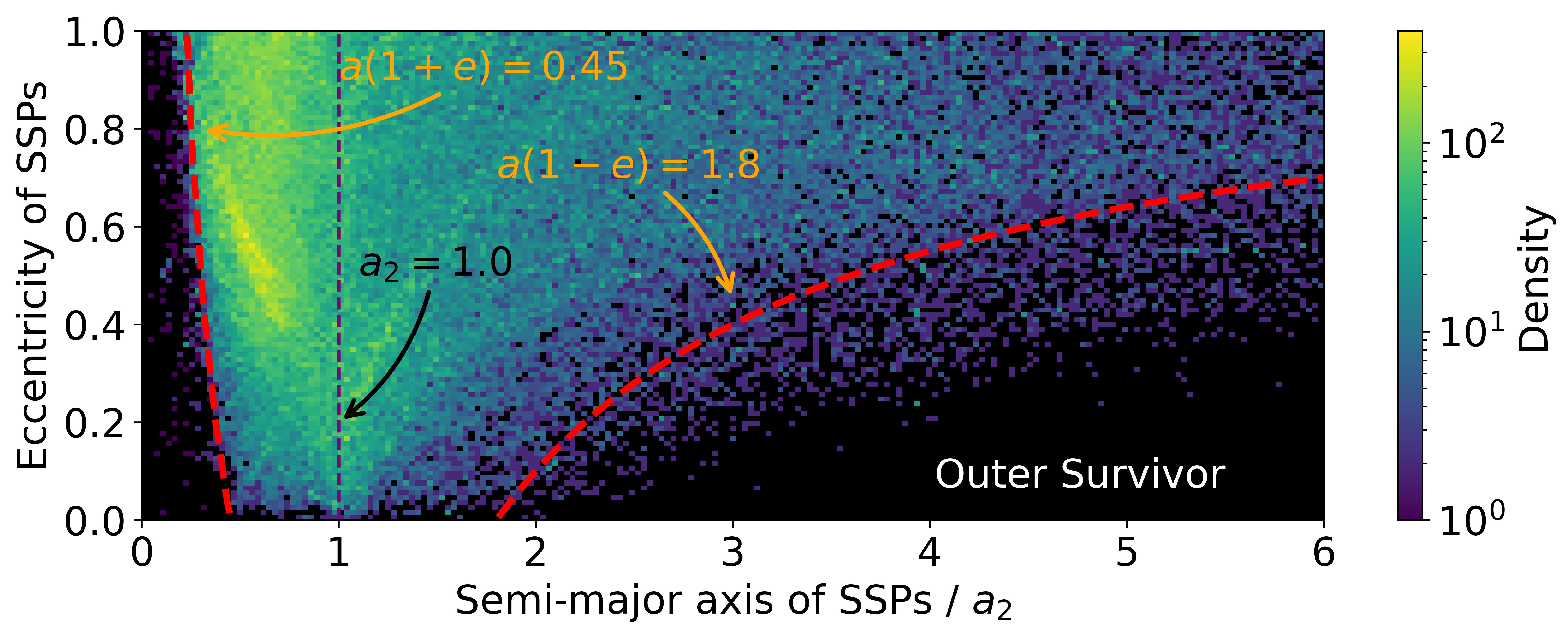}
    \end{minipage}
    
    \caption{
    Same as Fig.~\ref{Figs: Single planet system orbital parameter}, except for the runs with $V_\infty=0.1 v_{2i}$.}
    \label{Figs: Sv Single planet system orbital parameter}
\end{figure*}

\begin{figure*}[htbp]
    \centering
    \begin{minipage}[b]{0.49\linewidth}
        \centering
        \includegraphics[width=\linewidth]{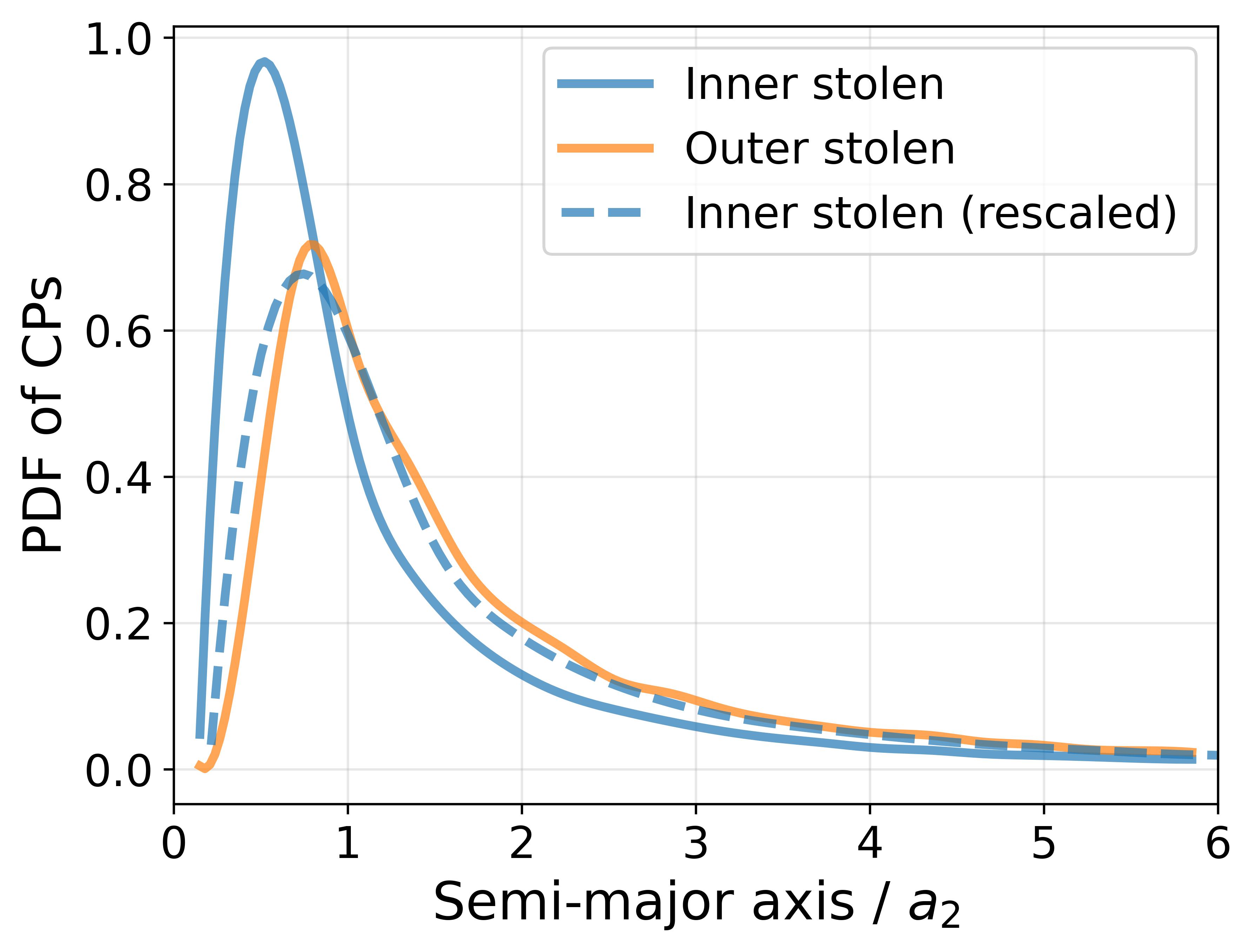}
    \end{minipage}
    \hfill
    \begin{minipage}[b]{0.49\linewidth}
        \centering
        \includegraphics[width=\linewidth]{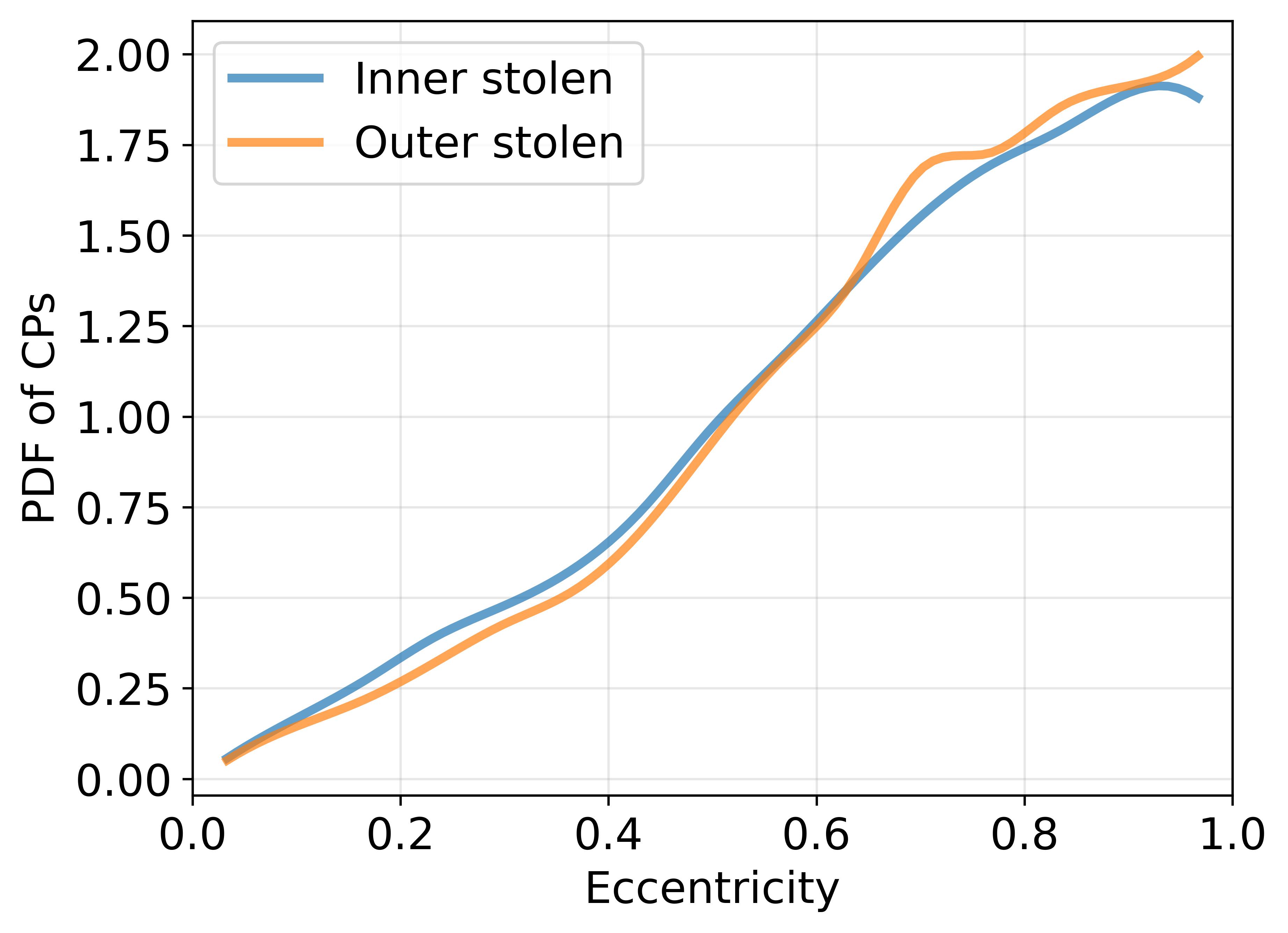}
    \end{minipage}
    
    \begin{minipage}[b]{0.49\linewidth}
        \centering
        \includegraphics[width=\linewidth]{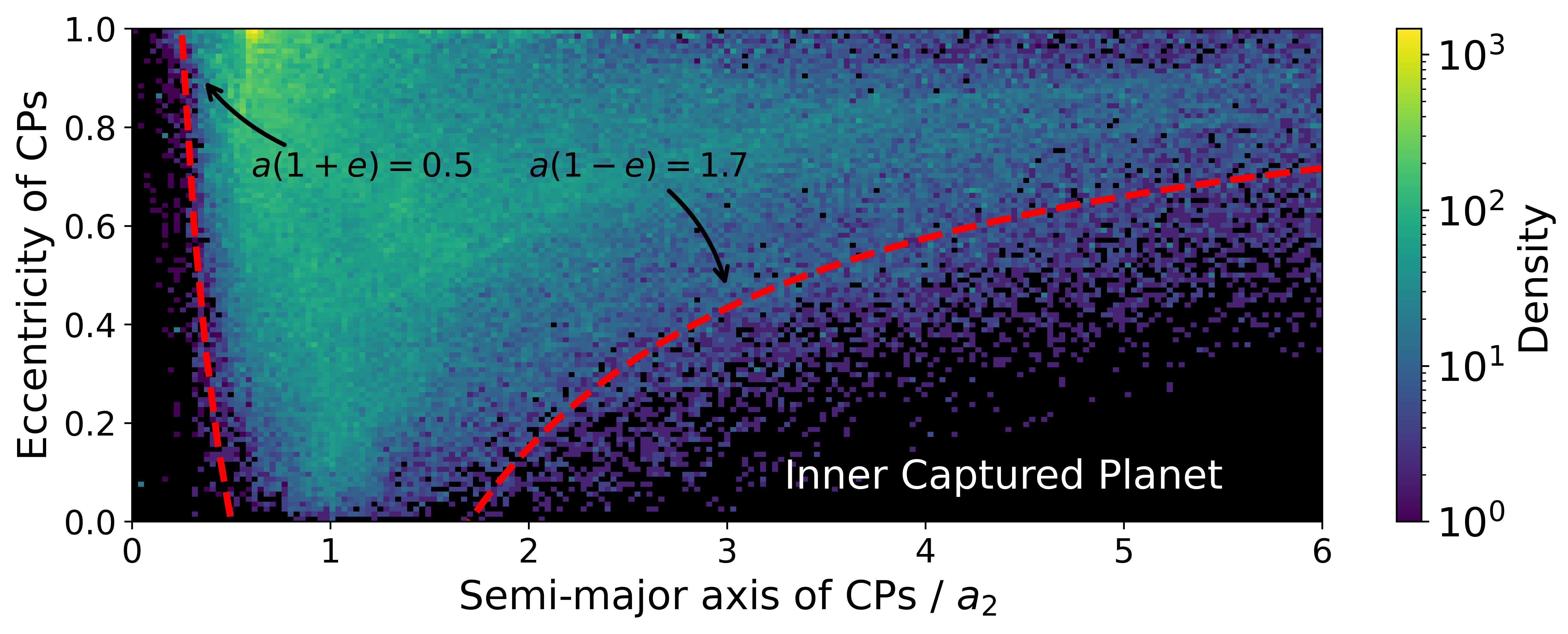}
    \end{minipage}
    \hfill
    \begin{minipage}[b]{0.49\linewidth}
        \centering
        \includegraphics[width=\linewidth]{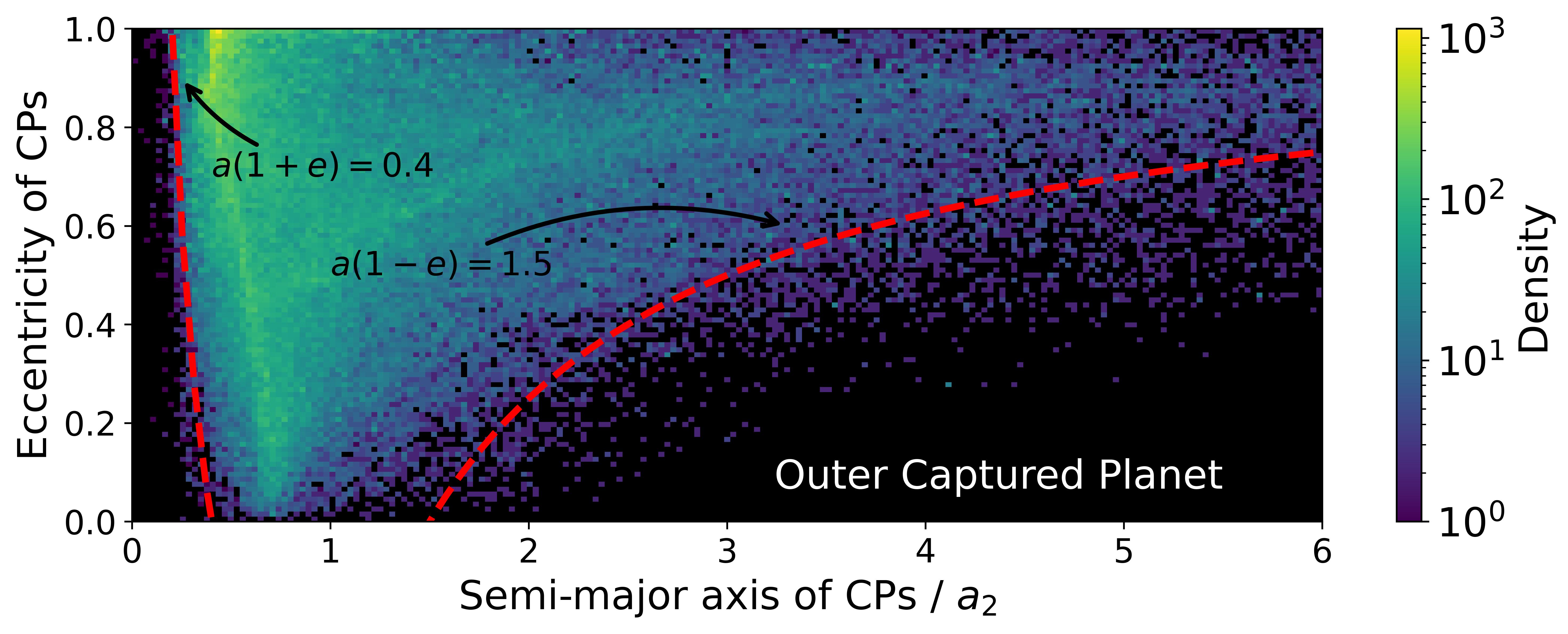}
    \end{minipage}
    \caption{
    Same as Fig.~\ref{Figs: Captured planet orbital parameter}, except for the runs with $V_\infty=0.1 v_{2i}$.}
    \label{Figs: Smallv Captured planet orbital parameter}
\end{figure*}

In Section~\ref{sec: additional runs_v}, we discuss the results for high-velocity flybys ($V_\infty = v_{2i}$). Here we show the results for low-velocity flybys ($V_\infty = 0.1v_{2i}$) in Figs.~\ref{fig:Sv_fraction_outcomes}-\ref{Figs: Smallv Captured planet orbital parameter}.

\bibliographystyle{aasjournal}
\bibliography{sample631}{}
\end{document}